\documentclass[aps,noshowpacs,nofootinbib,superscriptaddress,notitlepage]{revtex4-1}                                
\usepackage{xspace}
\usepackage{graphicx}
\usepackage[top=2cm,bottom=3cm]{geometry}
\usepackage[svgnames]{xcolor}
\usepackage[colorlinks=true,linkcolor=DarkBlue,citecolor=DarkBlue]{hyperref}
\usepackage{rotating}
\usepackage{units}
\usepackage{amsmath}
\usepackage{lineno}
\usepackage{listings} 
\usepackage[normalem]{ulem}
\usepackage{multirow}
\usepackage[section]{placeins}
\usepackage{SIunits}
\usepackage{hepunits}
\usepackage{hepparticles}
\usepackage{cancel}
\usepackage{hepnames}
\usepackage{epstopdf}
\usepackage{mathtools}
\usepackage[capitalise]{cleveref}
\usepackage{braket}
\usepackage{slashed}
\usepackage{url}
\usepackage{subfigure}
\usepackage[normalem]{ulem}

\newcommand{\nova}{NOvA\xspace}
\newcommand{\mb}{MiniBooNE\xspace}
\newcommand\nuance{\textsc{nuance}}
\newcommand{\minerva}{MINERvA\xspace}

\newcommand{\qq}{\ensuremath{Q^{2}}\xspace}
\newcommand{\enuqe}{\ensuremath{E_{\nu}^{\textrm{QE}}}\xspace}
\newcommand{\qqqe}{\ensuremath{Q^{2}_{\textrm{QE}}}\xspace}

\newcommand{\mares}{\ensuremath{M^{\mbox{\scriptsize{RES}}}_{\textrm{A}}}\xspace}
\newcommand{\maqe}{\ensuremath{M_{\textrm{A}}^{\mbox{\scriptsize{QE}}}}\xspace}

\newcommand{\ED}[1]{\textbf{\textcolor{blue}{[Edit: #1]}}} 

\begin{document}

\title{Comparisons and challenges of modern neutrino scattering experiments (TENSIONS2016 report)}

\date{\today}

\author{M. Betancourt}\affiliation{Fermi National Accelerator Laboratory, Batavia IL, USA}
\author{S. Bolognesi}\affiliation{IRFU, CEA Saclay, Gif-sur-Yvette, France}
\author{J. Calcutt}\affiliation{Michigan State University, Department of Physics and Astronomy, East Lansing, MI, 48824, USA}
\author{R. Castillo}\affiliation{Fermi National Accelerator Laboratory, Batavia IL, USA}
\author{A. Cudd}\affiliation{Michigan State University, Department of Physics and Astronomy, East Lansing, MI, 48824, USA}
\author{S. Dytman}\affiliation{University of Pittsburgh, Department of Physics and Astronomy, Pittsburgh, PA 15260, USA}
\author{B. Eberly}\affiliation{SLAC National Accelerator Laboratory, Menlo Park, California, USA}
\author{A.P. Furmanski}\affiliation{University of Manchester, School of Physics and Astronomy, Manchester, UK}
\author{R. Fine}\affiliation{University of Rochester, Department of Physics and Astronomy, Rochester, New York, USA}
\author{J. Grange}\affiliation{Argonne National Laboratory, Argonne, Illinois 60439, USA}
\author{L. Jiang}\affiliation{University of Pittsburgh, Department of Physics and Astronomy, Pittsburgh, PA  15260, USA}
\author{T. Katori}\affiliation{Queen Mary University of London, School of Physics and Astronomy, London, United Kingdom}
\author{J. Kleckner}\affiliation{University of Pittsburgh, Department of Physics and Astronomy, Pittsburgh, PA  15260, USA}
\author{J. Kleyklamp}\affiliation{University of Rochester, Department of Physics and Astronomy, Rochester, New York, USA}
\author{K. Mahn}\affiliation{Michigan State University, Department of Physics and Astronomy, East Lansing, MI, 48824, USA}
\author{B. Messerly}\affiliation{University of Pittsburgh, Department of Physics and Astronomy, Pittsburgh, PA  15260, USA}
\author{G. Perdue}\affiliation{Fermi National Accelerator Laboratory, Batavia IL, USA}
\author{L. Pickering}\affiliation{Imperial College London, Department of Physics, London, United Kingdom}
\author{J. P. Stowell}\affiliation{University of Sheffield, Department of Physics and Astronomy, Sheffield, United Kingdom}
\author{J. Sobczyk}\affiliation{University of Wroc\l{}aw, Institute of Theoretical Physics, Wroc\l{}aw, Poland}
\author{N. Suarez}\affiliation{University of Pittsburgh, Department of Physics and Astronomy, Pittsburgh, PA  15260, USA}
\author{H. Tanaka}\affiliation{University of Toronto, Department of Physics, Toronto, Ontario MSS 1A1, Canada}
\author{R. Tayloe}\affiliation{Indiana University, Bloomington, Indiana 47405, USA}
\author{R. T. Thornton}\affiliation{Indiana University, Bloomington, Indiana 47405, USA}
\author{M. Wilking}\affiliation{State University of New York at Stony Brook, Department of Physics and Astronomy, Stony Brook, New York, USA}
\author{C. Wilkinson}\affiliation{University of Bern, Albert Einstein Center for Fundamental Physics, Laboratory for High Energy Physics (LHEP), Bern, Switzerland}
\author{C. Wret}\affiliation{Imperial College London, Department of Physics, London, United Kingdom}
\author{G. P. Zeller}\affiliation{Fermi National Accelerator Laboratory, Batavia IL, USA}

\begin{abstract}
Over the last decade, there has been enormous effort to measure neutrino interaction cross sections important to oscillation experiments. However, a number of results from modern experiments appear to be in tension with each other, despite purporting to measure the same processes. The TENSIONS2016 workshop was held at University of Pittsburgh July 24-31, 2016 and was sponsored by the Pittsburgh High Energy Physics, Astronomy, and Cosmology Center (PITT-PACC).  The focus was on bringing experimentalists from three experiments together to compare results in detail and try to find the source of tension by clarifying and comparing signal definitions and the analysis strategies used for each measurement. A set of comparisons between the measurements using a consistent set of models was also made.  This paper summarizes the main conclusions of that work.
\end{abstract}

\maketitle

{\it Corresponding Author:} S. Dytman (dytman@pitt.edu)

\tableofcontents

\newpage
\section{Introduction} 
\label{intro}
\begin{figure}
\begin{center}
 \includegraphics[width=0.55\textwidth]{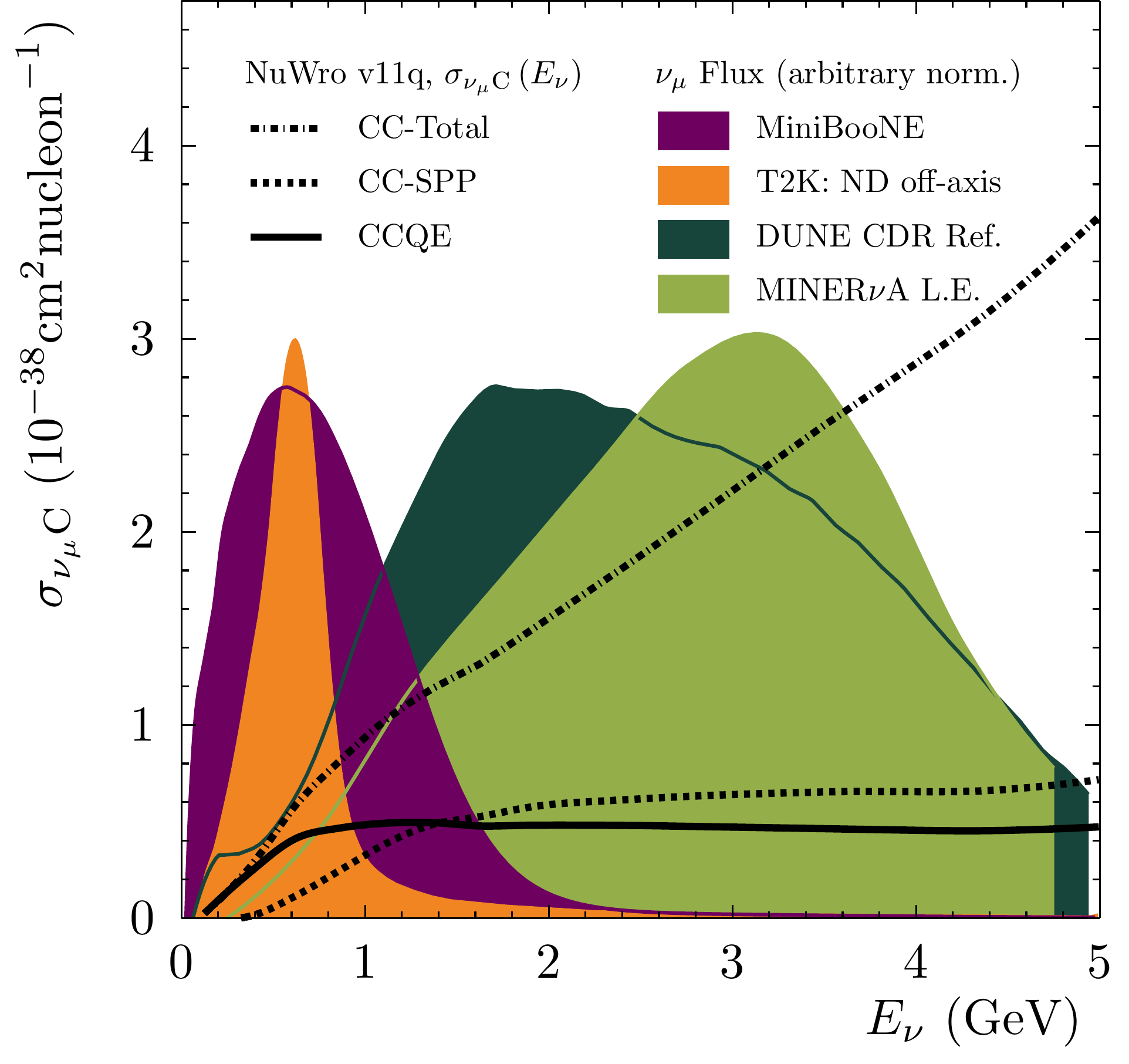}
\end{center}
\caption{NuWro~\cite{Golan:2012wx} $\nu_{\mu}-^{12}$C CC-inclusive cross section prediction, broken down into constituent modes, and compared with fluxes from current and planned neutrino oscillation~\cite{t2k-flux,Acciarri:2015uup,mb-flux} and cross-section~\cite{Aliaga:2016oaz} experiments.} 
\label{fig:xsec_vs_flux} 
\end{figure}

Current and planned neutrino oscillation experiments operate in the 0.1--10 GeV energy range, and use heavy nuclear targets (typically $^{12}$C, $^{16}$O or $^{40}$Ar). This energy range corresponds to a difficult ``transition region'' in the neutrino interaction cross section.  Interactions include (quasi-)elastic which occur on a constituent nucleon within the nucleus at lower energies and excitation of a nucleon resonance at intermediate energies which leads to meson production.  At higher energies, interactions occur predominantly on constituent quarks within the nucleons.  Numerous results from \mb~\cite{mb-nim}, \minerva~\cite{Aliaga:2013uqz}, and T2K~\cite{Abe:2011ks,Assylbekov:2011sh} are published or in progress.  A recent review~\cite{Alvarez-Ruso:2017oui} discusses the overall situation at length.  Figure~\ref{fig:xsec_vs_flux} shows the flux distributions from the \minerva cross section experiment and the T2K and \mb oscillation/cross section experiments used in this study and the planned DUNE experiment, compared to the charged current (CC) cross section taken from the NuWro neutrino interaction generator~\cite{Golan:2012wx}. The NuWro prediction is also shown broken down into exclusive channels, or modes, which have strong neutrino energy dependence. It is clear that modeling the transition between these interaction modes is important for neutrino oscillation experiments, which all have broad neutrino flux distributions.

At lower energies, where interactions occur with a nucleon rather than with a constituent quark, a number of nuclear effects complicate the modeling problem further. The initial momentum of nucleons within the nucleus and the energy required to liberate them are not negligible compared to the momentum and energy transfer in the interaction, so cannot be neglected. Furthermore, additional nuclear screening effects, or additional interactions where the initial hard scatter is with more than one nucleon further complicate the problem. 

Finally, Final State Interaction (FSI) effects modify the outgoing particle content and kinematics through subsequent interactions as the initial interaction products propagate through the dense nuclear medium. Because of FSI, it is not possible to unambiguously separate interaction modes in an experiment, it is only possible to measure the post-FSI final state particle content. For example, a charged current quasi-elastic (CCQE) interaction, $\nu_l + n \rightarrow l^{-} + p$, may only produce a single visible charged lepton in a detector, and is indistinguishable from a single pion production interaction where the pion is absorbed in the nucleus. As a result, defining a measurement as CC0$\pi$ (one lepton, no final state pions) may more accurately represent what is being measured. However, CC0$\pi$ is a more complicated signal definition, which may be more challenging for theorists to reproduce, not least because they have to include a model for all channels that can contribute through FSI, and for the FSI processes themselves. Additionally, different detector technologies have different capabilities, most notably due to hadron energy thresholds.  What is visible in one experiment may not be in another and a different signal definition may be required.  Requiring a proton in the signal can have significant consequences in backgrounds encountered and physics to which the measurement is sensitive.  Additionally, there may be holes in detector acceptance which may prevent background processes from being vetoed.


A number of recent results from \mb, \minerva, and T2K are all relevant to our current understanding of CCQE, but currently do not seem to present a consistent picture in the literature, a primary motivation for the current work. \mb provided the first high statistics measurements for charged current $\nu_\mu$~\cite{AguilarArevalo:2010zc} and $\overline{\nu}_\mu$~\cite{Aguilar-Arevalo:2013dva} and $E_\nu \sim$1 GeV, which first indicated a problem with CCQE, or CCQE-like modeling, as both shape and magnitude of the total CCQE cross section was in conflict with earlier NOMAD results~\cite{NOMAD_CCQE} at higher energies. Subsequent \minerva results at intermediate energies without protons~\cite{Fiorentini:2013ezn,Fields:2013zhk} and with protons~\cite{Walton:2014esl,Betancourt:2017uso} did not fully support the increase in the magnitude of the cross section seen by \mb. Although the \mb discrepancy was not settled, new high quality data at a broad range in $Q^2$ became available, motivating systematic investigations into the available models. In Ref.~\cite{Wilkinson:2016wmz}, members of the T2K collaboration used a number of models for CCQE and $2p2h$ processes which had been implemented in the NEUT generator, and fit the available $QE$-like data from \mb and \minerva. The two experiments were in considerable tension in the context of the models tested, indicating disagreement between the datasets. Adding to the confusion, both results seem to be consistent with a need for $2p2h$ with the preliminary calculations available.  Subsequently, T2K has published new measurements of CC0$\pi$~\cite{Abe:2016tmq} which uses a series of calculations (described in this document) to show the importance of $2p2h$. These measurements, and the apparent tension between them, are discussed in Section~\ref{sec:qe}.  Due to time constraints, we focus on the charged current $\nu_\mu$ results.

Resonant pion production processes are the dominant interaction process for energies above $E_\nu \approx 2$ GeV, and contribute significant strength to the oscillation analysis signal for MINOS~\cite{Adamson:2014vgd}, \nova~\cite{Adamson:2017qqn}, and DUNE~\cite{Acciarri:2015uup}. Measuring pion production cross sections in modern experiments poses similar challenges as the CCQE case; presently, CC$\pi$ experiments are based on detection of one or more pions and CCQE experiments are based on absence of pions.  The pion detected can be the result of resonant, nonresonant, deep inelastic, coherent, or FSI processes~\cite{Alvarez-Ruso:2017oui}, though at the neutrino energies of interest, $\Delta(1232)$ resonance production is dominant.

Most experiments measure single pion production processes where FSI processes significantly distort the outgoing pion kinematic distributions. Like the QE case, \mb published the first modern results for CC1$\pi^+$~\cite{MB_1pi} and CC1$\pi^0$~\cite{AguilarArevalo:2010xt} and \minerva have published measurements for CC1$\pi^\pm$  and CCN$\pi^\pm$ production with neutrino~\cite{Eberly:2014mra,McGivern:2016bwh} and CC1$\pi^0$ production with anti-neutrinos~\cite{Le:2015,McGivern:2016bwh}.  
In addition, T2K has published a result for $\pi^+$ from $H_2O$~\cite{Abe:2016aoo}.
However, as investigated by NuWro~\cite{Sobczyk:2014xza}, GENIE and MINERvA~\cite{Eberly:2014mra}, there are strong discrepancies in both shape and normalization between the MiniBooNE and MINERvA  $\nu_\mu$CC1$\pi^+$ production results. Attempts to reconcile this disagreement, discussed in various publications, did not point to any clear explanation~\cite{Eberly:2014mra,Sobczyk:2014xza,Wilkinson:2016wmz,gibuu_minervapi_15,gibuu_1picomp_17,Alvarez-Ruso:2017oui,Abe:2016aoo,Lalakulich:2013,Hernandez:2013,Rodrigues:2014jfa}.  These data sets are examined in Sect.~\ref{sec:1pi}.




Monte Carlo event generators are an important part of every experiment. They provide (in principle) a model for all possible interactions for a given measurement with which analyses can be developed, provide a way to estimate the efficiency for the signal selection, and provide a model for background processes, and tools to estimate systematic uncertainties on the models. Experimental publications compare results with theoretical models, although this is often limited to those available in the generator used to analyze the data. As experiments use different generators, comparison between experiments can be difficult. Because of the choice each experiment makes for the event generator, they must assess systematic errors resulting from the resulting model dependence.  The NUISANCE software~\cite{Stowell:2016jfr} has been developed to facilitate comparisons between all neutrino interaction generators and the available data, and has been used to make the model comparisons shown in this work. An introduction to the generators considered is given in Section~\ref{sec:generator}. Comparisons with both older and recent Monte Carlo codes are presented in Sections ~\ref{sec:qe} and ~\ref{sec:1pi}.

This report summarizes work done during the TENSIONS2016 workshop, which was held at the University of Pittsburgh July 24--31 2016, and was sponsored by the Pittsburgh High Energy Physics, Astronomy, and Cosmology Center (PITT-PACC). The workshop focused on possible reasons for the apparent disagreements between the measurements made in the QE-like and single pion production channels. It had a novel format where very few formal talks were scheduled and the emphasis was on producing ways to compare results.  This format was new to neutrino physics and the experiences are part of the report.  The primary aims of the workshop were to: clarify the signal definition, methodology and models used by each experiment; explore previously unappreciated model dependence in the results; and prepare new comparisons between interaction models to each experiment unavailable to the collaborations at the time the measurement was made. Members of three collaborations (T2K, \minerva, \mb) and experts from the generator community attended the workshop. Each experiment provided supplementary simulation files which included new information about the efficiency and selection cuts used for each measurement. Each experiment also provided their neutrino interaction flux information, which was used to generate various MC samples corresponding to the models used for each analysis, as well as a suite of new models unavailable at the time of the respective publications.

This work is somewhat unique because it is not a broad review paper.  Instead, the goal is to examine a particular set of problems in detail, giving significant background on each experiment and Monte Carlo event generator and discussing the results in a broader format.  Various recommendations are made for future experiments.

The structure of this report is as follows. The generator models used in the workshop studies are described in Section~\ref{sec:generator}.  The various QE-like and single pion production measurements considered here are reviewed in Sections~\ref{sec:qe} and~\ref{sec:1pi} respectively. The analysis approach, key details, and references are discussed in these sections. Emphasis is put on the details which were not appreciated or easily found prior to the workshop by those not on the experiment.  Sections~\ref{sec:compareqe} and~\ref{sec:compare1pi} discuss the role of selection cuts in each measurement and assess the differences and overall compatibility between measurements QE-like and single pion production measurements, respectively. The critical role of generators in efficiency estimation and possible impact on the experimental results is discussed.  New comparisons between data and generator models are shown in Sections~\ref{sec:qemod}, \ref{sec:qecomp}, and \ref{sec:1picomp}.  Finally, the workshop is summarized in Section~\ref{sec:summary}, which includes a discussion of issues raised by both the QE-like and single pion investigations, and comments on logistical problems encountered by the workshop and participants.

\section{Generator summary}
\label{sec:generator}

In this section, the generator models compared to data later in this work are briefly described. All generators which are commonly found in the current cross section literature are considered. The most widely used generators, GENIE~\cite{Andreopoulos:2009rq} and NEUT~\cite{neut}, are primarily maintained by experimentalists in addition to their normal work.  Generator models are taken from theoretical papers; the process can be greatly expedited when the theory model authors can offer advice.  Generator authors often make similar model choices, but the implementation can be very different because of parameter and model adjustments; Ref.~\cite{boydModelComparisons} includes many plots showing these problems. Because these generators are used for full detector simulations by experiments, emphasis is necessarily put on covering all regions of phase-space, and having many model choices available, even if this introduces some inconsistency. The necessity of including systematic uncertainties for experimental applications also makes it a challenge to include sophisticated new models quickly. \nuance~\cite{CASPER2002161} falls in the same category, although it is no longer actively maintained, and as such has not been used in recent years. GiBUU~\cite{gibuu_review_11} and NuWro~\cite{Golan:2012wx} have better theoretical bases, and are therefore more consistent models.  They are frequently used to benchmark the less sophisticated generators. NuWro typically introduces new interaction level models quickly, and NEUT and GENIE often benefit by migrating these new models into their frameworks. NuWro is being updated for use in full detector simulations, e.g. a parameter reweighting framework has been recently introduced~\cite{Pickering:2016icq}.  GiBUU differs from the other generators primarily by the treatment of the nuclear potential and intranuclear particle transport model. Detailed discussion of the models implemented in each generator can be found in the following subsections.

Model choices are briefly summarized for QE in Table~\ref{tb:qemodels}. Many event generators start with the Llewellyn-Smith~\cite{LlewellynSmith:1971uhs} model for free nucleons, and the Smith-Moniz Relativistic Fermi gas model~\cite{Smith:1972xh} for bound nucleons. Both are easy to implement in generators, and form factors can be modified as needed. Recently, generators have included local Fermi gas (LFG) momentum distributions, which are more realistic.  Note that some generators use the Llewellyn-Smith formalism for bound nucleons, drawing the initial state nucleon from a relativistic Fermi gas distribution, which is functionally identical to the Smith-Moniz model.  

Using the local Fermi gas nuclear model, the Valencia group has constructed a consistent model for QE-like interactions. The QE models by Nieves {\it et al.}~\cite{Nieves:2004wx,nieves_2011}, includes long-range nucleon-nucleon (RPA) correlations and Coulomb effects for the outgoing muon on single-nucleon ($1p1h$ or true QE) and multi-nucleon ($2p2h$) interactions.
The Valencia multi-nucleon interaction ($2p2h$) model~\cite{Nieves_2p2h_14} is widely used because Gran and Sanchez~\cite{Gran:2013kda} studied its features and application with the theory authors. It is included as a distinct interaction channel which explicitly incorporates additional nucleons in the final state.  Broader applicability is gained by deleting events for which $q_{3} > 1.2\mbox{ GeV}$, where $q_3$ is the magnitude of the momentum transfer. Final state nucleons are distributed via phase space~\cite{Sobczyk:2012ms}; the isospin decomposition of the nucleons in the final is an ongoing problem.  Although some theoretical models have interference between 1- and 2-body currents, this is not true for this model.

\begin{table} [h]
\centering
\label{tb:qemodels}
\caption{Brief summary of QE-like models implemented in various generators. LS denotes the Llewellyn-Smith model~\cite{LlewellynSmith:1971uhs}; S-M denotes the Smith-Moniz~\cite{Smith:1972xh} model; and RPA stands for random phase approximation, an effect due to long-range nucleon-nucleon correlations.}
{\renewcommand{\arraystretch}{1.2}
\begin{tabular} {cccccc}
\hline\hline
Generator & Nuclear & QE & \maqe (GeV) & $2p2h$ NN  & Long range NN\\
 & model & model & & correlations & correlations \\
\hline\hline
GENIE v2.6.3 & RFG & LS & 0.99 & none & none\\
GENIE v2.8.6 & RFG & LS & 0.99 & none & none\\
NUANCE v3 & RFG & S-M & 1.23 &  $\pi$-less $\Delta$\cite{oset_1987} & none \\
NEUT v5.1.4.2 & RFG & S-M & 1.48 & $\pi$-less $\Delta$~\cite{oset_1987,singh_1998} & none\\
NuWro 11q     & LFG & LS & 1.03 & Nieves~\cite{nieves_2011,Gran:2013kda} & RPA~\cite{nieves_2011} \\
GiBUU         & LFG+ & \cite[\S III A]{gibuu_interaction_model_06} & 1.03 & Ref.~\cite{gibuu_2p2h_16} & not explicit~\cite{gibuu_2p2h_16} \\
GENIE v2.12.0alt & LFG & Nieves~\cite{Nieves:2004wx} & 1.05 & Nieves~\cite{nieves_2011,Gran:2013kda} & RPA~\cite{Nieves:2004wx}\\
NEUT v5.3.6 & RFG & S-M & 1.21 & Nieves~\cite{nieves_2011,Gran:2013kda} & RPA~\cite{nieves_2011} \\
\hline\hline
\end{tabular}}
\end{table}

\begin{table}[h]
\centering
\label{tb:1pimodels}
\caption{Generator pion production models.  R-S means Rein-Sehgal~\cite{Rein-Sehgal}, B-Y means Bodek-Yang~\cite{Yang:2009zx}, and B-S is Berger-Sehgal~\cite{Berger-Sehgal}.}
{\renewcommand{\arraystretch}{1.2}
\begin{tabular} {ccccccc}
\hline\hline
\multirow{2}{*}{Generator} & \multirow{2}{*}{\parbox{2cm}{Resonance model}} & \multirow{2}{*}{\parbox{1.2cm}{\mares (GeV)}} & \multirow{2}{*}{\parbox{2cm}{$\Delta$ angular distribution}} & \multirow{2}{*}{\parbox{1.4cm}{$W$ limit (GeV)}} & \multirow{2}{*}{\parbox{1.5cm}{Nonresonant model}} & \multirow{2}{*}{\parbox{1.5cm}{$\pi$-FSI model}} \\
& & & & & & \\
\hline\hline
GENIE v.2.6.3 & R-S & 1.23 & Isotropic & 1.7 & Scaled B-Y~\cite{bodek-yang} & Data-driven~\cite{Andreopoulos:2009rq}\\
NEUT v5.1.4.2 & R-S & 1.41 & Isotropic & 2.0 & R-S I=1/2 & Salcedo-Oset~\cite{Salcedo-Oset}\\
NuWro & $\Delta$-only~\cite{Graczyk:2009qm} &  0.94 & ANL/BNL~\cite{Sobczyk:2014xza}& 1.6 & Scaled B-Y & Salcedo-Oset~\cite{Salcedo-Oset}\\
GENIE v2.12.0alt & B-S  & 1.23 & R-S & 1.7 & Scaled B-Y & Data-driven~\cite{Andreopoulos:2009rq}\\
NEUT v5.3.6 & B-S & 0.95 & Anisotropic & 2.0 & R-S I=1/2 & Salcedo-Oset~\cite{Salcedo-Oset}\\
\nuance & R-S & 1.10 & Anisotropic & 2.0 & R-S & Custom~\cite{CASPER2002161} \\
GiBUU & Ref.~\cite{[\S III A]{gibuu_interaction_model_06}} & N/A & Isotropic & 2.0 & Ref.~\cite{gibuu_interaction_model_06} & Ref.~\cite{gibuu_1picomp_17}\\
\hline\hline
\end{tabular}}
\end{table}

Model choices are briefly summarized for single pion production in Table~\ref{tb:1pimodels}. At the core of most generator resonance models is the Rein-Sehgal~\cite{Rein-Sehgal} (R-S) model, which was developed to model neutrino interactions over 30 years ago, and is convenient to implement in generators.  The resonance parameters have changed significantly as the data improved.  It uses a non-relativistic quark model~\cite{FKR} to derive helicity amplitudes to produce resonances, and then models the subsequent decay of those resonances. Berger-Sehgal~\cite{Berger-Sehgal} updated the R-S model to include effects due to lepton mass. Resonance parameters such as masses, decay widths, and form factors have been updated by all generator groups. GiBUU, which has a resonance model grounded in electron-scattering measurements~\cite{MAID_07} and NuWro~\cite{Graczyk:2009qm} have done independent fits to the nucleon data. There are a variety of ways to describe nonresonant pion production~\cite{Alvarez-Ruso:2017oui}; strength can come from the tail of DIS process which we label as scaled B-Y (Bodek-Yang~\cite{bodek-yang}) or via low order diagrams~\cite{nieves_2011}.  The scaled B-Y choice uses a factor that decreases model strength to get agreement with data; it then includes both resonant and nonresonant processes.  Interference between resonant and nonresonant amplitudes is difficult for all generators and ignored by most simulations.  It is difficult to describe these effects accurately.  Final state interactions are often modeled using the Salcedo-Oset~\cite{Salcedo-Oset} intranuclear cascade model, tuned to pion--nucleus and nucleon--nucleus scattering data. It includes medium corrections via a local density approximation. Again the exception is GiBUU, which has a more sophisticated treatment introduced below. Versions of GENIE code used here have a data-driven model~\cite{Andreopoulos:2009rq} which has no medium dependence.

\subsection{GiBUU overview } 

Unlike the other interaction simulations used here, GiBUU~\cite{gibuu_review_11} 
aims to simulate a large number of different nuclear processes ($\gamma\textrm{A}$, $\textrm{p}\textrm{A}$, $\pi\textrm{A}$, $\textrm{e}\textrm{A}$, $\textrm{A}\textrm{A}$, and most recently, $\nu\textrm{A}$ scattering) over a wide range of energies with a single, consistent physics model.
At the time of writing, the latest update to the neutrino interaction model was `GiBUU 2016', and is described in detail in Ref.~\cite{gibuu_2p2h_16}; many comparisons with recent data can be found therein.  The most comprehensive description of the model can be found in Ref.~\cite{gibuu_review_11}.  Extensive comparisons to MINER$\nu$A, \mb, and T2K pion-production measurements can be found in Ref.~\cite{gibuu_1picomp_17}.

The initial state nucleon momentum distribution is modeled as a local Fermi gas---the momentum distribution is a function of the local nuclear density at the particle's initial position~\cite[\S III A]{gibuu_interaction_model_06}. The nuclear density parameterization follows the calculation in Ref.~\cite{nuclear_density_param_93}, with values for specific nuclei taken from Ref.\cite{nuclear_density_data_74}. The initial density distribution---used to calculate the nuclear potential---is modified to achieve a constant removal energy across the nucleus~\cite[\S II A]{gibuu_2p2h_16}. 


The impulse approximation is used to model true charged current quasi-elastic (CCQE) neutrino interactions with single nucleons. A standard dipole form is used for the axial component, while the BBBA07~\cite{BBBA07} parameterization is used for the vector form factors. The calculation also 
applies a density-dependent Pauli-blocking of interactions based on the kinematics of the final state nucleons~\cite[\S II A]{gibuu_interaction_model_06}.
  
The non-QE component of the `true' CC0$\pi$ is motivated through fits to electron-scattering data~\cite{bosted_mec_12} and
includes contributions from short- and long-range $2p2h$ effects as well as RPA correlations~\cite[\S II C 1]{gibuu_2p2h_16}.
This empirical approach to nuclear effects is somewhat different than used in the other interaction simulations, which are based on microscopic $2p2h$ and RPA calculations.
  
While single pion production is dominated by the $\Delta$ resonance, the GiBUU model includes electroweak production of 13 higher nucleon resonances as well as a non-resonant 1$\pi$ `background' channel.
The vector component is determined from the MAID model for electro-production~\cite{MAID_07} and the axial components are related to these via an assumption of Partially Conserved Axial Current, or PCAC.
The GiBUU model contains no contribution from coherent pion production; the implications for comparisons to the \minerva data set are discussed in Ref.~\cite[\S III. A. 1]{gibuu_minervapi_15}.
Single pion production occurs for invariant masses ($W$) from the $W = \pi+N$ threshold up to about $W = 2$ GeV, at which point the DIS model takes over.


Final state baryons and mesons are propagated via a semi-classical cascade that includes particle decay, and two and three body reactions.
In contrast to the standard cascade model, tracked hadrons are transported simultaneously and can react with each other as the simulation progresses.
While the sophistication of the transport model is a key strength of GiBUU, a detailed description is far beyond the scope of this document. For details, the reader is directed to Ref.~\cite{gibuu_review_11}.

The model choices made here are informed by the latest GiBUU analyses \cite{gibuu_2p2h_16, gibuu_1picomp_17}.
However, unlike the majority of previous GiBUU analyses, the model used in Ref.~\cite{gibuu_1picomp_17} and here does not include collisional broadening of the $\Delta$ resonance decay width~\cite{coll_broad_87}.
Although this choice reduces the pion production rate, recent pion electro-production data disfavors this model component\footnote{Private communication with U. Mosel}.



\subsection{GENIE overview} 
GENIE~\cite{Andreopoulos:2009rq} evolved from Neugen, the primary event generator for the MINOS experiment~\cite{Gallagher:2002sf}. The work described here uses three versions: GENIE 2.6.3, GENIE 2.8.6, and 2.12.0alt. Version 2.6.3 has the same models as v2.6.2 which used for almost all \minerva measurements in low energy mode (and for all those considered in this work), but works on more modern operating systems.  Version 2.8.6 featured improvements to the FSI model and some bug fixes. Version 2.12.0 supplies a large number of alternate models.  The models used for these studies are described below; this set of models will be labeled GENIE v2.12.0alt\footnote{The user can build this version by choosing a configuration with the model set given above rather than the default models.}.

In GENIE, the Bodek-Ritchie~\cite{BodekRitchie} Relativistic Fermi gas (RFG) nuclear model, which includes short range nucleon-nucleon correlations, is used in all default code versions as of December, 2017. The local Fermi Gas (LFG) is used as the nuclear model in all v2.12.0alt comparisons shown in this work, which has a more physical initial state nucleon momentum distribution, and the removal energy depends on the local nuclear density. 

The Llewellyn-Smith model~\cite{LlewellynSmith:1971uhs} is used for primary CCQE processes for all default GENIE versions. A dipole axial form factor, and BBBA07 vector form factors\cite{BBBA07} are used for all versions. For nuclear targets, Pauli-blocking is applied, based on the requirement that the momentum of the outgoing nucleon exceed the Fermi momentum $k_F$ for the nucleus in question. The CCQE model in GENIE v2.12.0alt uses the full Nieves QE model, which includes Random Phase Approximation (RPA) in-medium propagator effects and Coulomb effects\cite{nieves_2011}.  The $2p2h$ process is not included in v2.6.3 or v2.8.6 of GENIE. In v2.12.0alt, the Valencia $2p2h$ model is used~\cite{nieves_2011, Gran:2013kda}; the implementation in GENIE is fully described in Ref.~\cite{Schwehr:2016pvn}.  

Although all GENIE resonance models are based on the Rein-Sehgal model~\cite{Rein-Sehgal}, a variety of changes have been implemented, e.g. regular updates for new resonance masses and widths.  For all versions, the effect of the lepton masses on the allowed region of phase space is taken into account.  v2.12.0alt fully includes Berger-Sehgal lepton-mass corrections\cite{Rein:2006di} and the pion-pole diagram.  In default versions, the axial and vector form factors are the modified dipole forms as in the original Rein-Sehgal model; in GENIE v2.12.0alt, the $\Delta$ form factors have been updated from fits to \mb data~\cite{MB_1pi}.  While the $\Delta\rightarrow\pi$ decay is isotropic for early versions of GENIE, v2.12.0alt uses the angular distribution from Rein-Sehgal~\cite{Rein-Sehgal}. Note that in all versions of GENIE, interferences between resonances are neglected, the non-resonant contribution to single pion production comes from scaled versions of the Bodek-Yang~\cite{bodek-yang} model, with hadronization described by the custom AKGY model~\cite{Yang:2009zx}.

GENIE has a unique FSI model~\cite{Andreopoulos:2009rq} which uses a single interaction to simulate the multiple steps in traditional cascade models.  This has been tuned to hadron-induced data for a wide range of nuclei.  GENIE v2.12.0alt includes a variation of this model that improves the $A$ dependence through data for a wide range of nuclei.  

\subsection{NEUT overview} 
NEUT~\cite{neut} was written for the Kamiokande detectors and remains
the primary event generator for T2K.
For this publication we used NEUT 5.3.6 and NEUT 5.1.4.2, the versions used for recent T2K publications.
NEUT uses the Smith-Moniz model to simulate quasi-elastic scattering assuming a relativistic Fermi gas with values for the binding energy and Fermi momentum taken from electron scattering data fits~\cite{Moniz:1971mt}. A dipole is used for the axial form factor, and the BBBA05 parameterizations~\cite{bbba05} of the vector form factors are used. Additionally, the Nieves model RPA correction~\cite{nieves_2011} is applied in NEUT v5.3.6.

Additional CCQE-like $2p2h$ interactions were simulated differently in the two versions of NEUT studied. In NEUT 5.1.4.2 the effect of $\pi$-less $\Delta$-decay interactions was modeled based on Ref.~\cite{oset_1987}, where 20\% of $\Delta$ resonances produced decay without producing a pion at the interaction vertex. This results in a reduction of the single pion production cross section, and an increase in the CCQE-like cross section. Additional nucleons were simulated at the vertex to model the $\pi$-less $\Delta$-decay events. In NEUT 5.3.6, the Valencia multi-nucleon interaction model~\cite{nieves_2011, Gran:2013kda} is included as a distinct interaction channel which explicitly includes additional nucleons in the final state. The cross-section for the Nieves model is interpolated from $p_{\mu}-\cos\theta_{\mu}$ lookup tables with a cut on $2p2h$ events with $q_{3} > 1.2\mbox{ GeV}$. Final state nucleons are simulated using a basic model described in Ref.~\cite{Sobczyk:2012ms}, which simply conserves energy and momentum.

For NEUT v5.3.6, an alternative CCQE-like model was considered, including the sophisticated spectral function model by Benhar {\it et al.}~\cite{Benhar:1999bg} for the initial state nucleon momentum and removal energy distribution and the Nieves $2p2h$ model. This model was not chosen because of the inability to fit the \mb and \minerva CC0$\pi$ data consistently~\cite{Wilkinson:2016wmz}.

Both v5.1.4.2 and v5.3.6 use the Rein-Sehgal model for the resonant processes. v5.1.4.2 uses a large $M_A^{res}$ to get agreement with data and has non-interfering nonresonant background from the original $I=1/2$ Rein-Sehgal model. 
v5.3.6 uses the Rein-Sehgal model~\cite{Rein-Sehgal,Rein} with Graczyk-Sobczyk form-factors~\cite{Graczyk:2009qm} and Berger-Sehgal lepton mass corrections~\cite{Berger-Sehgal}, including the additional pion-pole diagram.  It includes resonance-resonance interference terms 
and a rescaled version of the non-resonant contribution of v5.1.4.2.  


The ``free parameters'' of the model are the axial mass $\mares = 0.95$, the normalization of the $C_{5}^{A}$ form factor $C_{5}^{A}(0) = 1.01$, and a non-resonant $I_{1/2}$ scaling parameter $= 1.30$. 
These are tuned to ANL and BNL deuterium data~\cite{Wilkinson:2016wmz} using corrections from Ref.~\cite{Wilk-Rodr}. Uncertainties are artificially inflated to cover discrepancies in $\nu-A$ scattering data once a nuclear model is included.

The pion final state interactions for $p_{\pi} < 500\text{ MeV/c}$ use the Salcedo-Oset~\cite{Salcedo-Oset} cascade model with in-medium corrections from Seki et al~\cite{Seki}. For higher momentum pions, the interaction probabilities are extracted from $\pi-N$ scattering data. The pion interaction probabilities 
are tuned to match $\pi-N$ scattering data~\cite{t2k_2015}.

\subsection{NUANCE overview} 

The \nuance\ \textsc{v3} event generator~\cite{CASPER2002161} was updated for the MiniBooNE experiment.  About ten years ago, it was the best event generator at low neutrino energies.  
The MiniBooNE collaboration made extensive modifications to fit $\mathcal{O}\left(1\,\mathrm{GeV}\right)$ neutrino cross section data from \mb and other experiments.  This document will use this version of the code~\cite{AguilarArevalo:2010zc} and it will be labeled \textsc{v3(\mb)}.

The RFG nuclear model is used for all processes in \nuance, with binding energy and Fermi momentum parameters taken from fits to electron-scattering data~\cite{Moniz:1971mt}.
An empirical Pauli-blocking multiplicative factor ($\kappa$) was added to match \mb low momentum transfer CCQE data~\cite{AguilarArevalo:2007ab,AguilarArevalo:2010zc}.
The parameters $\maqe = 1.23\pm0.20$ GeV and $\kappa=1.019\pm0.011$ were set to the results found in Ref.~\cite{AguilarArevalo:2007ab} and not the values determined from Ref.~\cite{AguilarArevalo:2010zc}. 
The BBA03 parametrization~\cite{Budd:2003wb} was used to describe the non-dipole behavior of the vector form factors.
Additionally, $\pi$-less $\Delta$-decay is implemented, as was described for NEUT.  A fraction of $\Delta^+$ and $\Delta^0$ ($\Delta^++$ and $\Delta^-$) resonance production events 20\% (10\%) were added to the CCQE-like events~\cite{CASPER2002161}.

The Rein-Sehgal model~\cite{Rein-Sehgal} is used to simulate pion production for invariant masses $W \leq 2$ GeV, including interference between resonances and the non-interfering Rein-Sehgal non-resonant background contribution, as well as the Rein-Sehgal form factors. The resonant axial mass, \mares was tuned to the available neutrino data from the time, to give $M_A^{1\pi} = 1.10\pm0.28$ GeV~\cite{AguilarArevalo:2010zc}.

The final state interaction (FSI) probabilities are based on the nuclear density of carbon and derived from $\pi-N$ and $N-N$ data~\cite{Flaminio:1983cz,Flaminio:1983fw,Flaminio:1984gr}. Details of the pion absorption and intranuclear pion charge exchange can be found in Ref.~\cite{AguilarArevalo:2010zc}.


\subsection{NuWro overview} 

NuWro~\cite{Golan:2012wx} was started at University of Wroc\l{}aw and has become an important `sandbox' for other generators, introducing new theoretical models which are used for testing before being adopted by NEUT and GENIE.  NuWro simulations for this work were done with a default combination of models in version 11q. CCQE events use the Llewellyn-Smith~\cite{LlewellynSmith:1971uhs} model with BBBA05~\cite{bbba05} vector and dipole axial vector form factors.  Simulations shown here have RPA corrections included in CCQE events as described in Ref.~\cite{Graczyk:2003ru}.  The default version uses the Valencia $2p2h$ model.


For resonance events, only the $\Delta$ resonance is explicitly included with nucleon-$\Delta$ form factors taken from Ref.~\cite{Graczyk:2009qm} with free parameter
values $C^5_A(0)=1.19$. Non-resonant background is added incoherently as described in Ref.~\cite{Juszczak:2005zs}. 

In the DIS region ($W>1.6$~GeV) with the Bodek-Yang prescription~\cite{Bodek:2002vp}.  Hadronic final states are generated using PYTHIA fragmentation routines~\cite{Sjostrand:2006za}, with some modifications as described in Ref.~\cite{Nowak:2006sx}.  For DIS events, formation zone effects are included \cite{Golan:2012wx}.  In the region $W\in (1.4, 1.6)$~GeV, NuWro employs a linear transition between the resonance and the DIS cross sections using Bodek-Yang and Pythia.


 
In the NuWro cascade model pions and nucleons are propagated through the nucleus with a realistic density profile~\cite{Golan:2012wx}. 
Pion interactions are modeled using the SAID phase shift solution for experimental pion-nucleon observables with medium corrections according to Salcedo-Oset~\cite{Salcedo:1987md}.
Nucleon interactions are modeled using free nucleon-nucleon cross sections with in-medium modifications taken from Ref. \cite{Pandharipande:1992zz}.

\section{QE-like measurements}	
\label{sec:qe}


Neutrino beam experiments have many similarities to electron beam experiments. The main theoretical formalism uses $W^\pm, Z$ exchange for weak interactions and $\gamma$ exchange for electromagnetic interactions.  Therefore, there are many similarities in kinematic quantities used.  The exchanged boson has a 4-vector $q$ with energy ($q_0=\nu$) and 3-momentum ($|\vec{q}|=q_3$) components.  The invariant mass of the transferred boson is $Q^2= -|q_\mu|^2=q_3^2-q_0^2$ and the total hadronic mass in the rest frame is $W$.
Neutrino experiments use beams with broad distributions in neutrino energy, which means that the neutrino energy must be calculated for each event, as opposed to being known a priori, as in the case of electron-scattering experiments.  
All measurements discussed in this work are charged current (CC), i.e. exchanged $W^\pm$.  All experiments measure the muon energy and angle in the lab; a growing number of experiments measure the properties of the hadrons in the final state.  The beam energy ($E_\nu$), 4-momentum transfer ($Q^2$), and its components ($q_0, q_3$) must all be calculated from the measurements of the final state.  Model dependence is unfortunately an important part of every experiment and therefore an  important part of the studies in this report.  All experiments estimate a large set of systematic errors which are meant to account for these effects.  


\subsection{\mb} 


The \mb detector is a spherical mineral oil Cherenkov detector, where relativistic charged particles produced by a neutrino interaction are detected by photomultiplier tubes (PMTs) that cover the inner wall of the spherical chamber with 11\%~\cite{mb-nim} photocathode coverage. The \mb measurement considered in this workshop was the CCQE-like cross section for muon neutrino interactions~\cite{AguilarArevalo:2010zc}. The source was the Booster Neutrino Beam (BNB) at Fermilab, operating in neutrino-enhanced mode. The predominantly muon neutrino flux has a peak energy of about 700 MeV~\cite{mb-flux}. 

The analysis sample was selected by requiring a single, contained, and well-reconstructed muon, and no final state charged, or neutral, pions. Containment was enforced by requiring the observation of a decay Michel electron found near the end of the projected stopping point of the muon candidate, and no activity in the outer veto region. The detection efficiency is around 30\%; kinematic acceptance losses are dominated by the containment requirement on muon direction and energy. 
The detector is spherically symmetric, and the detector modeling and reconstruction algorithms were validated against the ubiquitous muon flux provided by cosmic rays both traversing and stopping inside the detector, as described in Ref.~\cite{Patterson:2009ki}. Final-state positively charged pions are tagged and excluded by the observation of an additional decay electron. Neutral pions decay promptly to two photons which produce Cherenkov rings, which are also rejected. However, the majority of protons are below Cherenkov threshold and so are not controlled in the analysis. Therefore, the definition for CCQE-like interactions in \mb analyses is a single muon, no pions, and any number of outgoing nucleons.

The most model independent result of the \mb CCQE-like measurement is flux-integrated double-differential cross sections with respect to muon production angle, $\theta_{\mu}$ and kinetic energy, $T_{\mu}$. The paper also presented a CCQE-corrected measurement, where the contribution of single pion interactions were removed according to the NUANCE simulation used in the analysis. We note that many theoretical comparisons are made with this CCQE-corrected sample, but it is more dependent on the NUANCE generator, so we strongly encourage future users of the data to use the less model dependent result for model comparisons. In particular, a basic model for pionless delta decay was subtracted, which is a key component of multinucleon neutrino interaction ($2p2h$) models, so comparisons of these models to the CCQE-corrected data is difficult to interpret.


\begin{description}
\item [Signal definition] 1 (negatively charged)  muon, any number of nucleons (neutrons or protons), no charged or neutral pions. Signal definition is defined for particles in the final state (after final state interactions (FSI)).  
\item [Observables] Two dimensional: muon kinetic energy, angle relative to neutrino beam direction.
\item [Flux] Ref.~\cite{mb-flux}, with digital version available from \mb data release page:~\cite{mb_datarelease}. Only neutrino interactions (not anti-neutrino or electron neutrino) were considered as signal, hence the charge requirement in signal definition.
\item [Target material] CH$_2$ (mineral oil)
\item [Default generator for analysis] \nuance\ \textsc{v3(\mb)}.
\end{description}

\subsection{\minerva} 
\label{sec:minerva_qe}
The \minerva detector sits on-axis in the NuMI beamline at Fermilab, and for all measurements considered in this report, ran with the ``low energy'' configuration of NuMI, which has a beam peaked at 3 GeV. The $\nu_{\mu}-$CH CCQE-like measurement~\cite{Fiorentini:2013ezn} described here used $9.42\times 10^{19}$ protons-on-target in neutrino-enhanced mode. Although the publication~\cite{Fiorentini:2013ezn} used an older flux calculation, the results here have been adjusted to use the updated flux~\cite{Aliaga:2016oaz}.  The cross section is given as a function of \qqqe.

\begin{equation} 
  \label{eq:ccqe-enuqe}
  \enuqe = \frac{2M'_{i}E_{\mu}-(M'^{2}_{i}+m_{\mu}^{2}-M^{2}_{f})}{2(M'_{i}-E_{\mu}+\sqrt{E_{\mu}^{2}-m_{\mu}^{2}}\cos{\theta_{\mu}})},
\end{equation}
\begin{equation}
  \label{eq:ccqe-q2qe}
  \qqqe = -m_{\mu}^{2}+2\enuqe(E_{\mu}-\sqrt{E_{\mu}^{2}-m_{\mu}^{2}}\cos{\theta_{\mu}}),
\end{equation}
where $E_\mu$ is the muon energy, $m_{\mu}$ is the muon mass, $M_i$ and $M_f$ are the initial and final nucleon masses, respectively, and $M'_{i} = M_{i} - V$ where $V = 34$ MeV is the binding energy of carbon assumed in the analysis.

The event selection for this analysis was as follows. First, events which originate within the \minerva detector fiducial volume are required to have a muon in the \minerva detector that matches spatially and temporally to a muon in the MINOS near detector, which is used for muon spectroscopy. This requirement imposes a geometrical restriction on the acceptance of events into the analysis sample. 
Next, a cut is made to exclude events with reconstructed neutrino energy outside of the signal range, 1.5 GeV$<\enuqe<$10 GeV, where \enuqe was defined in Equation~\ref{eq:ccqe-enuqe}. At the time of the analysis, the reference GENIE MC had no model for $2p2h$, so the signal efficiency was calculated for true, nucleon-level, GENIE CCQE only. In order to retain  $2p2h$ events in the selection, the analysis is blind to a region around the vertex wide enough to contain protons and neutrons (pions) with kinetic energy less than 225 (100) MeV. Outside this region, a maximum of two isolated groups of spatially contiguous energy deposition are allowed, which can be created either by protons exiting the region or through secondary interactions. A cut on the total calorimetric energy deposited outside the vertex energy region (the recoil energy) is implemented as a function of \qqqe to reject inelastic events, but retain 95\% of true CCQE events for all \qqqe bins. In the neutrino selection, the recoil energy must be $<0.05$ GeV at low \qqqe, $<0.410$ GeV at high \qqqe, and $<-0.05+0.64$ \qqqe--0.22 $\left(\qqqe\right)^{2}$ in between \footnote{In the antineutrino selection, the recoil energy must be $<0.03+0.3$ \qqqe for all \qqqe.}. Events where a pion is produced, but is absorbed through final state interactions are still observable by the energy transferred to the hadronic system, and are removed by the recoil energy cut. The analysis therefore rejects events where a pion is produced at the interaction vertex, but is unaffected by the presence of additional low-momentum protons. 
At the time of the analysis, there was no  $2p2h$ model available to \minerva; the general insight at the time was that  $2p2h$ would produce low-momentum protons, and so the analysis was designed to avoid sensitivity to that region. 
In this work, we compare CCQE+$2p2h$ models to the MINERvA dataset. An important caveat is that \qqqe is misreconstructed for  $2p2h$ events, which might result in a different efficiency through the sliding recoil energy cut, which has not been explicitly studied.

\begin{description}
\item [Signal definition used for generator comparisons] true CCQE+ $2p2h$ at the vertex, for $1.5 \leq E_{\nu} \leq 10$ GeV.
\item [Observables] one dimensional, \qqqe (defined in Equation~\ref{eq:ccqe-q2qe}).
\item [Flux] Ref.~\cite{Aliaga:2016oaz}, with digital version available from \minerva data release page~\cite{minerva_datarelease}.
\item [Target material] CH.
\item [Default generator for analysis] GENIE v2.6.2 (modern implementation is v2.6.3).
\end{description}

\subsection{T2K}\label{sec:t2kqe} 
The cross-section measurement considered here is described fully in Ref.~\cite{Abe:2016tmq}. The signal is defined as CC$0\pi$, including all the neutrino interactions which do not produce a pion in the final state. Therefore, the signal includes pure Quasi-Elastic interactions, as well as interactions with
pairs of correlated nucleons and $\Delta$ pion-less decays (collectively known as $2p2h$). The signal also includes events with pions produced at the interaction point but which are subsequently absorbed before leaving the nucleus due to final state interactions (FSI).
The analysis was performed using NEUT 5.1.4.2 as the reference Monte Carlo, with both GENIE and NuWro used for full detector Monte Carlo simulation for comparisons, bias tests, and cross-checks, as discussed later.
The selection is 90\% pure, and control samples are used to constrain the background simulation in the analysis.
 
Events are reconstructed and selected using the T2K tracking near detector. Only interactions which take place in the first scintillator target (FGD) are considered as signal in this analysis. 
The FGD is followed by Time Projection Chambers (TPCs) with particle identification capability and where the momentum and charge of the 
tracks can be precisely reconstructed. To reach the TPCs, the tracks need to have a large enough momentum 
in order not to lose all their energy in the FGD and need to have relatively forward polar angle. 
Using the FGD alone and the calorimeters surrounding the tracker, some
high angle and low momentum tracks, escaping the TPC, can be reconstructed with limited efficiency and precision.  

The analysis selection is subdivided into four topologies: 
events with a muon reconstructed in the TPC and no other tracks;
events with a muon in the TPC and a proton in the TPC; events with a muon in the TPC and a proton in the FGD;
and events with a muon in the FGD and a proton in the TPC.
The four topologies cover different phase space regions, as can be seen in Fig. 2 of Ref.\cite{Abe:2016tmq}.
In particular, this is the first T2K analysis capable of selecting high angle and low momentum muons,
thanks to the fourth topology. An improved reconstruction algorithm with better efficiency for high angle and backward tracks has been recently developed in T2K but was not available at the time
of this analysis. 

The cross section is evaluated as a function of muon kinematics, double differential in momentum and angle, including all four topologies. No attempt is made to measure or use proton kinematic information.

Events with two reconstructed protons, which can come from multi-nucleon interactions or final state interactions, are explicitly rejected, but the threshold for proton reconstruction is around 500 MeV.  Therefore, the probability of having two reconstructed protons above that threshold is very small (although not well known). Using the $2p2h$ simulation available at that time in NuWro, the amount of expected events with two protons reconstructed is expected to be $<$1\%, which is negligible compared with errors $\gtrsim$15\% (statistically dominated) in each bin of the differential measurement  and an error of $\sim$11\% on the integrated cross section.

\begin{description}
\item [Signal definition] CC0$\pi$, 1 negatively charged muon, any number of nucleons (neutrons or protons), no charged or neutral pions in the final state.
\item [Observables] double differential in muon momentum, $p_{\mu}$ and the cosine of the muon angle relative to neutrino beam direction $\cos\theta_{\mu}$.
\item [Flux] Ref.~\cite{t2k-flux}, with digital version available from Ref.~\cite{t2k_datarelease}. Only neutrino interactions (not anti-neutrino or electron neutrino) were considered as signal. 
\item [Target material] CH (C$_8$H$_8$, plastic scintillator)
\item [Default generator for analysis] NEUT 5.1.4.2.
\end{description}

\subsection{QE-like measurement comparisons}
\label{sec:compareqe}
We first note that the signal definitions differ between the T2K, \minerva and \mb QE-like measurements, motivated by the different detector designs. The T2K and \mb signal definition is based on the topology of CC0$\pi$, with no charged or neutral pions, but any number of nucleons in the final state. For both detectors there is a high threshold for proton detection, or neutrons through secondary interactions in the detector which would produce a visible proton.  Since final state pions can be detected, CC0$\pi$ is a natural signal choice. One advantage of this insensitivity to final state nucleons is that the efficiency for $1p1h$ and $2p2h$ processes should be similar; this is useful because we know that $2p2h$ processes should contribute.  However, there were no appropriate models implemented in generators when the analyses considered here were carried out. For the calorimetric \minerva detector, the choice of signal definition is less simple. 
The response of the detector to CC1$\pi^+$ interactions with and without pion absorption is similar, as the hadronic (non-vertex recoil) energy is reconstructed as the calorimetric sum regardless of final state particle content. If MINERvA had tried to do an analysis like MiniBooNE, where the signal definition included CC1$\pi^+$ with pion absorption and CCQE, the analysis would have mixed two very different signal populations with different efficiencies, which can bring in model dependence in the efficiency correction.
In addition, the cuts on the hadronic energy were designed to allow $2p2h$ events into the signal selection, but with no $2p2h$ model in the generator, the $1p1h$ efficiency had to be assumed. As such, the \minerva signal is $2p2h$ enhanced, but the validity of describing the signal as $1p1h+2p2h$ is not clear.

The capabilities of each detector place further restrictions on the signal definition. \minerva and T2K can select the sign of the muon, although for \minerva this introduces an angular restriction on the signal events of $\theta_{\mu} \leq 20^{\degree}$ as charge selection is done using the magnetized MINOS near detector, downstream of \minerva. \mb has no ability to distinguish the sign of the final state muon, so in principle the measurement is sensitive to both neutrino and antineutrino interactions. However, the $\sim$1\% $\bar{\nu}_{\mu}$ contamination is subtracted using the MC model, introducing a small amount of additional model dependence~\cite{AguilarArevalo:2010zc}.

\subsection{Model dependence}
\label{sec:qemod}
The CCQE signal could be considered easy to detect because the principal interaction is two-body, therefore simple. However, identification of that final state is masked by FSI and detector effects.  Detecting only the muon results in confusion with the pion production kinematics.  Further, a pion produced in a nucleus which is later absorbed inside the same nucleus must be considered signal in most single arm measurements.  Two strategies to deal with this complexity include redefining the signal at CC0$\pi$ and/or detecting the outgoing nucleon (proton for $\nu_\mu$ experiments). 
We note there are challenges with both approaches. Using a signal using only final state particles avoids the necessity for model dependent FSI corrections.  Using a CC$0\pi$ signal makes it difficult for theoretical models (not true for generators) as the contribution of CC1$\pi$ with pion absorption through FSI must be added to the prediction. 


The \mb measurement was based on a single muon-like track.  Both true CCQE and CC$0\pi$ measurements were published.  These data provided the primary impetus for applying the $2p2h$ models.  The first \minerva measurements had muon-only~\cite{Fiorentini:2013ezn} and muon+proton~\cite{Walton:2014esl} signals.  Although no $2p2h$ model was included in the simulation, attempts were made to deemphasize its importance (see Sect.~\ref{sec:minerva_qe}).  Although the model comparisons in the original paper showed inconsistent agreement with the data, later models applied correctly have shown consistency.

T2K was able to make significant advances, partly due to the benefit of previous work.  Analyses were designed to be, as much as possible, model independent.  By defining the signal as CC$0\pi$, corrections for FSI effects are avoided.  A double differential cross section is measured as a function of muon kinematics (momentum and angle); efficiency corrections then depend only on the detector reconstruction capabilities and not on the fundamental process (e.g. $Q^2$ or $E_\nu$) which produced the outgoing muon.  Therefore the efficiency corrections are, to a large extent, independent of the MC used to evaluate them, and this is discussed in the following paragraphs. The bins of muon kinematics with very low efficiency may still be potentially affected by model-dependence, but they are well separated by the other bins and comparison to models can be done in restricted parts of the phase space.

The model-dependence due to efficiency corrections is significant for measurement signals defined using `true' variables (like $Q^2$ or $E_\nu$) calculated in the nuclear medium, less so for `composed' variables calculated from reconstructed variables (like $Q_{QE}^2$ or $E_\nu^{QE}$ which are computed from reconstructed muon kinematics).
For true variables, the model-dependence is evident because we rely on a particular MC simulation in order to map the observed muon momentum and angle to a particular value of the `true' variable. The most obvious problem comes from incomplete identification of the final state, e.g. $1p1h$ vs. $2p2h$ processes or pion emission followed by absorption.  The reconstructed energy can have errors in reconstructed $E_\nu$ of a few hundred MeV for pion absorption and this error depends on both model choice and reconstruction technique.
The model dependence of `composed' reconstructed variables is more subtle. The efficiency of muon detection is typically not flat as a function of muon momentum and angle due to detector limitations. `Composed' variables mix muons with different kinematics in the same bin with different efficiencies.   In this case it's practically impossible to separate the regions of very low efficiency 
from the rest of the phase space.  This was discovered to be an issue for the T2K analyses with a 1D $Q_{QE}^2$  distribution.  There is also a related issue for measurements which are not completely multi-differential but integrated over one or more of the relevant variables. For instance a one-dimensional measurement as a function of muon momentum involves an integration over muon angle.  This will mix events with forward and backward muons which have different efficiency corrections.  So, the overall efficiency correction in a given bin of muon momentum depends on the distribution of muon angles inside that bin which must be provided by the MC simulation.
The way to avoid this kind of issue is to define restricted regions of phase space in fundamental measured variables (like the muon kinematics) where the efficiency is reasonably large and constant, and/or to bin in as many dimensions as is feasible. 

\begin{figure}[htbp]
\centering
\includegraphics[width=0.7\textwidth]{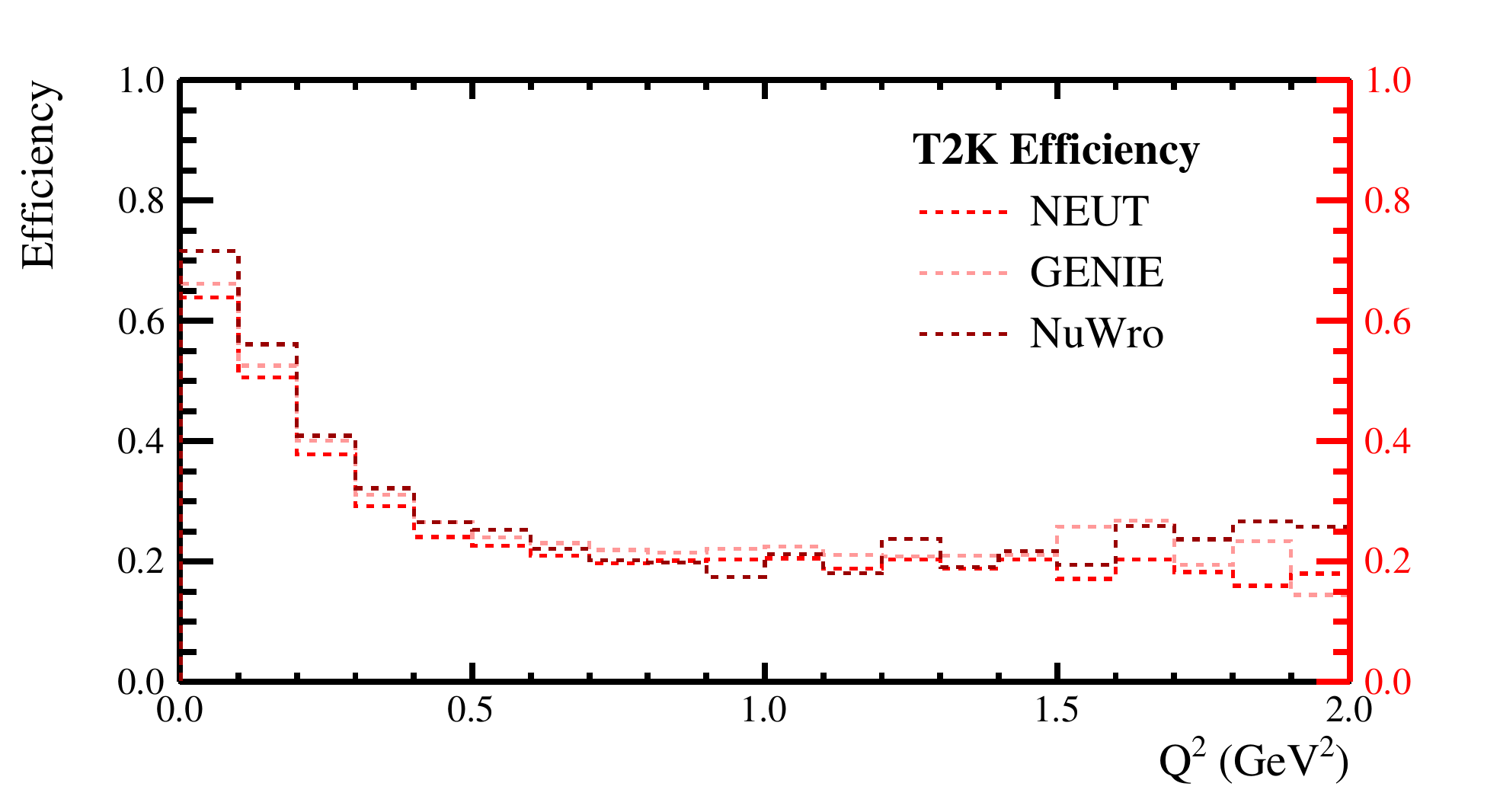} 
\includegraphics[width=0.7\textwidth]{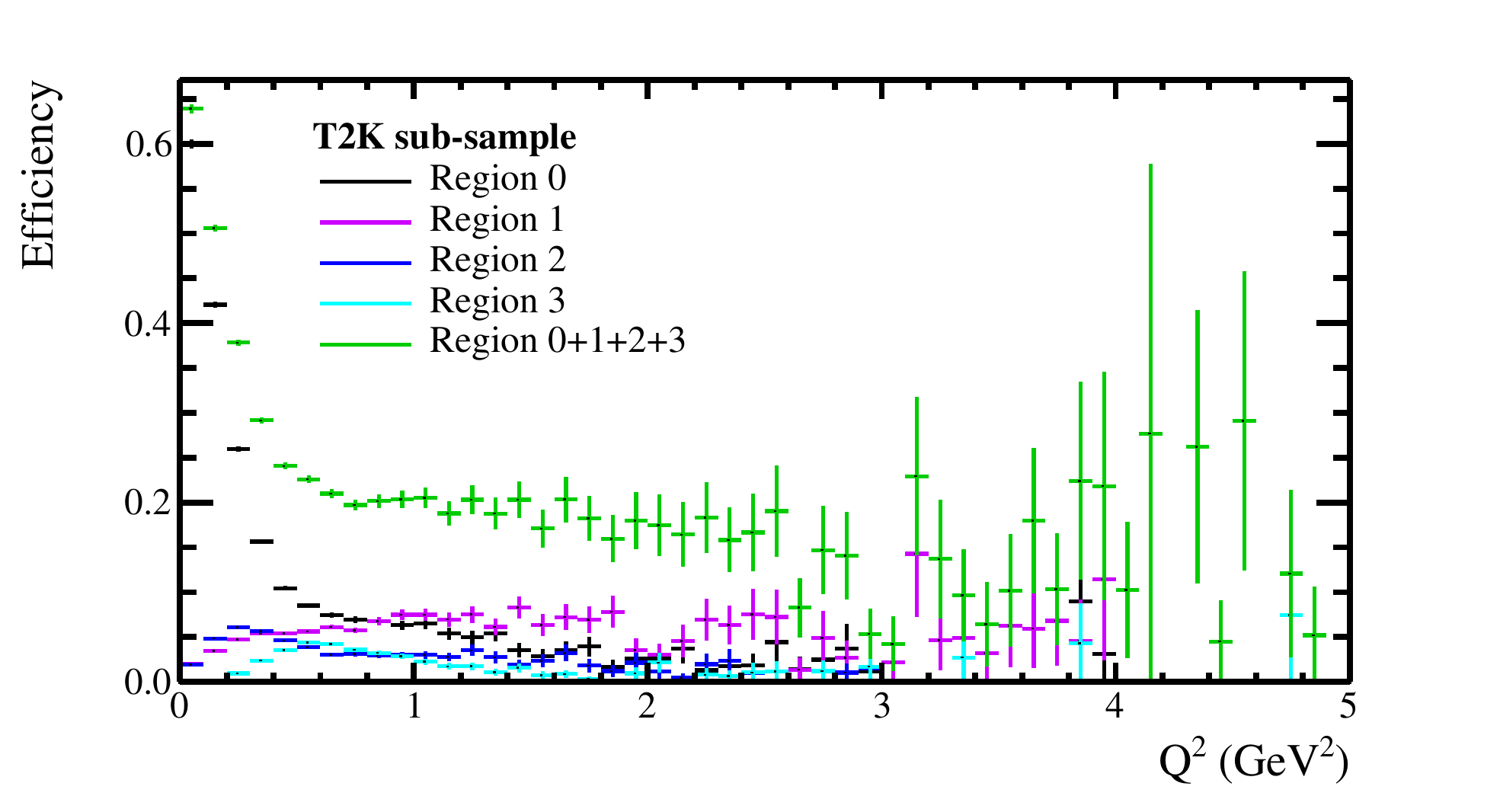} 
\caption{Top: efficiency of the T2K CC$0\pi$ selection as a function of $Q^2$ for the three generators used in the analysis: GENIE v2.8.3, NEUT 5.1.4.2 and NuWro v11o. Bottom: efficiency, separated by T2K sub-sample selections (see Sect.~\ref{sec:t2kqe} for a description) using NEUT v5.1.4.2.}
\label{fig:T2KCC0pi_eff}    
\end{figure}

Particular care should be taken regarding `hidden' variables, i.e. variables which may affect your reconstruction efficiency in an indirect way and which are integrated out in the final measurement. One example is the proton kinematics in the CC$0\pi$ T2K analysis. No attempt is made to compute the cross-section as a function of proton kinematics but for backward going muons the presence of a reconstructed forward proton is needed to select the event.  
For the T2K analysis, the explicit model dependence of the efficiency was studied, using three neutrino interaction generators (NEUT v5.1.4.2, GENIE v2.8.3, and NuWro v11o). Each were simulated with the full detector response to check for areas in which the primary generator might be biasing the result (see Fig.\ref{fig:T2KCC0pi_eff} top).  The difference between the Monte Carlo calculations are non-negligible and this must be covered in the analysis by large signal modeling systematics (Fig. 5 of Ref.\cite{Abe:2016tmq}), which are especially important in regions where the efficiency is low (high angle and low momentum).  In the T2K analysis, the influence is small because for a low momentum backward muon the presence of a forward proton can be simulated by energy conservation in the approximation of small nuclear recoil. 




\begin{figure}[htbp]
\centering
\includegraphics[width=0.7\textwidth]{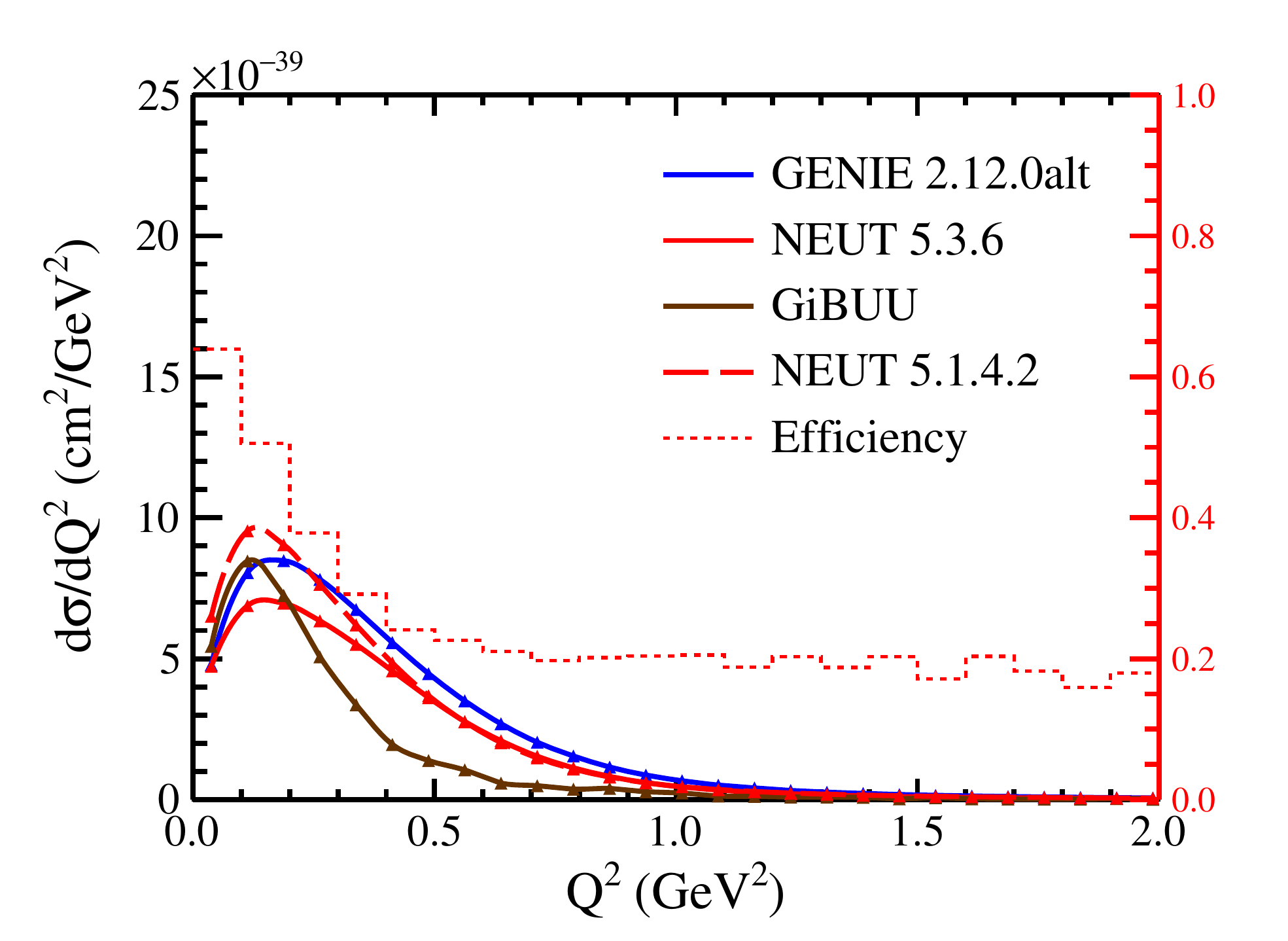}  
\caption{Efficiency of T2K CC$0\pi$ selection as a function of the $Q^2$ overlaid with NEUT 5.1.4.2 and the modern generators.}
\label{fig:t2k-q2-effall}
\end{figure}


Fig.~\ref{fig:t2k-q2-effall} shows the efficiency of the T2K CC0$\pi$ selection overlaid with the generator predictions of interest for the workshop. The data is compared with efficiency based on NEUT 5.1.4.2, the cross section of NEUT 5.1.4.2, and additional modern generator models (NEUT 5.3.6, GENIE 2.12.0alt, and GiBUU).  We note that NuWro is not included because it is almost identical to NEUT 5.3.6. The efficiency for $Q^2<0.3$ GeV in Fig.\ref{fig:T2KCC0pi_eff} is changing rapidly due to difficulties in reconstruction.  
If T2K had produced a measurement unfolded in $Q^2$, this would lead to unacceptable variations in results using different generators.
In contrast, measurement of the cross-section as a function of muon kinematics is more robust against model variations. Provided that the efficiency is flat in a given $p_\mu-\theta_\mu$ bin, the efficiency will be largely independent of the model used in the analysis.
In the larger picture, the fast variation of efficiency as a function of $Q^2$ makes the extraction of model-independent physics difficult. For example, the fraction of $2p2h/1p1h$ population in the measured cross-section depends strongly on the $Q^2$ distribution of $2p2h$ and $1p1h$ processes which have their own model dependence. Such a measurement must then be compared to models only through a full detector simulation (as T2K has done with Fig.\ref{fig:T2KCC0pi_eff}).

\begin{figure}[htbp]
\centering
\includegraphics[width=0.7\textwidth]{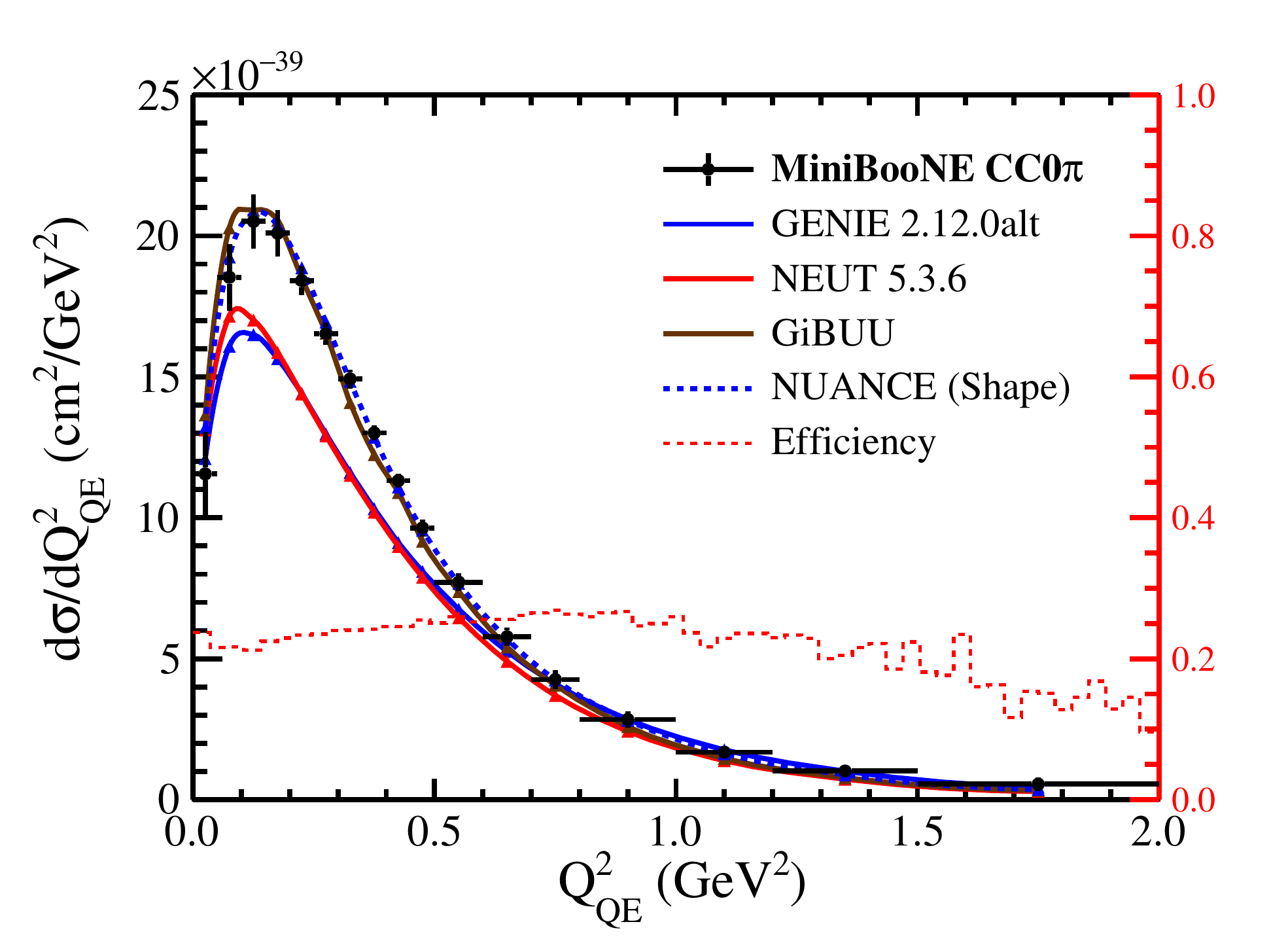} 
\caption{Efficiency of the \mb CC$0\pi$ selection (with CCQE, $2p2h$, and 1$\pi$-absorption contributions) as a function of \qq, overlaid with the CC$0\pi$ prediction NUANCE v3 and more recent generator models. The NUANCE calculation is normalized to the data.}
\label{fig:mbeff}    
\end{figure}

In a surprising contrast to the T2K analysis, the \mb efficiency for CC0$\pi$ events was flat across both $q_0$-$q_{3}$ and \qq. The latter is shown in Fig.~\ref{fig:mbeff}.
Note that the efficiency is calculated relative to the number of contained events.  
Furthermore, it was found that the \mb selection loses efficiency only for lower momentum/kinetic energy (KE) of the muon candidate; this is due to preferential selection of muon-like tracks over electron-like tracks.

\begin{figure}[htbp]
\centering
\includegraphics[width=0.7\textwidth]{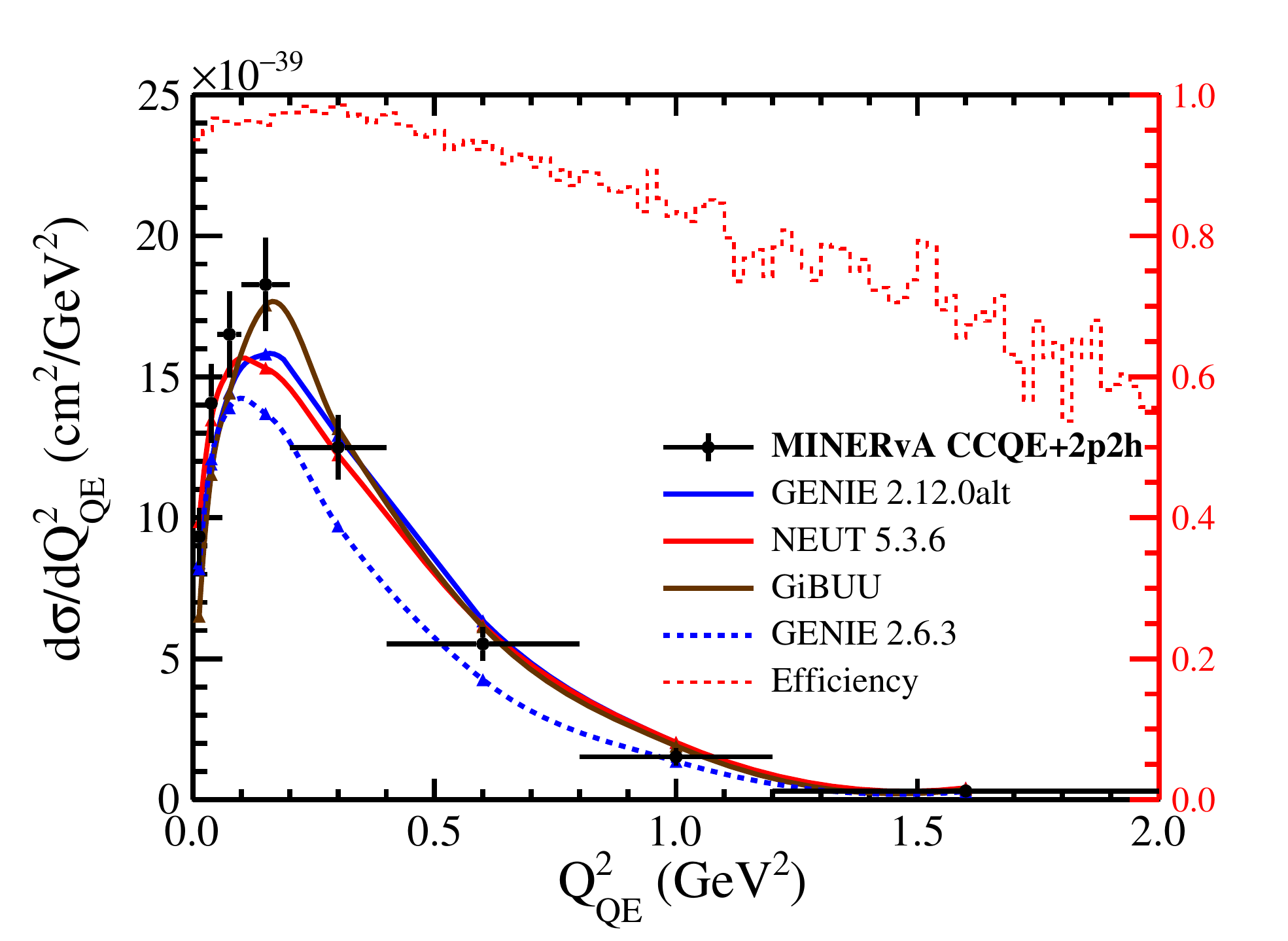} 
\caption{Efficiency of \minerva CCQE+$2p2h$ selection as a function of $Q^2$ based on the GENIE 2.6.3. Also shown are the \minerva data overlaid with the predictions from a selection of modern generators considered at the workshop and the generator used in the analysis (GENIE 2.6.3).}
\label{fig:mineffall}    
\end{figure}

Finally, we compare the \minerva selection efficiency to the generator predictions of interest in Figure~\ref{fig:mineffall}. We note that the efficiency does not include acceptance effects akin to T2K as it is only the relative efficiency after the MINOS-matched selection (sufficient momentum and angle to match to MINOS).   While the lowest $Q^2$ region has a relatively high and flat efficiency, there is a steady reduction above 0.5 GeV$^2$.  At higher $Q^2$, as the efficiency is changing, the analysis may be more susceptible to mis-modeling in the extraction of the cross section. 

We note that all three experiments assign significant systematic uncertainties to cover generator mis-modeling; these choices are not readily checked with the investigations described here. The primary conclusion is that future analyzers should closely scrutinize the model uncertainties assigned in regions where efficiency changes rapidly. 

\begin{figure}[htbp]
\centering
\includegraphics[width=0.495\textwidth]{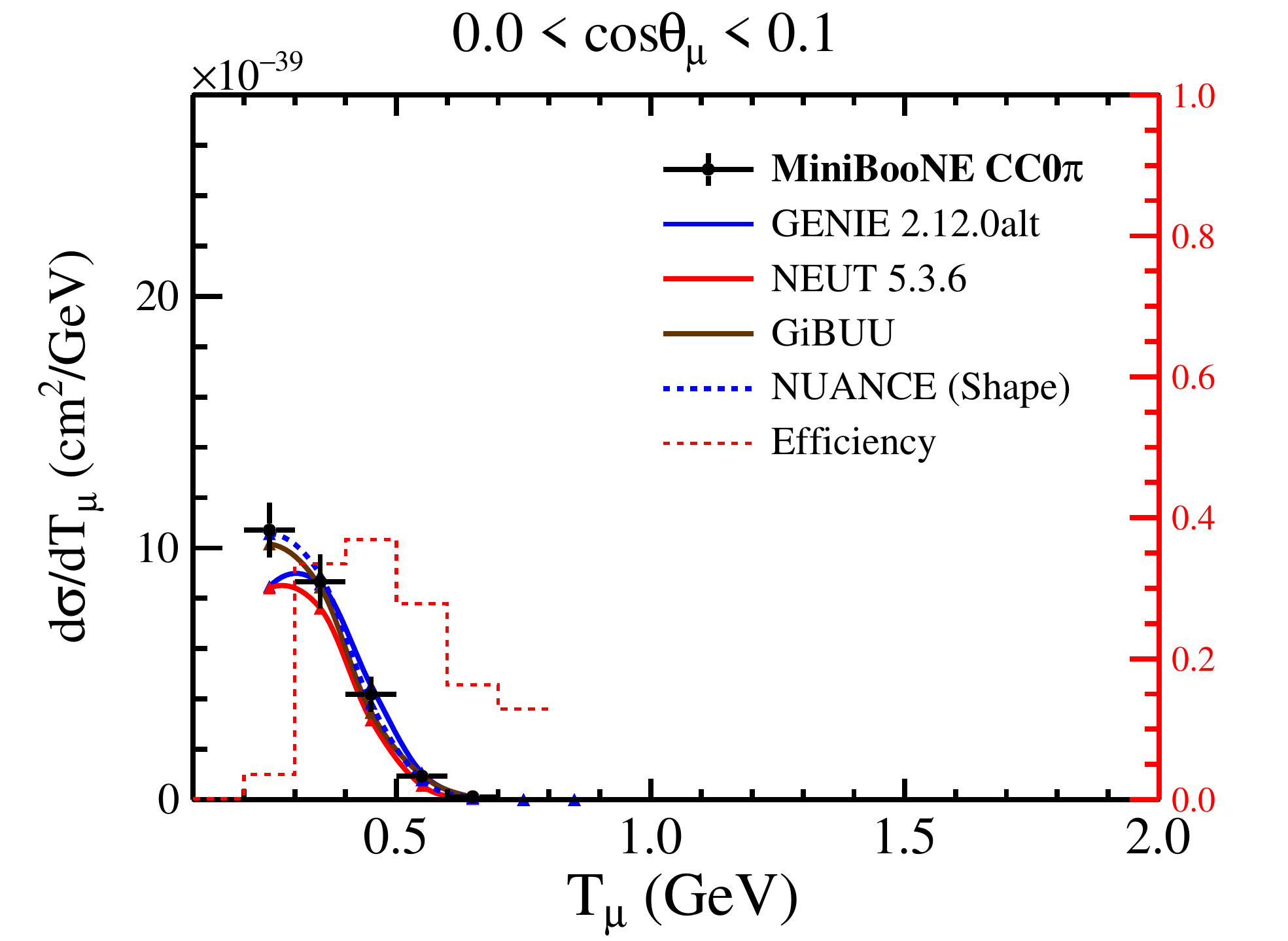} 
\includegraphics[width=0.495\textwidth]{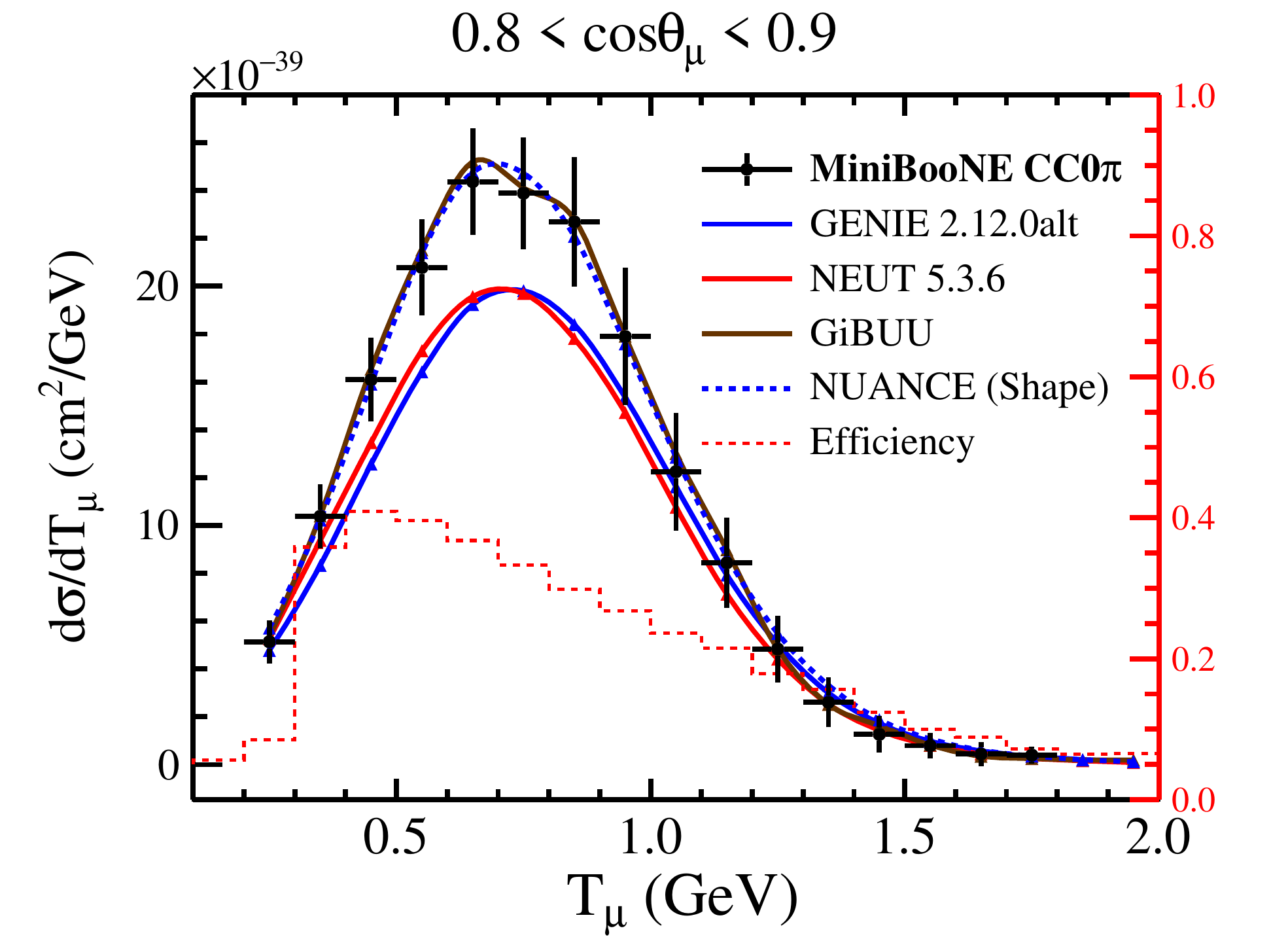}
\caption{\mb CC$0\pi$ double differential cross section data with efficiency as calculated using \nuance\ and the \nuance\ prediction.  In addition, other more recent generator results are shown.}
\label{fig:mbeffall}    
\end{figure}

\begin{figure}[htbp]
\centering
\includegraphics[width=0.495\textwidth]{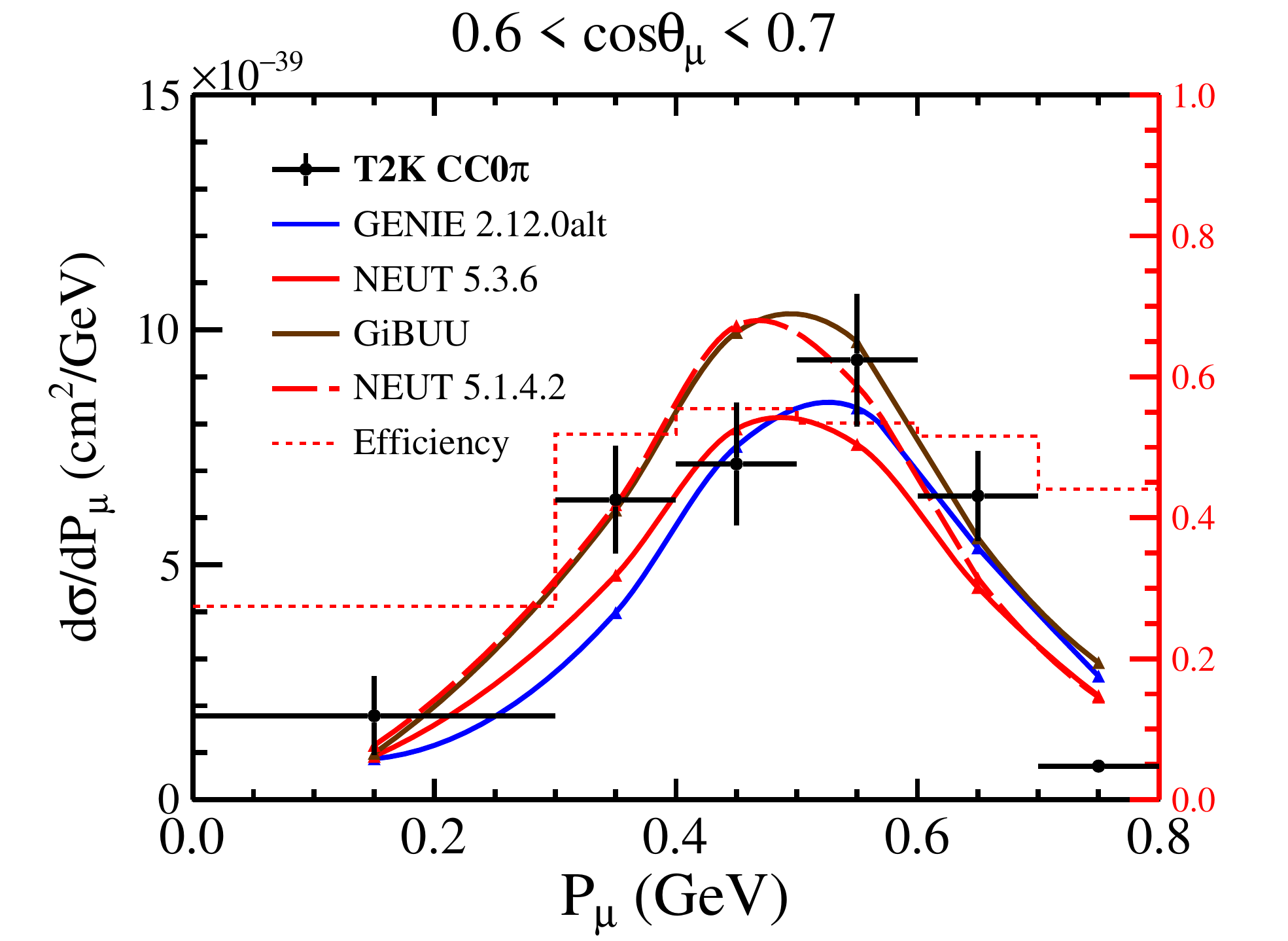} 
\includegraphics[width=0.495\textwidth]{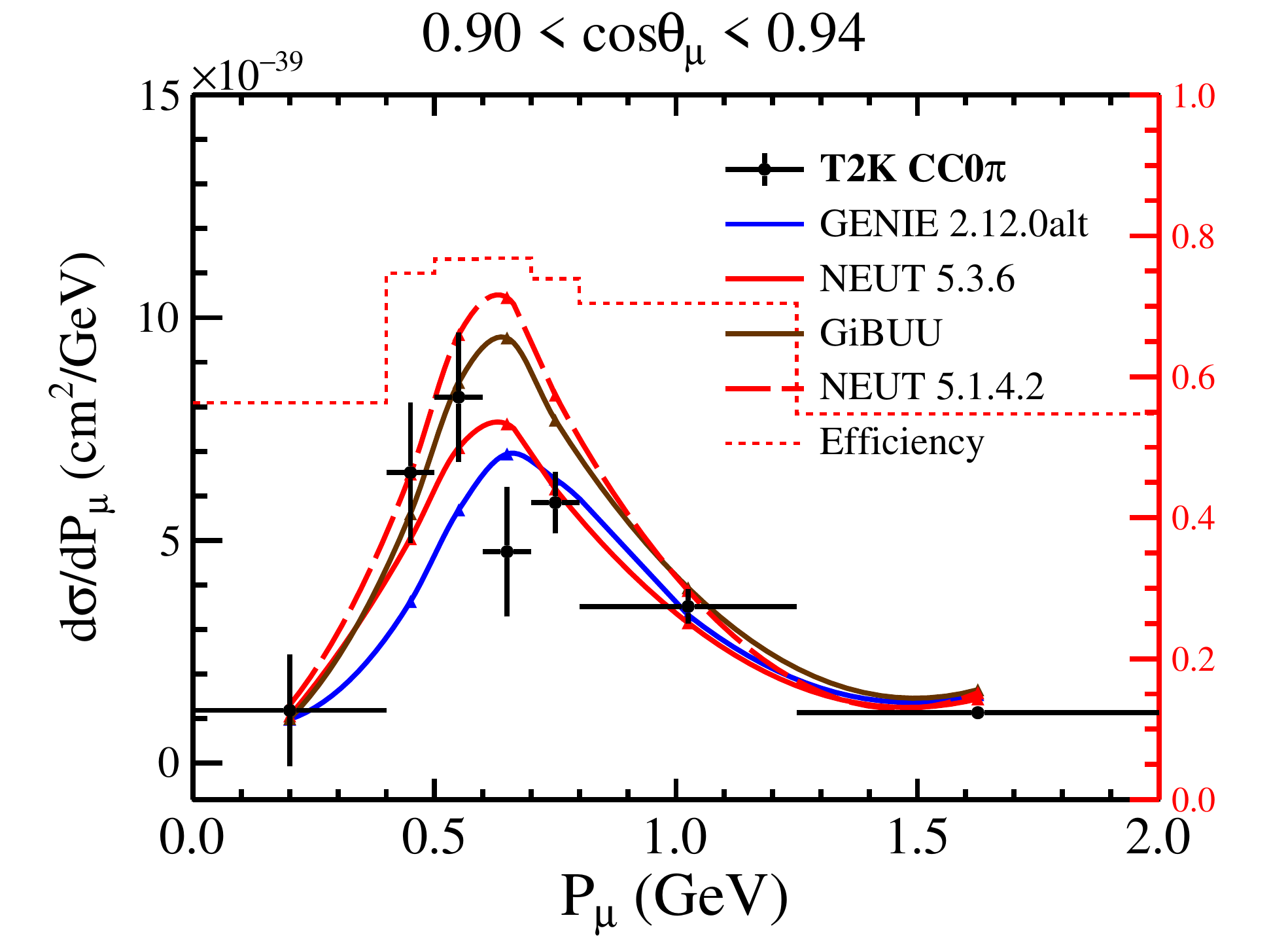}
\caption{T2K CC$0\pi$ double differential cross section data with efficiency as calculated using NEUT v5.1.4.2 and the NEUT prediction.  In addition, other more recent generator results are shown.}
\label{fig:t2keffall}
\end{figure}

In addition to issues related to overall acceptance, there were some specific selection choices which impacted the underlying physics. The \minerva analysis includes restrictions on both reconstructed $E_\nu^{QE}$ and true $E_\nu$ in either signal or selection\footnote{The effect of the cut ($E_\nu^{QE}<10$ GeV) is negligible due to the lack of flux in that region. } (described in Section~\ref{sec:minerva_qe}) which are intended to reflect the higher average energy of the selection due to the requirement of a matched track to the MINOS detector. The use of true $E_\nu$ as a signal definition may add complication and/or dependence on the MC used by MINERvA, when the desired quantity is reconstructed momentum and angle at MINOS.  A similar issue comes with the \mb 1$\pi$ selection where the muon-pion PID failed because of near equality of detector response; in that case, a cut on the underlying physics ($W$) was used. It is preferable to put signal selection requirements on variables such as the muon momentum where the smearing between true and reconstructed quantities can be well defined.

\begin{figure}[htbp]
\centering
\subfigure[T2K]{\includegraphics[width=0.45\textwidth]{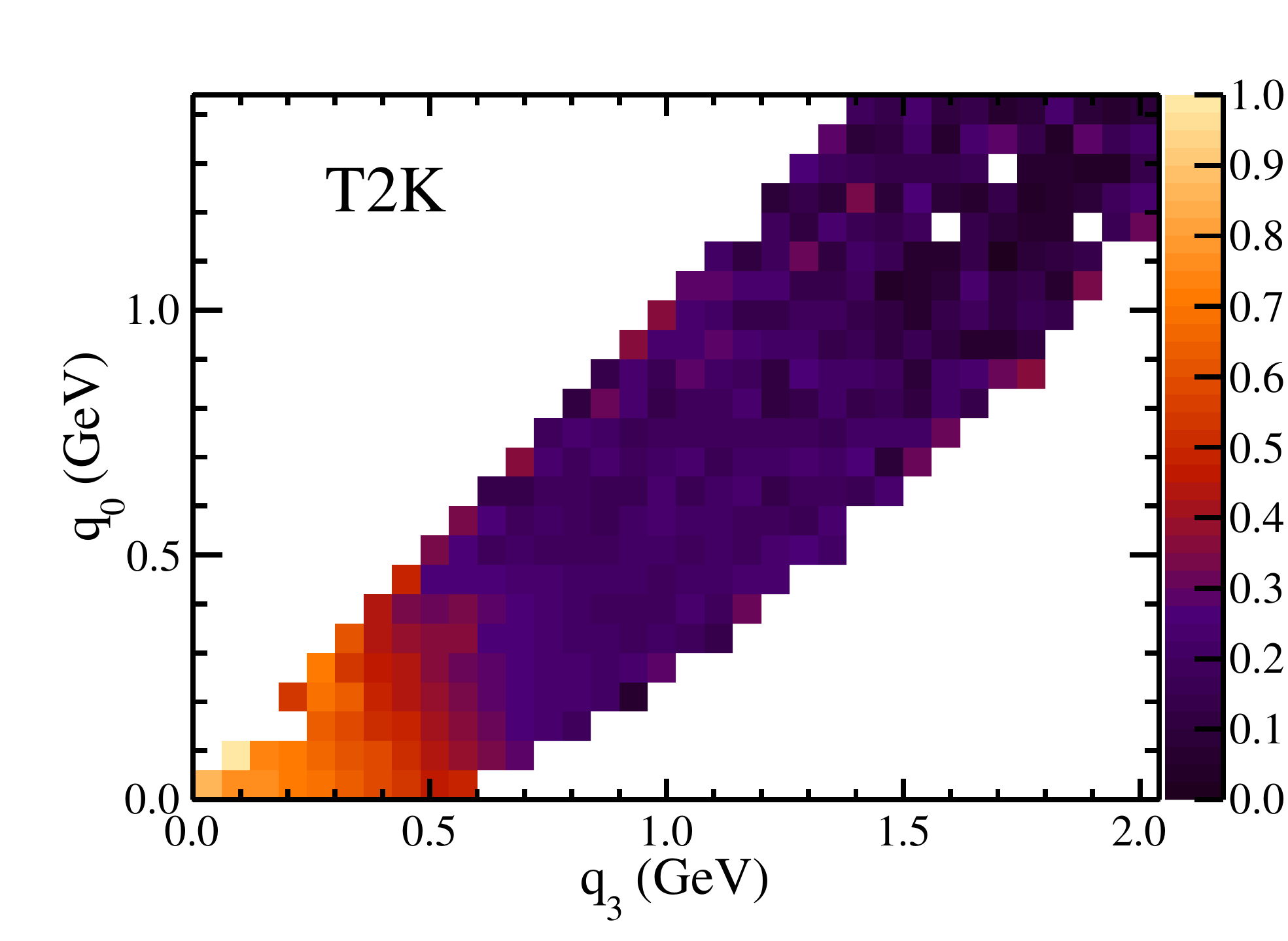}} 
\subfigure[\mb]{\includegraphics[width=0.45\textwidth]{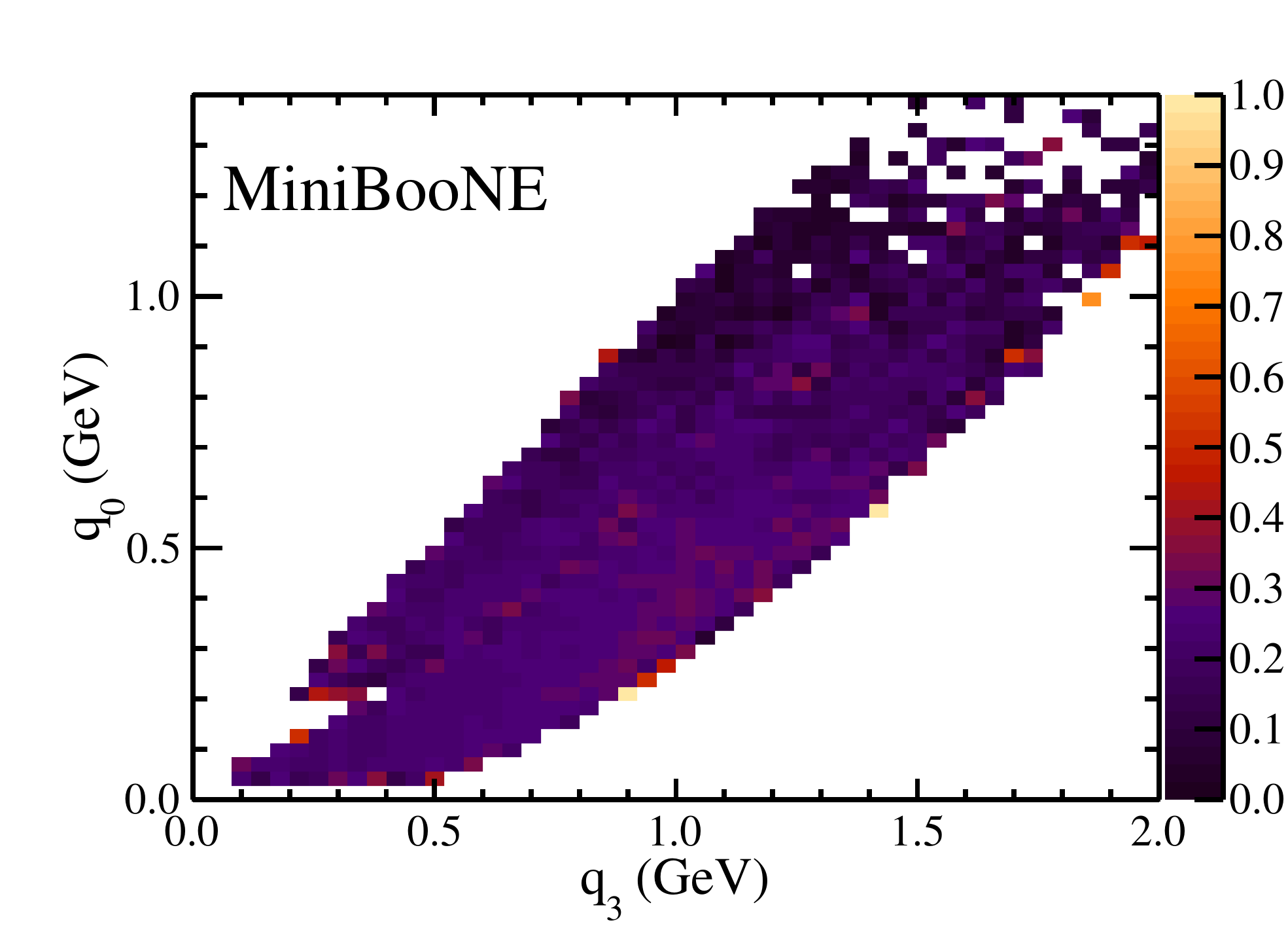}} 
\subfigure[\minerva]{\includegraphics[width=0.45\textwidth]{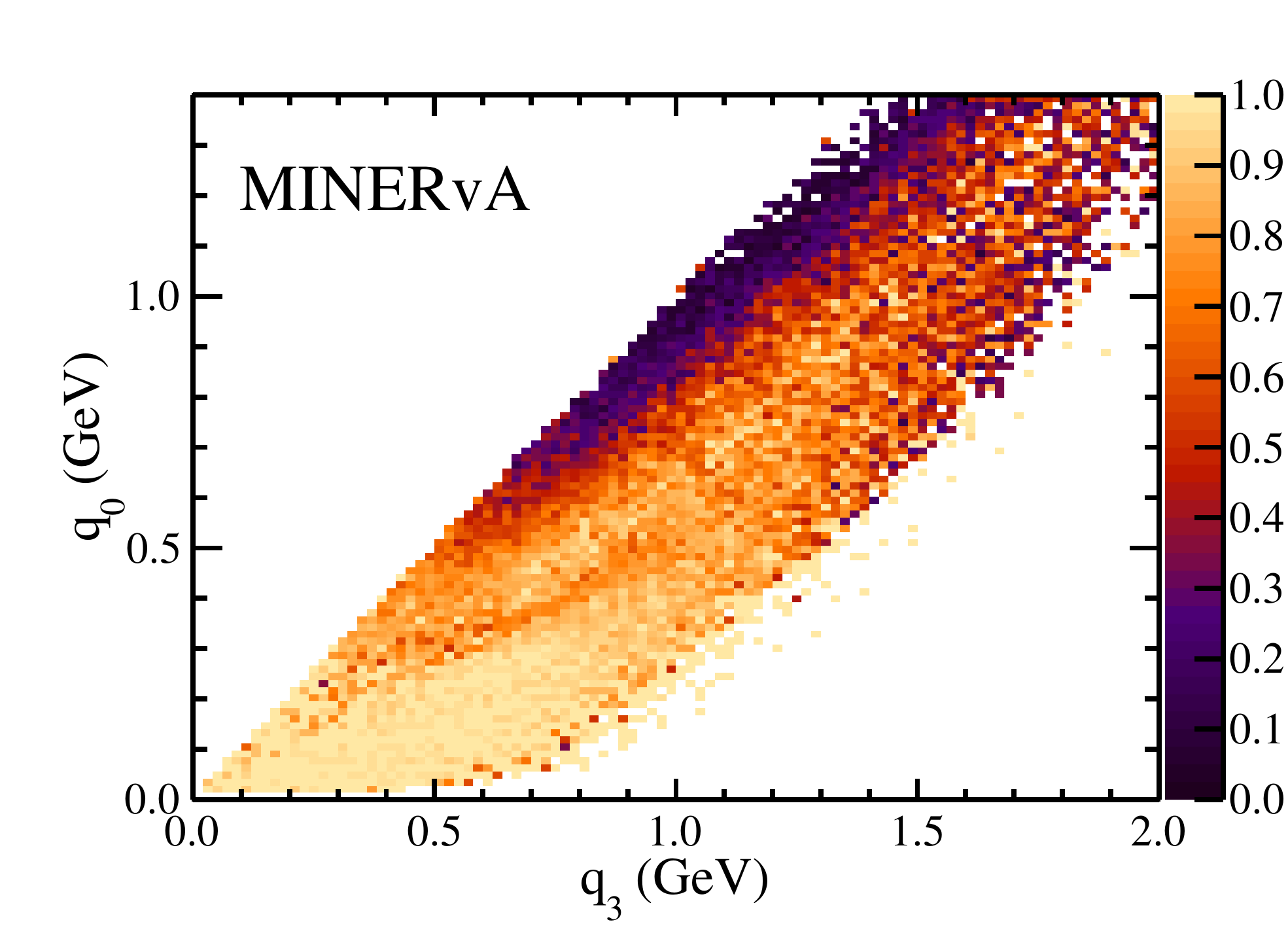}}
\caption{The efficiency of CC$0\pi$ topologies as a function of $q_0$-$q_3$ for the experiments considered here. Note: MINERvA's signal definition is CCQE(+$2p2h$), so this is not an exact comparison, but the inefficiency at high $q0$ affects $2p2h$ significantly, despite attempts to minimize this in the analysis. }
\label{fig:t2kmbminq0q3}    
\end{figure}

Digging deeper into the physics of CC$0\pi$ experiments, Fig.~\ref{fig:t2kmbminq0q3} shows how efficiency for a CC$0\pi$ signal changes for the three experiments across $q_0-q_3$ phase space. This plot includes the effect of acceptance and specific cuts. \mb had a 4$\pi$ detector and indeed has relatively flat efficiency across $q_0-q_3$.  The T2K efficiency is not flat due to acceptance. For \minerva, acceptance and cuts on the hadronic state significantly shaped the $2p2h$-RES region as compared to QE. In particular, though a region around the vertex was blinded, the efficiency still changes rapidly for $2p2h$ events. This kind of selection cut may make the analysis sensitive to the $2p2h$ model used.   \minerva faced two unique challenges which T2K and \mb did not. First, for the higher beam energies, \minerva has a more difficult time isolating QE-like interactions due to significant presence of other processes and (at the time) no Michel tag. Second, the lack of a $2p2h$ model at the time made it nearly impossible for \minerva to avoid all possible $2p2h$ model sensitivity.

Both \mb and \minerva selections have reduced purity for $\qqqe < 0.2$ GeV$^{2}$ relative to the rest of the underlying physics region; in the case of \minerva, this was driven by the masking of the vertex region due to concerns about (unknown) $2p2h$ contribution. This is interesting as the low $Q^2_{QE}$ region has historically been difficult to model well. It is helpful to look at purity in reconstructed as well as true variables, to avoid unfolding-related concerns and the (presumed) origins of the backgrounds. For example, studies found the mapping between \qqqe and true $Q^2$ to be biased for a T2K selection, and so, a projection of only true $Q^2$ would not have been informative.


\subsection{QE-like generator considerations}
\label{sec:qecomp}

Quasielastic measurements discussed here use different event generators: \nuance\ for \mb, GENIE for \minerva, and NEUT for T2K. Although the models used are similar, different choices in implementation or choice of parameters can modify the predictions considerably. While GENIE and NEUT were easily usable, \nuance\ was hard to resurrect.  As a result, only shape results for \nuance\ QE were available.  One notable difference is the values of \maqe used, which affects the \qq distribution through the quasielastic form factor.  While GENIE uses the value obtained from fits to deuterium bubble chambers (close to 1.0 GeV), NEUT uses a value of 1.21 GeV based on fits to K2K data, and broadly consistent with fits to modern heavy target data when nuclear effects are neglected.  NUANCE uses the value extracted from a shape-only fit, i.e., 1.23 GeV.  In recent years, inflated axial mass values have been replaced by proper models for $2p2h$ processes as an explanation of the apparent excess in data.  This shift was in progress during \mb and \minerva analysis and  none were available in the generators at the time of the analyses discussed. Additionally, pion absorption is treated differently in each program, which modifies the contribution from CC$1\pi$ events that migrate into CC0$\pi$ samples.

\begin{figure}[htbp]
\centering
\includegraphics[width=0.495\textwidth]{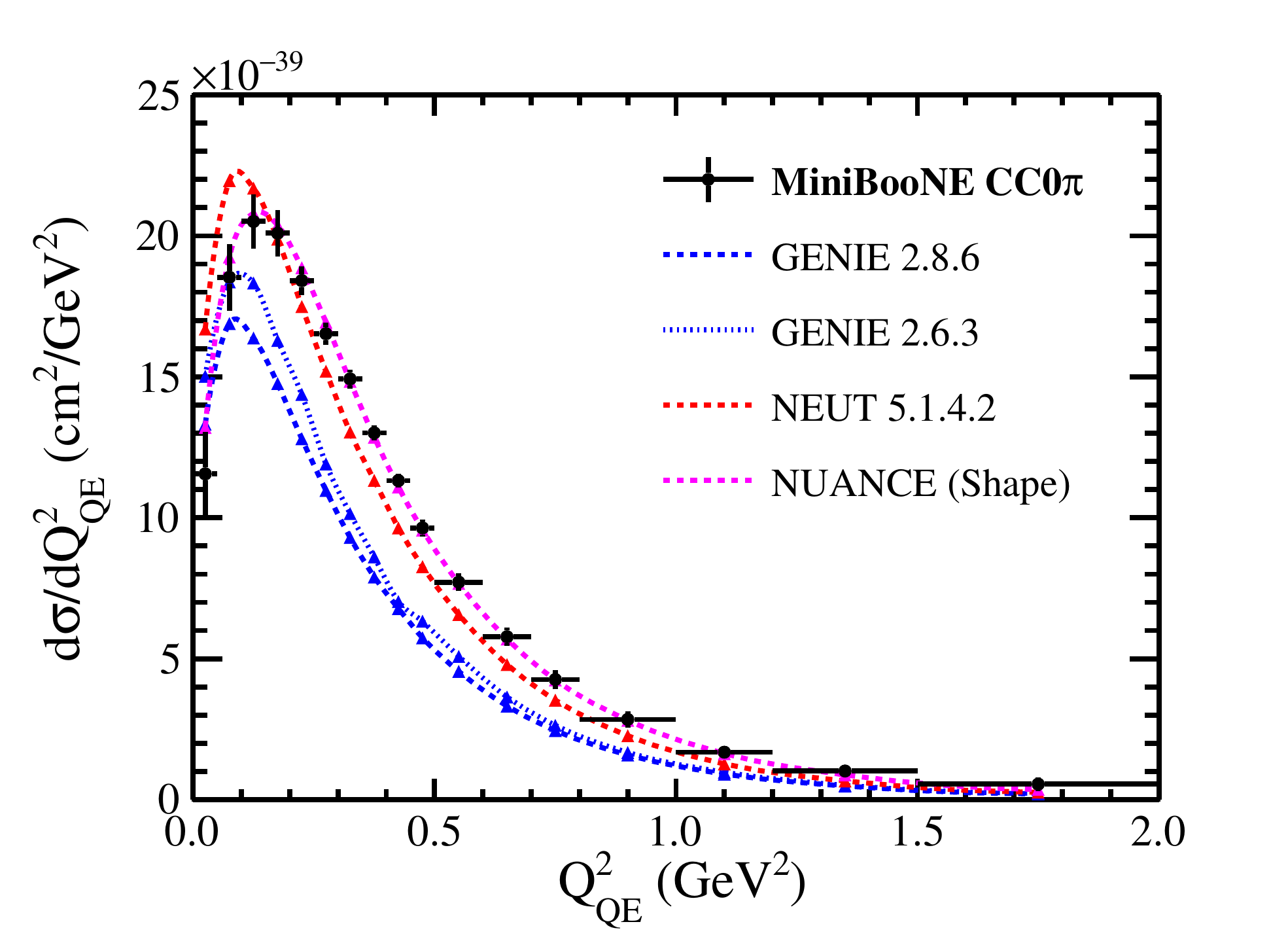} 
\includegraphics[width=0.495\textwidth]{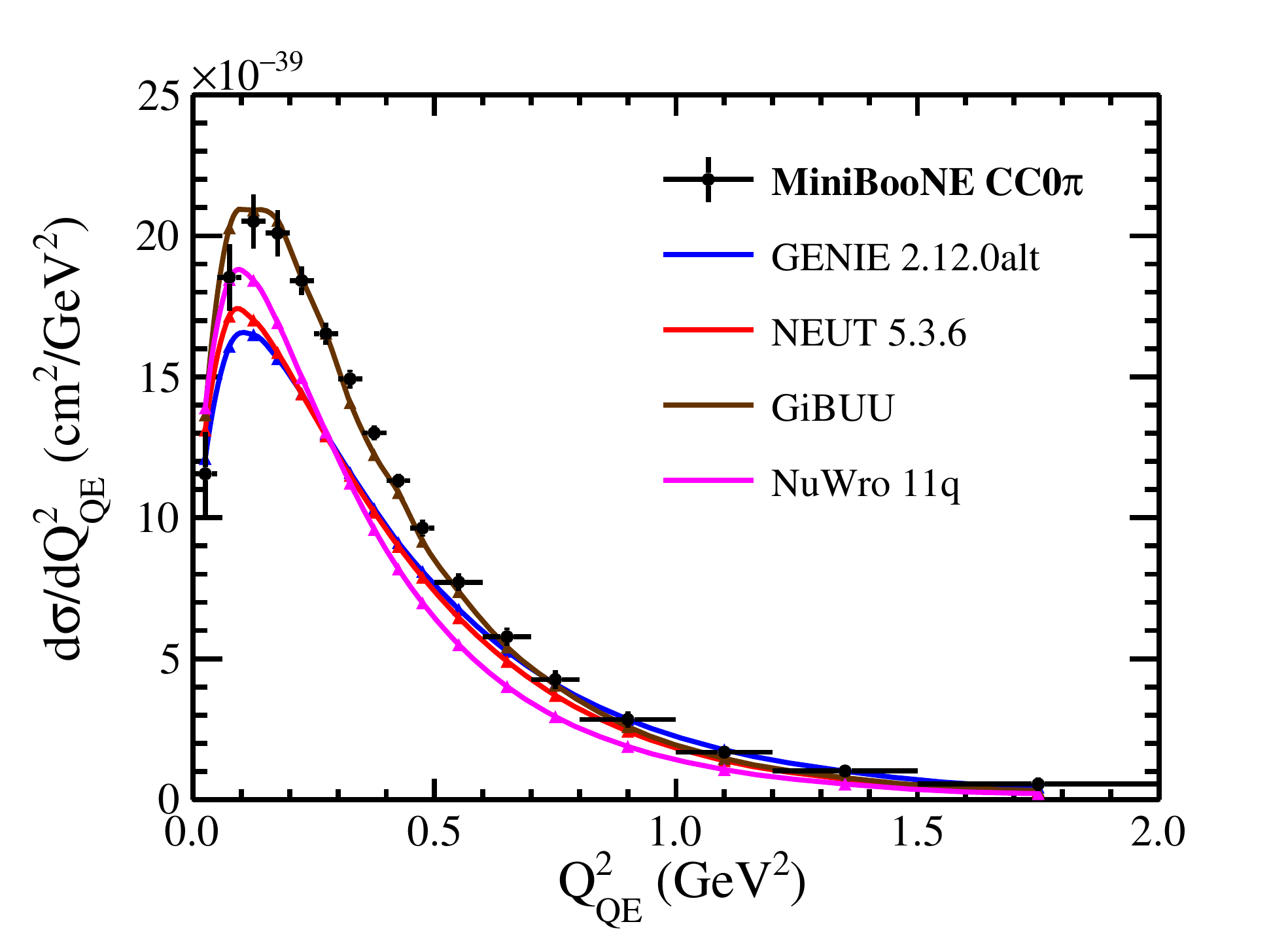}
\caption{\mb CCQE-like data for \qqqe.}
\label{fig:miniboone-qe-comp}    
\end{figure}

\begin{figure}[htbp]
\centering
\includegraphics[width=0.495\textwidth]{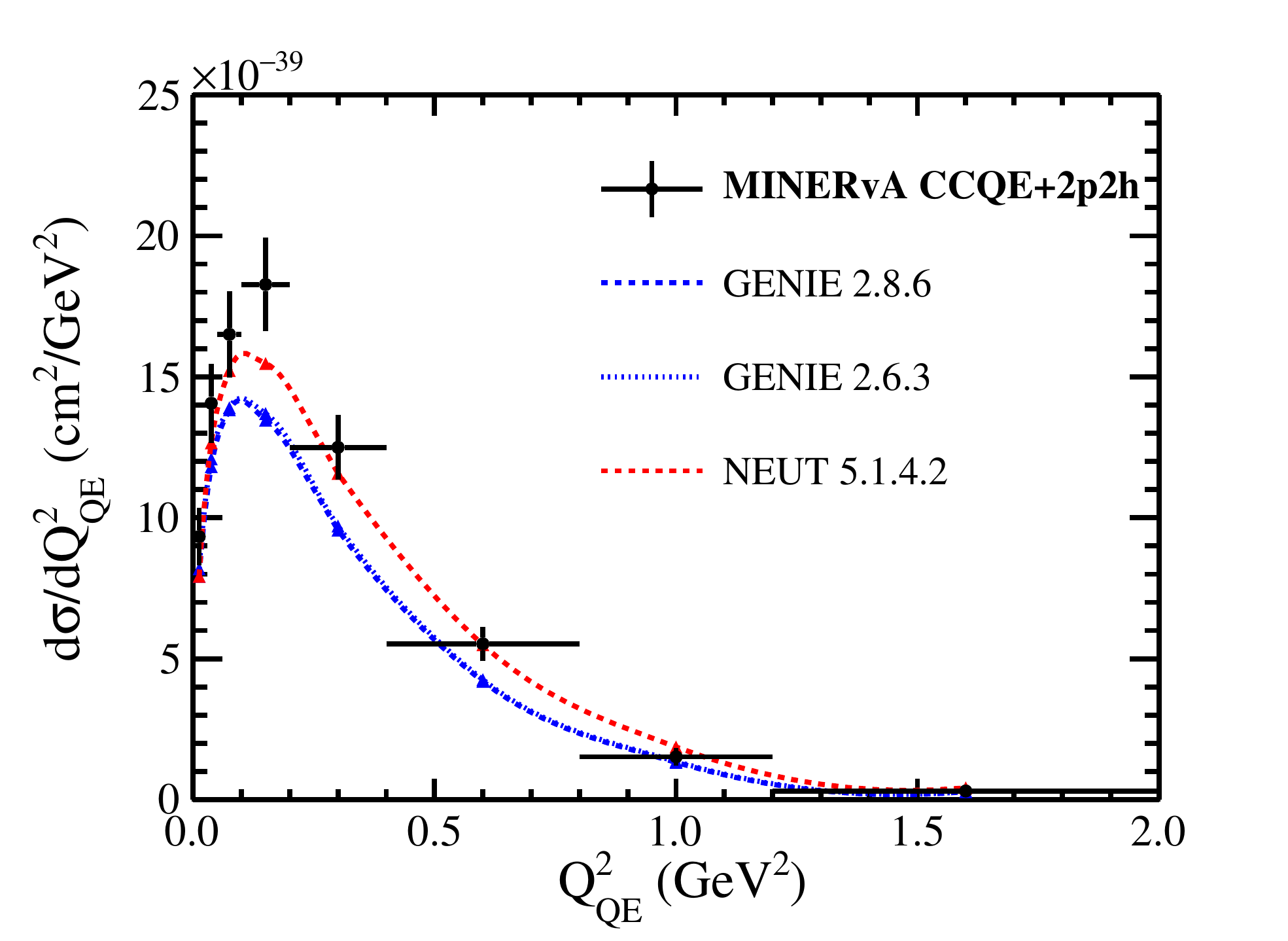} 
\includegraphics[width=0.495\textwidth]{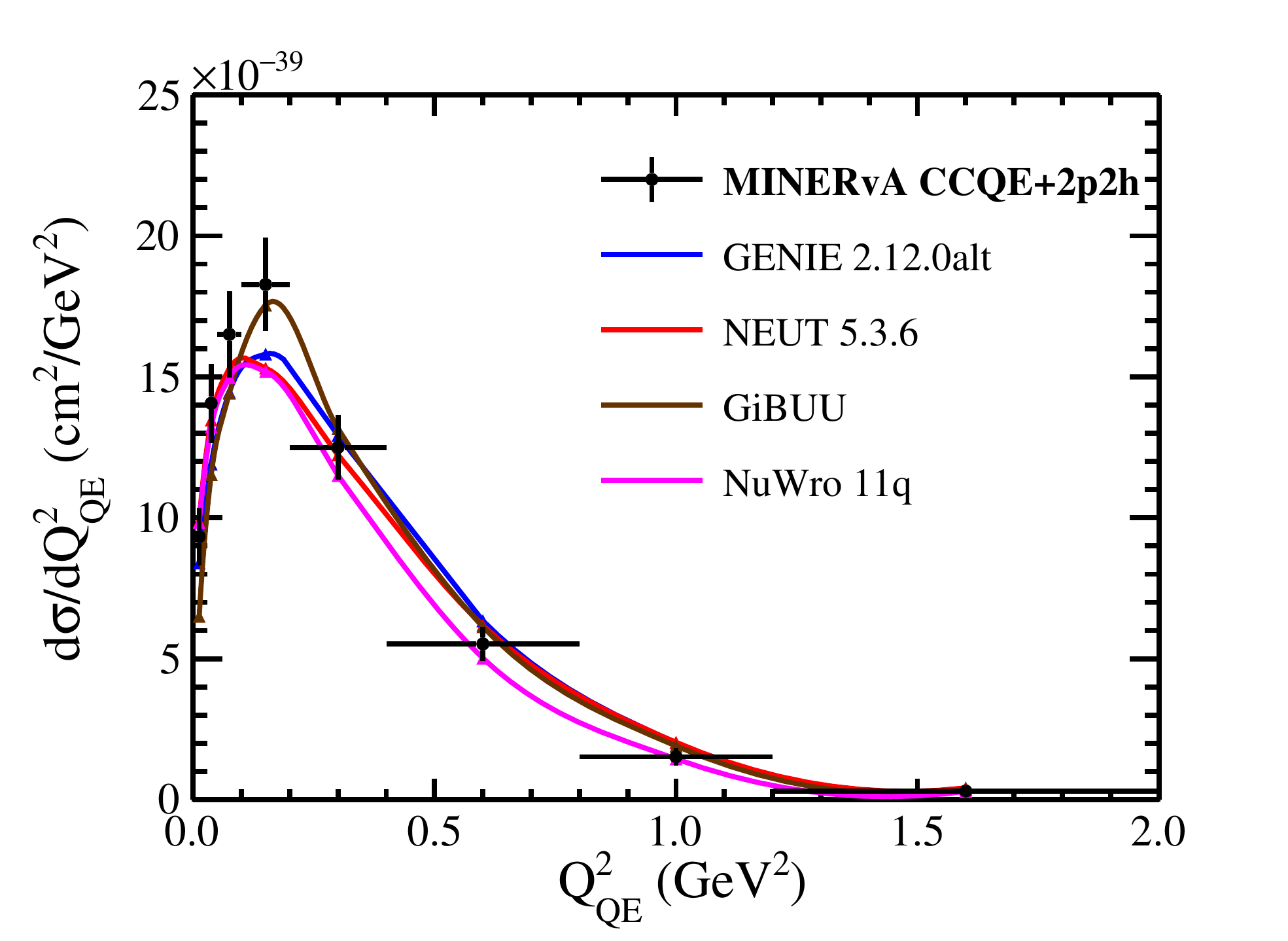}
\caption{\minerva CCQE data for \qqqe.}
\label{fig:minerva-qe-comp}    
\end{figure}

Codes have undergone significant change in the last few years due to the efforts of nuclear theorists to adapt models used for electron scattering to neutrino interactions and efforts by the generator authors to include them. The nuclear model was changed from relativistic Fermi gas (RFG) to local Fermi gas (LFG) or Spectral Function (SF).  Nucleon-nucleon correlations of short, medium ($2p2h$) or long (RPA) range are now known to be important, and at low neutrino energies Coulomb effects can also be important. Fig.~\ref{fig:miniboone-qe-comp} shows the \mb CC0$\pi$ \qqqe data with both old and new generator models.  The range of cross section prediction was large for the older models because of different attempts to match the \mb data. However, recent models use theoretical models which were motivated by or tuned to this data, so unsurprisingly, there is less variation between them.  In Fig.~\ref{fig:minerva-qe-comp}, the same comparison is made for the \minerva CCQE-like data.  It is important to note that the data shown here have been updated to include the improved NuMI flux calculations first presented in fall, 2015~\cite{Aliaga:2016oaz}.  This increases each data point by about 15\%.  As a result, agreement of recent generator versions with both \mb and \minerva data sets has improved significantly.  This is consistent with the findings of a theoretical group~\cite{Megias:2016fjk}.

The new models in Figs.~\ref{fig:miniboone-qe-comp} and \ref{fig:minerva-qe-comp} show less variation because all have included a $2p2h$ model.  GENIE, NEUT, and NuWro all use the Valencia model~\cite{Nieves_2p2h_14} and GiBUU has a separate version.  GENIE 2.12.0alt further adopts the Nieves CCQE model~\cite{Nieves:2004wx} and is therefore about 10\% less than the \mb data, same as was shown for the Valencia theoretical calculation.  

\begin{figure}[htbp]
\centering
\includegraphics[width=0.495\textwidth]{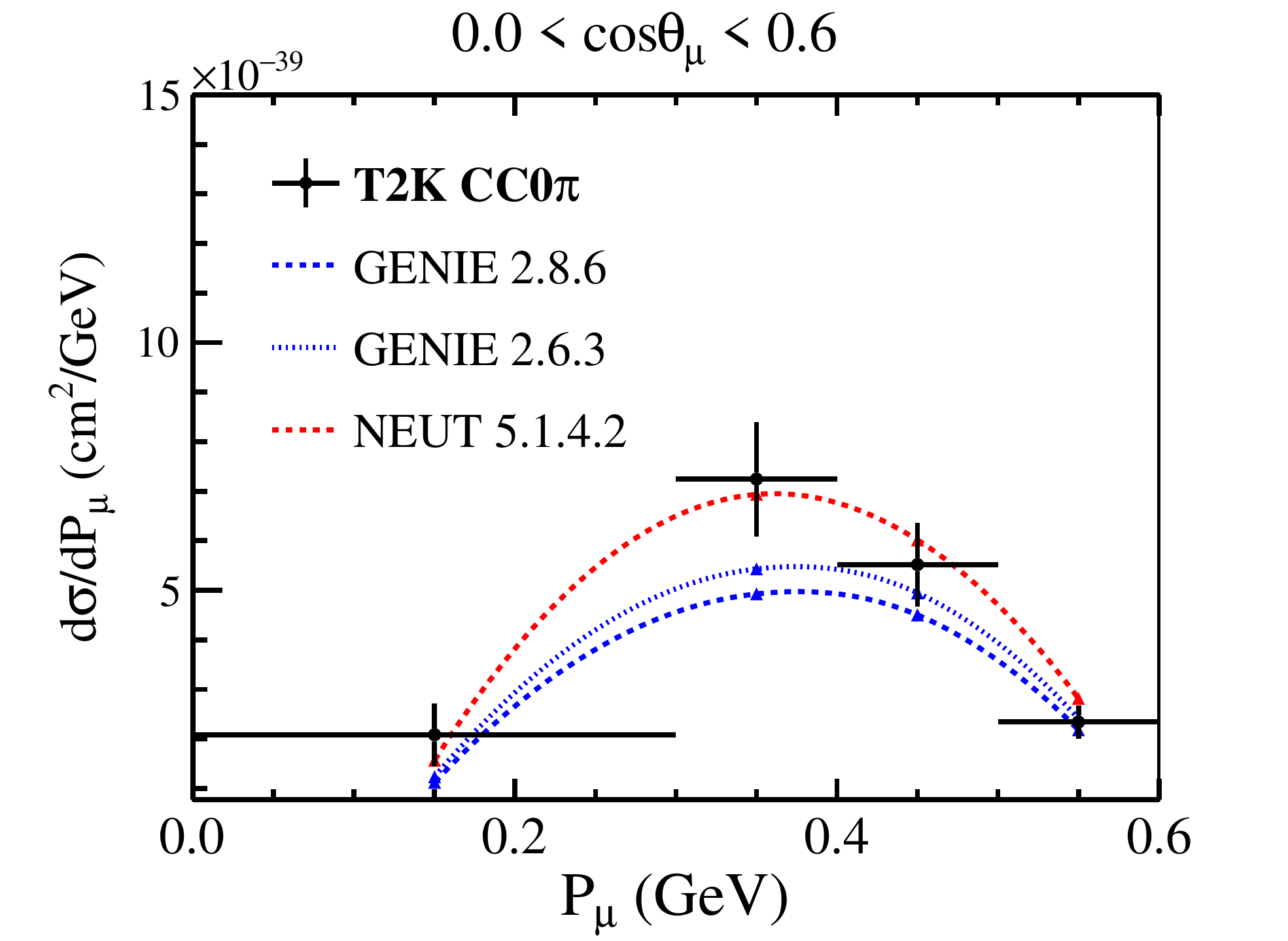} 
\includegraphics[width=0.495\textwidth]{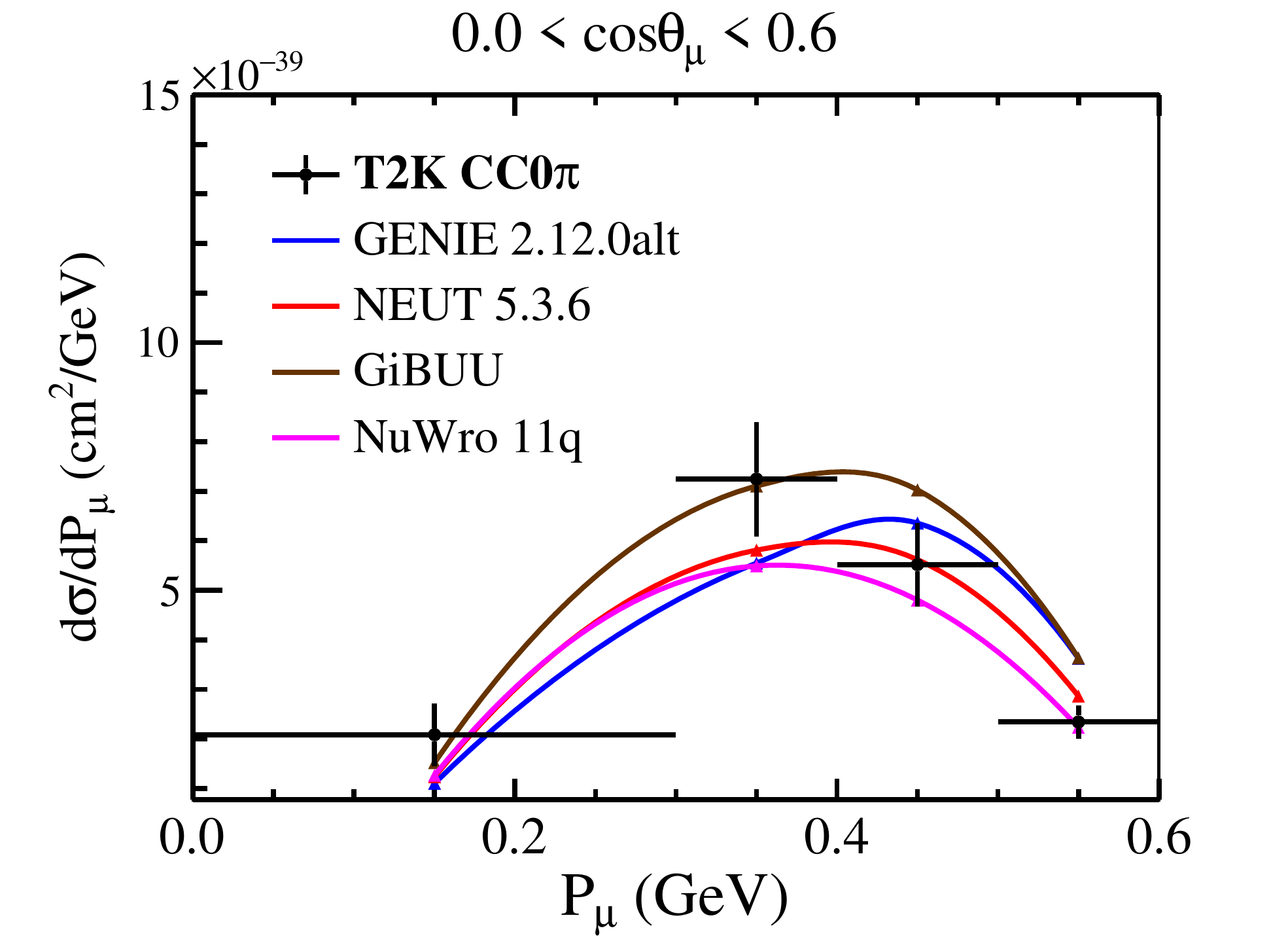}
\caption{T2K CC0$\pi$ data as a function of $p_{\mu}$ for angles 53$^\circ$-90$^\circ$.}
\label{fig:t2k-qe-comp1}    
\end{figure}

\begin{figure}[htbp]
\centering
\includegraphics[width=0.495\textwidth]{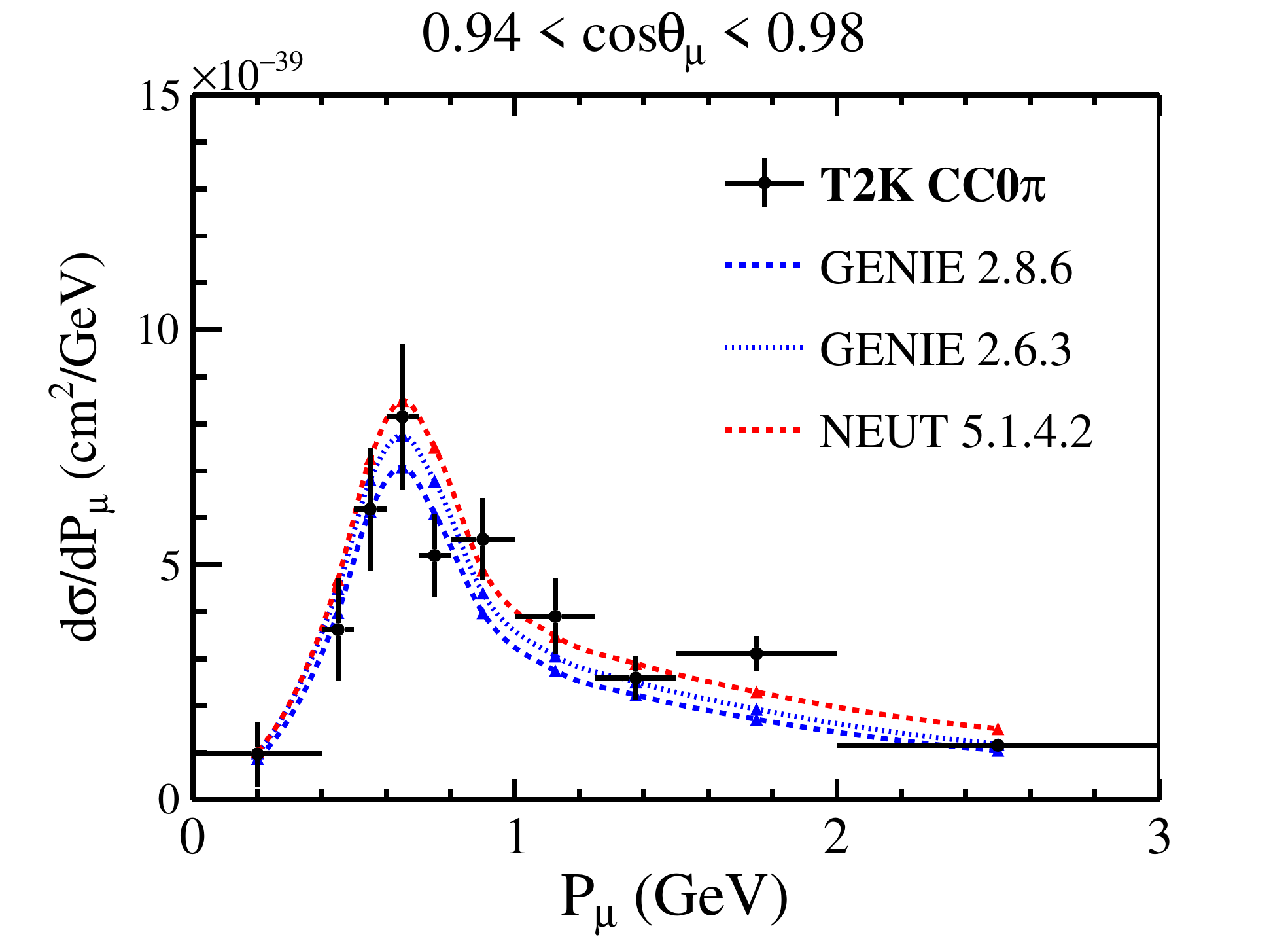} 
\includegraphics[width=0.495\textwidth]{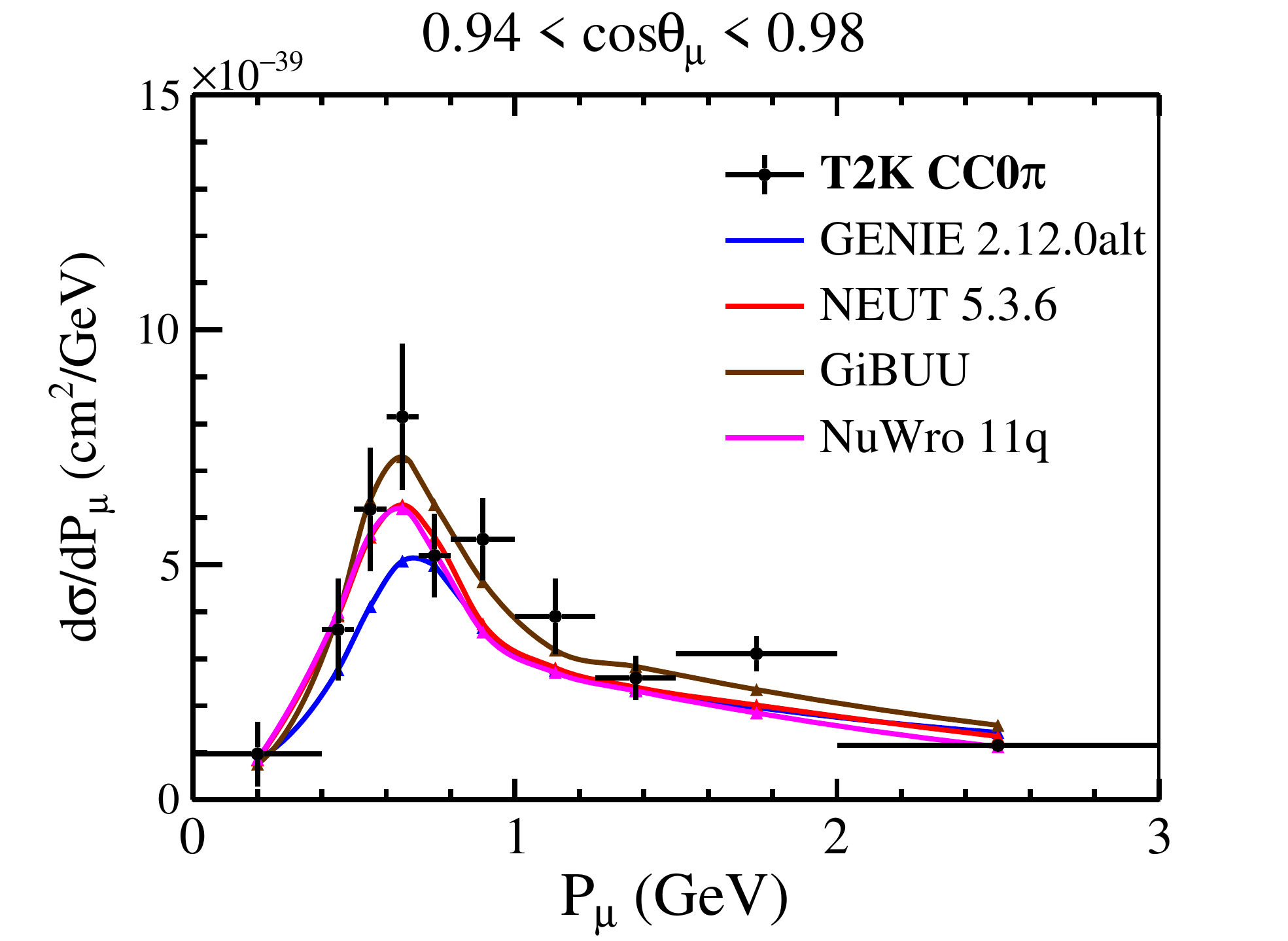}
\caption{T2K CC0$\pi$ data as a function of $p_{\mu}$ for angles 11.5$^\circ$-20$^\circ$.}
\label{fig:t2k-qe-comp2}    
\end{figure}

Figs.~\ref{fig:mbeffall}, ~\ref{fig:t2keffall}, \ref{fig:t2k-qe-comp1},and \ref{fig:t2k-qe-comp2} show the muon differential cross sections.  These have less model dependence and provide more detail. For \mb, the GiBUU calculation is in good agreement with both angular bins shown.  Recent T2K data adds to our knowledge. Figs. ~\ref{fig:t2k-qe-comp1} and \ref{fig:t2k-qe-comp2} shows the muon momentum spectrum for two different angular ranges. The NuWro v11q results are close to the NEUT results, as expected due to sharing of generator models. In general, newer models are in better agreement with T2K data than older models, and forward angular bins are described less well than those with larger angles.  The most forward angle T2K data is at very low $Q^2$.  It is interesting that the newer calculations tend to agree with the data in shape.  In magnitude, the event generators are lower than GiBUU and GiBUU has the best agreement with all data sets.  The most interesting feature is that GENIE, NEUT and NuWro disagree with the data by constant factors that would be approximately constant for each code.

Reproduction of data at very low $Q^2$ ($Q^2<0.2 (GeV/c)^2$) involves many competing effects mainly due to nuclear structure.  Data from \mb (Fig.~\ref{fig:miniboone-qe-comp}) and \minerva (Fig.~\ref{fig:minerva-qe-comp}) both have sharp dips of similar shape.  Recent models all have strong dips due to nucleon-nucleon correlations.  Although they disagree in shape with respect to both \mb and \minerva, disagreement with the T2K $p_\mu$ spectra (Fig.~\ref{fig:t2k-qe-comp2}) is more in magnitude.  In each case, better data would provide stronger constraints on models.  At high $Q^2$ ($Q^2>1 (GeV/c)^2$), short-range correlations become important.  The \mb data has the most accuracy and is described well by all newer models.

\section{Single pion production measurements} 
\label{sec:1pi}

\subsection{\mb overview}   


The signal for \mb's CC1$\pi^{+}$ measurement required an event with a single outgoing $\mu^-$ and a single $\pi^+$. 
No other mesons and any (including zero) number of photons and nucleons were allowed. No kinematic cuts were applied on the signal definition. \mb made no attempt to correct for nuclear effects to probe the initial interaction vertex. The measurement is hence an ``observable CC1$\pi^{+}$ cross section" and does not aim to isolate a particular interaction mode.  
The final sample in publication~\cite{MB_1pi} contained 48,322 events with a purity of $90.0\%$ and selection efficiency of $12.7\%$. We note that a similar but not identical sample was used by MiniBooNE to constrain backgrounds to the MiniBooNE CC0$\pi$ measurement discussed in Section~\ref{sec:qe}.

Because of their similar masses, pions and muons produce almost indistinguishable Cherenkov rings, leading to very similar likelihoods for muon and pion hypotheses. However, pions are much more likely than muons to hadronically interact in the detector. The \mb CC1$\pi^{+}$ reconstruction took advantage of this and looked for events with kinked tracks, signifying a change in the particle direction coming from hadronic scattering, i.e. elastic or inelastic scattering.  The fitter searched for events with three tracks: one from the upstream muon, one from the upstream pion, and one from the downstream (re-scattered) pion. The reconstruction simultaneously fit the kinked track ($\pi$ candidate) and straight track ($\mu$ candidate). The downstream pion candidate ring was required to have less energy than the upstream pion candidate ring. Additionally, three separate events in time were required for the CC1$\pi^+$ candidate: the first sub-event from the muon and pion discussed above; the second and third coming from decay (``Michel'') electrons  from the stopped muon and pion. 
Overall, the pion energy reconstruction performs best in the region $75 < T_\pi < 150 \text{ MeV}$. 
Although the pion energy reconstruction performs best at low energies, the opposite is true for the angular reconstruction. Below $70 \text{ MeV}$, Cherenkov-emitting pions propagate less than $10 \text{ cm}$ in \mb, which is insufficient to provide a track direction. $16\%$ of the generated NUANCE events containing pions populate this low-momentum space, compared to only $1\%$ of generated muons. The efficiency shape at higher $T_\mu$ primarily comes from the containment criteria for the muon: higher energy muons will not stop in the \mb detector so will not be selected. This is a very similar case to the MiniBooNE CC0$\pi$ case. 

The $M_{\pi N} < 1350 \text{ MeV}$ selection cut was designed to eliminate events likely to mis-reconstruct the muon as the pion and vice versa. 
This cut is present in the \mb publication~\cite{MB_1pi} but not M. Wilking's thesis~\cite{MW_thes}. Correctly matched events increase from $78.6\%$ to $88.0\%$ by introducing this cut. The $\pi+N$ mass was reconstructed using both the muon and pion kinematic variables. The calculation assumed a stationary target nucleon, in which the only outgoing particles include only a single muon, a single charged pion, and a single nucleon. In reality, Fermi-like motion of nucleons in nuclei smear the initial state energy distribution and FSI effects may give rise to secondary nucleons and/or pions as well as absorption. The $\pi+N$ mass is calculated ``post-FSI" and is not a ``true hadronic mass": it does not correspond truly to a resonance-like or DIS spectra.  The NUANCE calculation was used for $M_{\pi N} > 1350 \text{ MeV}$.

Importantly, no covariance matrix was provided with the measurement, although one was used internally. Each data point has a large uncorrelated systematic error, making data/simulation agreement difficult to assess.  \\

\begin{description}
\item [Signal definition] 1 (negatively charged) muon, any number of nucleons (neutrons or protons), one positively charged pion, no neutral pions. Signal definition is after final state interactions (FSI). 
\item [Observables] 1D $T_{\pi}$ and $T_\mu$ (and various in 2D comparisons including $cos(\theta_\mu)$ and cos($\theta_\mu$), are at their web site.  Model-dependent interaction-level variables, such as $\sigma(E_\nu)$ and $Q^2$, are also available.
\item [Flux] Ref.~\cite{mb-flux}, with digital version available from \mb data release page:~\cite{mb_datarelease}. Only neutrino interactions (not anti-neutrino or electron neutrino) were considered as signal, hence the muon charge requirement in signal definition. 
\item [Target material] $CH_2$ (mineral oil which has 2.08 H atoms per C atom)
\item [Default generator for analysis] \nuance\ \textsc{v3(\mb)}
\end{description}


\subsection{\minerva overview} 

The signal definition of the \minerva single charged pion result used in this study required exactly one charged pion ($\pi^+$ or $\pi^-$) produced in a charged current, $\nu_\mu$ interaction.  In addition to the single charged pion, any other baryons and mesons (including $\pi^0$s) were allowed. 

The signal definition also included restrictions on $E_\nu$ and $W$.  The calculated neutrino energy was required to be in the range $1.5 < E_\nu < 10 \text{ GeV}$, and invariant hadronic mass $W$ so that $W < 1.4 \text{ GeV}$.  The purpose of the restricted range for $E_\nu$ was to exclude muons that couldn't be detected in MINOS and higher energy neutrino flux which was poorly modeled at the time.  The $W$ restriction was chosen to enable comparison with lower energy measurements, e.g. MiniBooNE.  Like almost all pion production measurements, the sample includes a few percent coherent pion production events.

\minerva signal definitions have evolved in the last 3 years.  The first published analysis~\cite{Eberly:2014mra} presented both $W<1.4$ GeV (1$\pi$) and $W<1.8$ GeV (N$\pi$) analyses both with a signal of $\theta_\mu < 20 \degree$ and the full range of $\theta_\mu$.  A definition of $W$ before FSI ($W_{true}$) was used in the signal definition.  Since then, the flux has been significantly improved~\cite{Aliaga:2016oaz} and the choice of the signal $W$ was changed to the value after FSI ($W_{exp}$) which is same as an experiment would measure.  New results for $W_{exp}<1.8$ GeV were published~\cite{McGivern:2016bwh} with the updated flux and $W$ definition using the full $\theta_\mu$ range in the signal; that analysis included events with more than one charged pion.  The updated $W_{exp}<1.4$ GeV results for full $\theta_\mu$ were approved by the collaboration and posted in the \minerva data base~\cite{minerva_datarelease}.  At present, these are considered the best results and will be used in this study.



  Events must contain a charged long track (presumed to be the muon) in both the \minerva detector and the MINOS near detector (located downstream of \minerva). The tracks must match in time and space.  The charge is accurately determined from the MINOS track, an important feature of \minerva data.  This MINOS requirement effectively restricts the muon angle to $\lesssim 20\degree$.  Analyses are then further subdivided into two different signal definitions, one with $\theta_\mu < 20 \degree$ and the other with no restriction on the muon angle.  The more narrow definition limits the $Q^2$ range available in the experiment and the broader definition requires Monte Carlo to fill in the gap.

Additional cuts were applied to improve the purity.  The muon and pion must have the
same vertex.  Any pion candidate track must pass a $\frac{dE}{dx}$ quality cut with a Michel electron matched to the endpoint of the track.  With no magnetic field in the \minerva  detector, $\pi^+$ and $\pi^-$ are distinguished only by the Michel electron.  Since the $\pi^-$ come only from FSI in a CC interaction, they are a few percent of the signal and very unlikely to have a Michel electron. 

Reconstructed neutrino energy is calculated from $E_\nu = E_{recoil} + E_\mu$, where $E_{recoil}$ is the total final-state hadronic energy measured via calorimetry~\cite{Aliaga:2013uqz}.  All non-muonic energy is matched to the GENIE-GEANT detector response, accounting for the passive materials in a conventional parameterization. 
The reconstructed invariant hadronic energy, $W$, was calculated assuming a target nucleon at rest: $W^2 = {m_p}^2 - Q^2 + 2m_pE_{recoil}$ and $Q^2 = 2E_\nu(E_\mu - |\vec{p_\mu}|cos(\theta_{\mu\nu})) - {m_\mu}^2$. \minerva and \mb use the same physics formulas but \minerva measures the recoil energy. 
  
The calculation of $W$ assumes the struck nucleon is at rest, causing a model dependent smearing of about 8\% for this measurement. 
Finally, events must have at least one, but no more than two non-muonic tracks. Multi-hadron events make up a small fraction of the single charged pion signal, so this cut mostly serves to remove events with zero hadrons. \\

\begin{description}
\item [Signal definition] 1 negatively charged muon, any number of nucleons (neutrons or protons), one charged (positive OR negative) pion, any number of neutral particles (including $\pi^0$). In addition, the signal requires $1.5 < E_\nu < 10 \text{ GeV}$ and $W$(exp)$<1.4 GeV$.  Parallel analyses for $\theta_\mu<20\degree$ and full $\theta_\mu$ range were presented.
\item [Observables] 1D $T_{\pi}$ and $\theta_\pi$ are provided. Also provided are muon kinematic variables and (model-dependent) interaction-level variables, such as $E_\nu$ and $Q^2$.
\item [Flux] Ref~\cite{Aliaga:2016oaz}, with digital version available from \minerva data release page~\cite{minerva_datarelease}. Only neutrino interactions (not anti-neutrino or electron neutrino) were considered as signal, hence the muon charge requirement in signal definition.
\item [Target material] $CH$ (plastic scintillator).  Measurements are presented as cross section per target nucleon to include additional elements (e.g. TiO$_2$ scintillator coating) in a reasonable way.
\item [Default generator for analysis] GENIE v2.6.2
\end{description}

\subsection{T2K overview}  

In the T2K analysis, the signal is defined as $\nu_\mu$ charged current events on carbon or hydrogen with one positive pion in the final state, CC$1\pi^+$, and any number of nucleons in the final state. The distributions presented here correspond to events selected from the NEUT Monte Carlo sample.  This analysis is being prepared for publication.


The event selection selects inclusive CC $\nu_{\mu}$ interactions in the fiducial volume of the upstream FGD (scintillator).  Through tracking in the TPC inside a magnetic field, the charge and momentum of tracks can be determined.
The final CC$1\pi^+$-like selection cuts require the presence of one and only
one positive pion and a negative muon in the event, rejecting events with electron, positron or negative pion candidates.  For the sample shown here, the muon candidate is identified as the highest-momentum negatively-charged track.  The pion is identified as a {\it pion-like} track in the TPC based on d$E$/d$x$ measurements and curvature.  Rejection of  events with $\pi^0$s uses the TPC d$E$/d$x$ measurements and shower identification in side calorimeter detectors. 



The biggest contamination in the sample corresponds to CC events with more than one positive pion, or at least one negative or neutral pion. This contamination is mostly composed by DIS processes. For the signal, the main component at neutrino interaction level according to NEUT is the pion resonant production.

Secondary interactions of pions in the detector material posed a significant challenge for the T2K (and also MINERvA) analysis; approximately 30\% of candidate events underwent interactions. However, it was noted that the detector simulation for T2K and MINERvA are similar. GEANT4 version 4.9.4 was used for T2K, and GEANT4 4.9.4p2 for MINERvA.

In this measurement we use an improved version of the MiniBooNE formula~\cite{MB_1pi} for neutrino energy reconstruction, adding terms depending on
the pion direction and the binding energy of the target nucleon. 
The updated prescription is defined as:

\begin{equation}
E_{\nu}= \frac{m_p^2 - (m_p - E_{bind} - E_{\mu} - E_{\pi})^2 + |\overrightarrow{p}_{\mu} + \overrightarrow{p}_{\pi}|^2}{2(m_p - E_{bind} - E_{\mu} - E_{\pi} + \frac{\overrightarrow{p}_{\nu}(\overrightarrow{p}_{\mu} + \overrightarrow{p}_{\pi})}{E_{\nu}})}
\end{equation}

The full kinematics of the muon and pion are sufficient 
to fix the neutrino energy with the assumption of a single
recoiling nucleon.  This is a good assumption when the
neutrino energy is low enough that only the $\Delta(1232)$
resonance is excited.
The 4-momentum transfer, $Q^2$, is then determined as 
  $Q^2=-q^2=(p_{\mu}-p_{\nu})^2$,
where p$_{\mu}$, and p$_{\nu}$ are the 4-momentum vectors of the muon, and neutrino, respectively.
The hadronic invariant mass, W, is then
$W^2 = ((E_{\nu}+m_p) -E_{\mu})^2 - (|p_{\nu}| - |p_{\mu}|)^2$. 

\begin{description}
\item [Signal definition] 1 negatively charged muon, any number of nucleons (neutrons or protons), one positive charged pion, any number of neutral particles. 
\item [Observables] 1D $T_{\pi}$ and $\cos \theta_\pi$ are shown here. Distributions in $p_{\mu}$ and $\cos \theta_\mu$ are also provided. Model-dependent distributions such as $\sigma(E_\nu)$ and $Q^2$ and angular distributions in the $\Delta$ rest frame are also provided.
\item [Flux] Ref~\cite{t2k-flux}, with digital version available from Ref.~\cite{t2k_datarelease}. Only neutrino interactions (not anti-neutrino or electron neutrino) were considered as signal. 
\item [Target material] $CH$ (plastic scintillator)
\item [Default generator for analysis] NEUT 5.1.4.2 
\end{description}

\subsection{1$\pi$ measurement comparisons}
\label{sec:compare1pi}

Pion production analyses are complicated by the addition of a second track of a particle with a mean free path less than the nuclear size.  This complicates acceptance because topology and  kinematics are significantly altered, and possible confusion of pions and muons must be considered.  Each experimental technology has different issues.  It is interesting to note that the primary background for the \minerva measurement was pion production events above $W=1.4$ GeV feeding down into the signal region.  The primary problem in \mb was the difficulty in reconstructing low momentum particles and confusion between pions and muons at the highest energies.  For T2K, momentum reconstruction is much better than in \mb or \minerva because of the TPC; however the pion-muon separation is difficult. By using MINOS to detect muons, \minerva was able to avoid these issues.  We construct several 2-D distributions of the selected MC sample for the most relevant observables. These distributions include the efficiency (with respect to true variables) and the purity of CC$1\pi^+$ events (with respect to reconstructed variables). 

\begin{figure}[htbp]
\centering
\includegraphics[width=7.5cm]{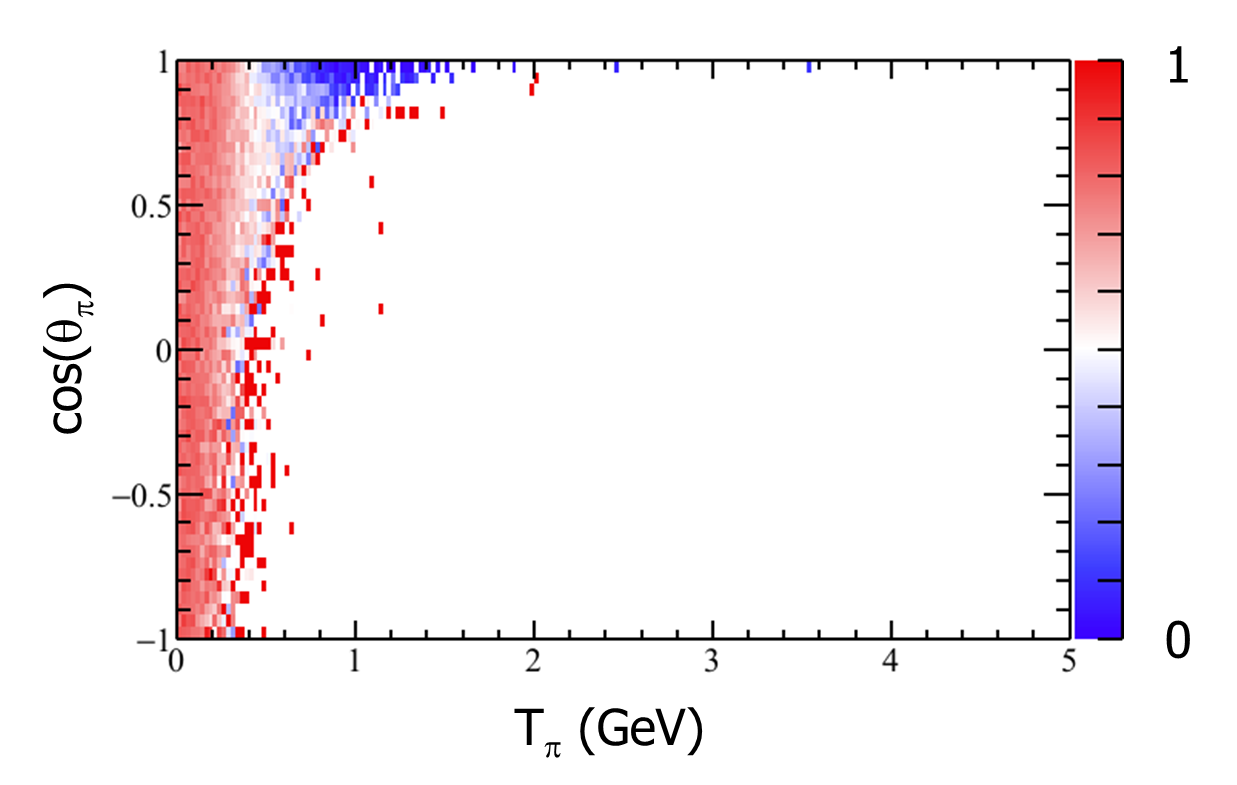}
\includegraphics[width=7.5cm]{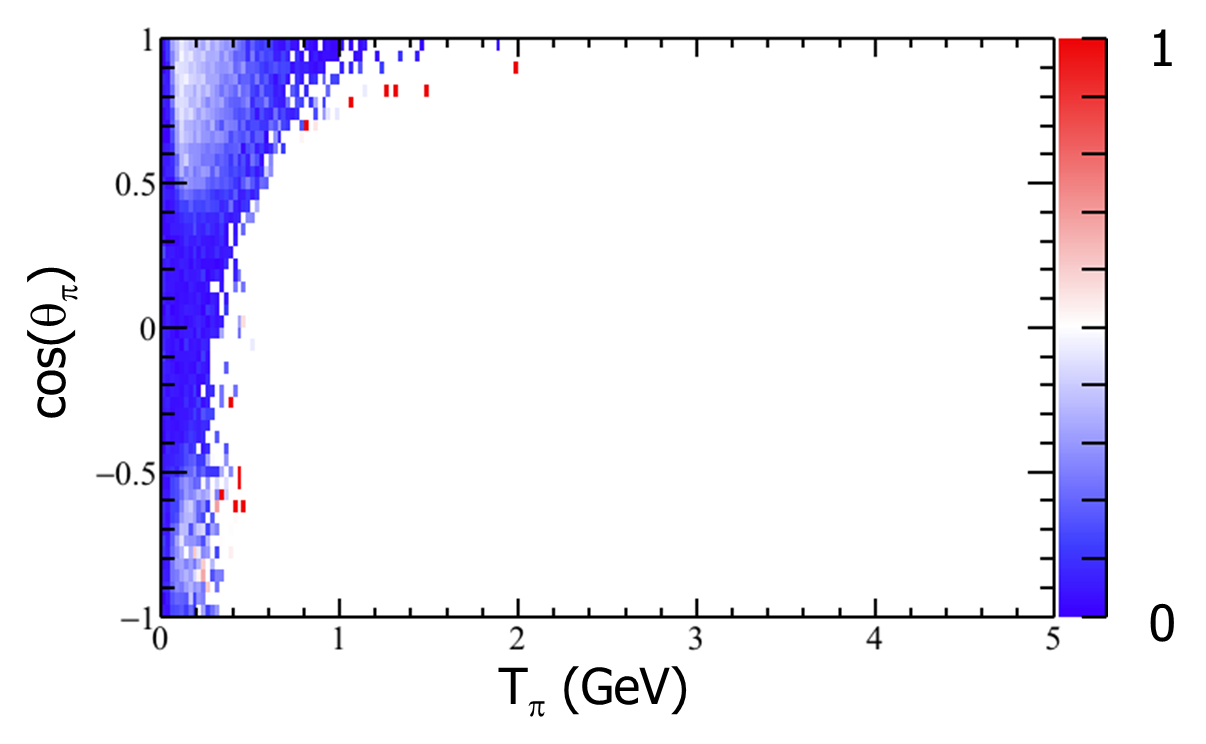}
\caption{\minerva\ 1$\pi^\pm$ measurement cut progression.  All plots show efficiency as a function of $\pi$ kinetic energy and cos($\theta_\pi$) in lab frame. (left) After establishing an event vertex in the scintillator section and properly detecting the muon in both \minerva and MINOS, the efficiency falls monotonically as the pion energy increases.  (right)  After tracking cuts have been applied, difficulties with vertical tracks are manifested as a significant dip in efficiency.}
\label{fig:minervacuts}    
\end{figure}

Fig.~\ref{fig:minervacuts} shows two $T_\pi$ vs. $\cos(\theta_\pi)$ plots at different stages of the cut application for the \minerva $1\pi$ measurement.  $T_\pi$ and $\theta_\pi$ were the primary observables of the measurement~\cite{Eberly:2014mra,McGivern:2016bwh}.  Since the sample used is Monte Carlo, some data based backgrounds are not included. The left side of Fig.~\ref{fig:minervacuts} is the result after the $W_{exp}<1.4$ GeV cut, giving a smooth dependence on these variables.  The result after application of $E_\nu$, tracking and pion $dE/dx$ cuts is shown in the right side of Fig.~\ref{fig:minervacuts}.  The main effect is a significantly lower efficiency at about 90$\degree$ due to scintillator geometry, giving potential for model dependence.  Unlike the QE case, where we found that selection choices and/or acceptance sculpted the physics (Fig.~\ref{fig:t2kmbminq0q3}), the $1\pi$ case showed that the efficiency was fairly flat for both \minerva and T2K across $Q^2$ and $W$. However, we note that both selections ended up with appreciable efficiency for certain physics regions. 

\begin{figure}[htbp]
\centering
\includegraphics[width=7.5cm]{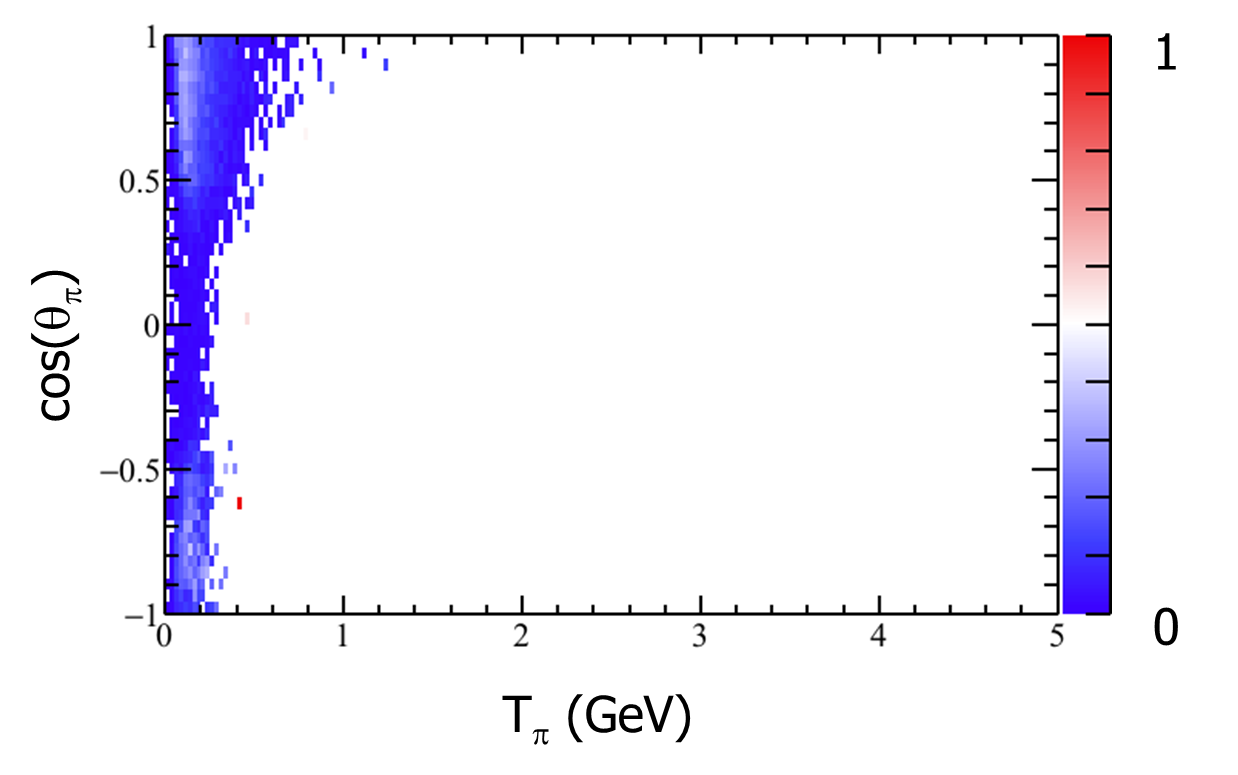}
\includegraphics[width=7.5cm]{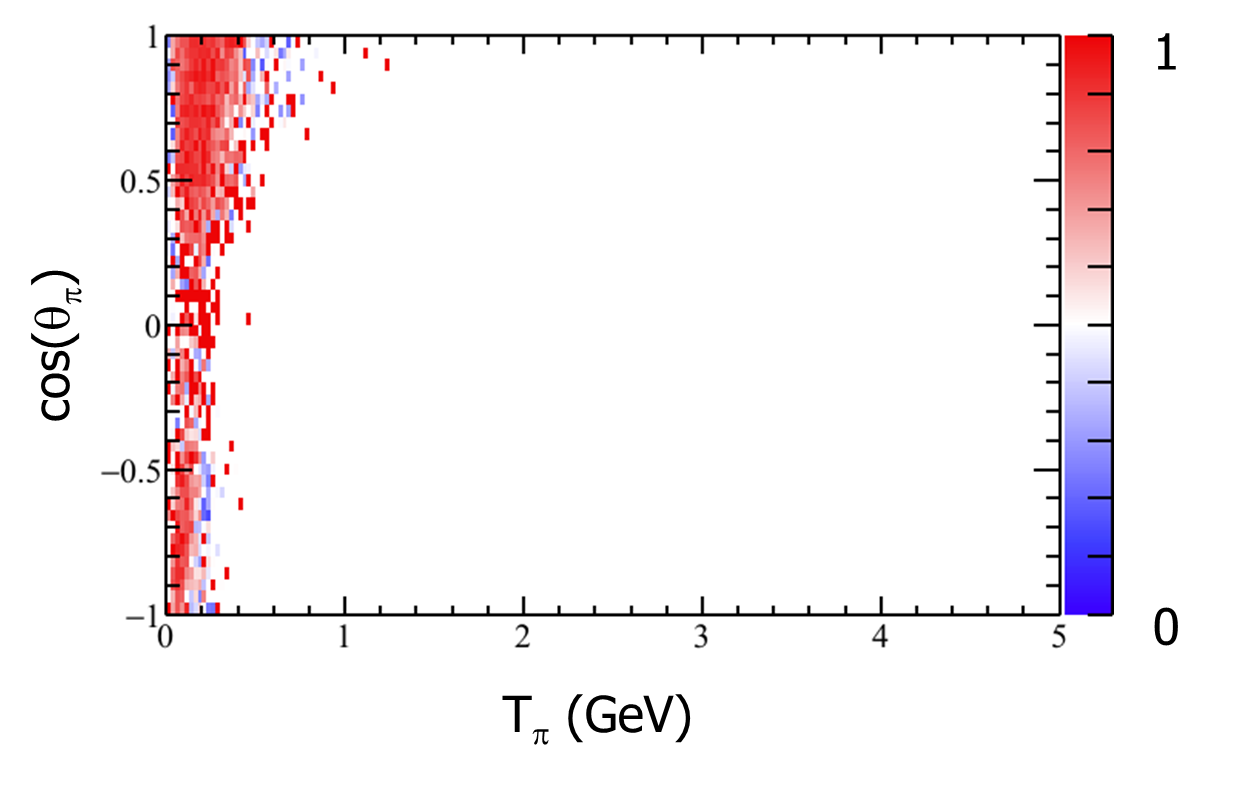}
\caption{Final Monte Carlo event sample for the \minerva\ 1$\pi^\pm$ measurement: efficiency (left) and purity (right) for events after all cuts as a function of $\pi$ kinetic energy and cos($\theta_\pi$) in lab frame.  The efficiency plot can be compared with the right side of Fig.~\ref{fig:minervacuts} to see the effects of adding particle identification.  The shape is unchanged by these cuts.}
\label{fig:mnv_1pi_effpur}    
\end{figure}
\begin{figure}[htbp]
\centering
\includegraphics[width=7.5cm]{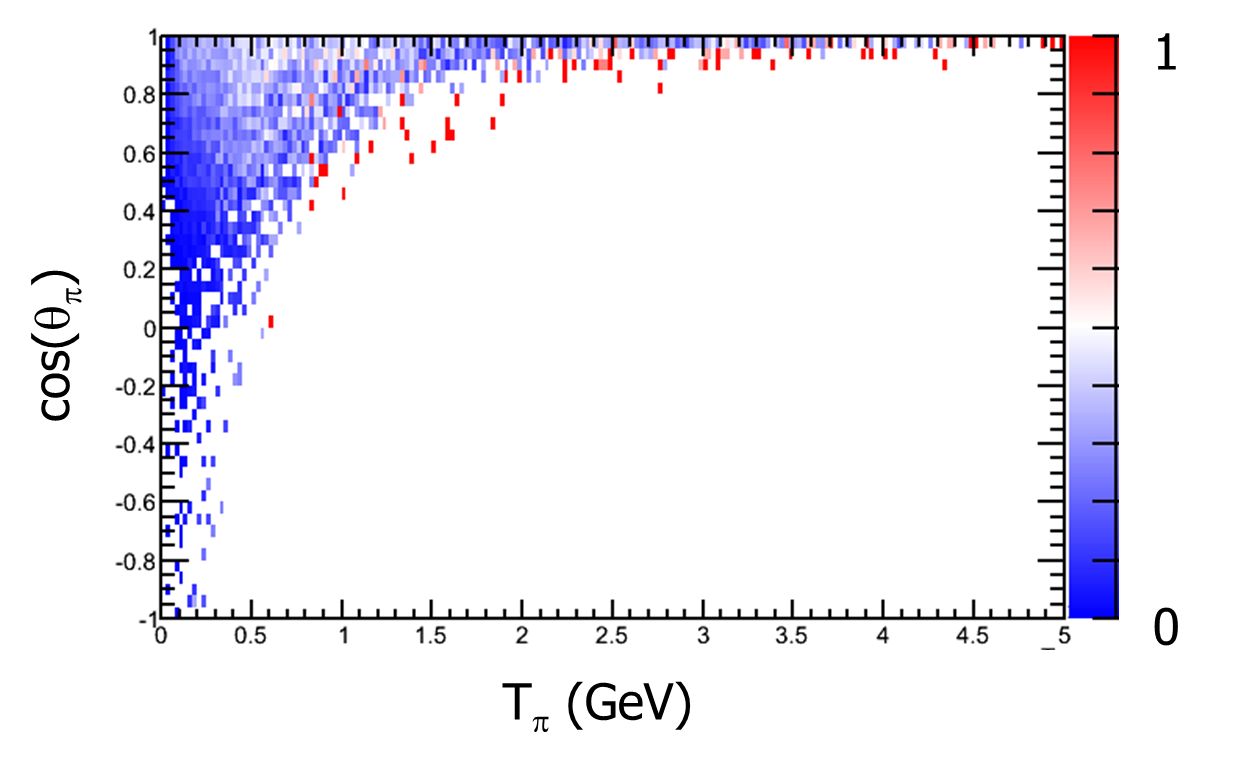}
\includegraphics[width=6.7cm]{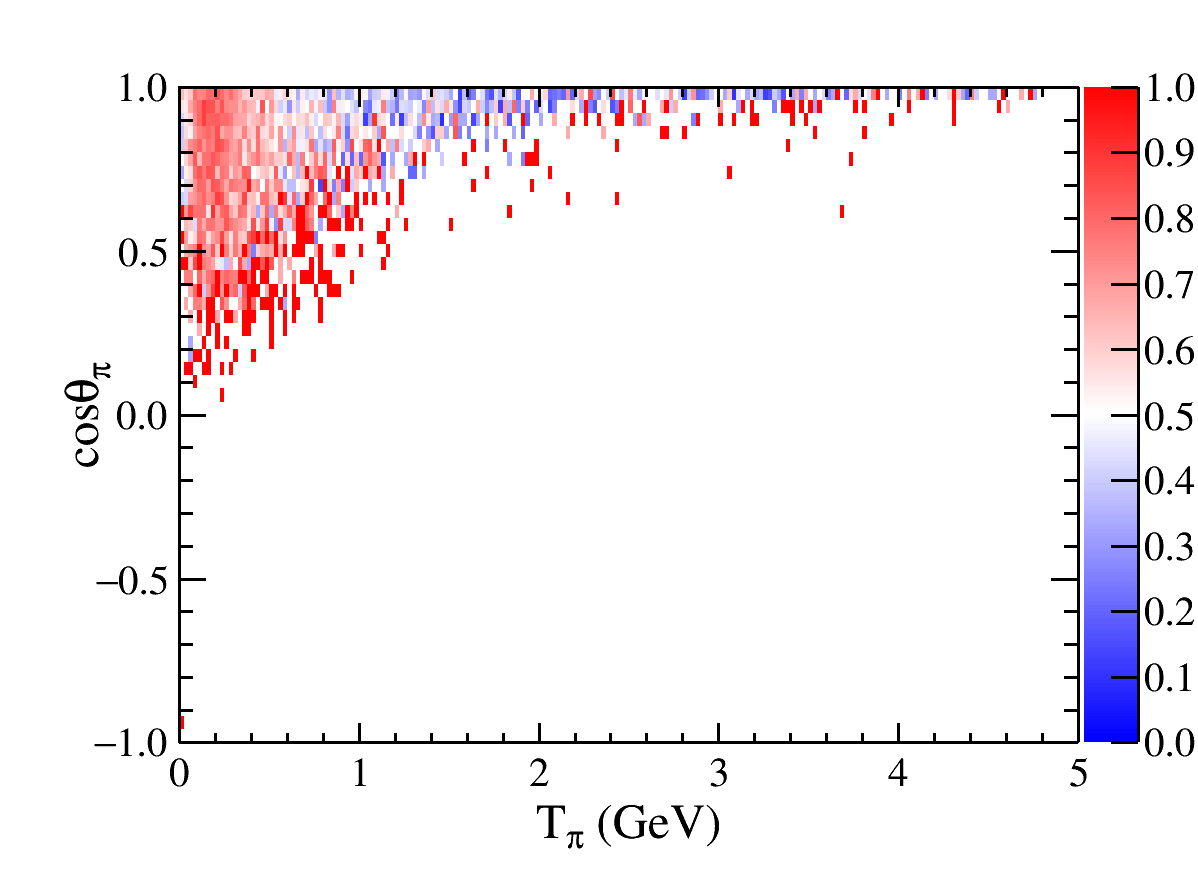} 
\caption{Final Monte Carlo event sample for T2K 1$\pi^+$ measurement: efficiency (left) and purity (right) for events after all cuts in lab frame, plotted as a function of $\pi$ kinetic energy and cos($\theta_\pi$).  It should be noted that the published \minerva data is in the $\theta_\pi$ variable and the low efficiency range applies to a smaller proportion of the range.  The difficulty in disentangling $\pi$ and $\mu$ in the TPC makes interpretation of the purity plot more complicated.}
\label{fig:t2k_1pi_effpur}    
\end{figure}
\begin{figure}[htbp]
\centering
\includegraphics[width=7.5cm]{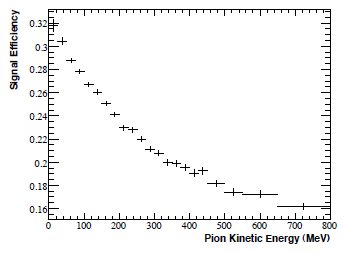}
\includegraphics[width=7.5cm]{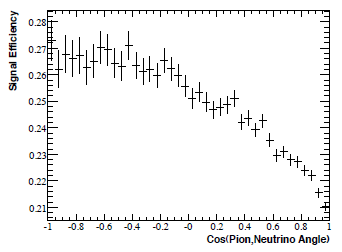}
\caption{\mb 1$\pi^+$ measurement: (left) efficiency for final data in lab frame, plotted as a function of $\pi$ kinetic energy~\cite{MW_thes}.  (right) efficiency for cos($\theta_\pi$)~\cite{MW_thes}.}
\label{fig:mb_1pi_effpur}    
\end{figure}

Figure~\ref{fig:mnv_1pi_effpur} (\minerva) and Figure~\ref{fig:t2k_1pi_effpur} (T2K) show efficiency and purity for $T_\pi$ and $cos(\theta_\pi)$ after all cuts.  The overall low efficiency is due to the cutoffs in $\theta_\mu$ (see above).  For \minerva, difficulties in reconstruction of pions that interact in the scintillator detector are significantly suppressed by requiring a Michel electron.  The inability to track particles at angles close to 90$\degree$ in \minerva is seen in the very small efficiency there.  Although \minerva has larger purity overall because of the Michel electron requirement, T2K has larger kinematic coverage.  While we were unable to obtain directly the simulation used for the MiniBooNE analysis to prepare efficiency overlays, efficiency plots from Ref.~\cite{MW_thes} are shown in Fig.~\ref{fig:mb_1pi_effpur}.  The main feature is the more uniform dependence on these variables due to the large size and isotropic acceptance of the \mb detector.  Nevertheless, reconstruction of long tracks at large pion energies are a problem.

\begin{figure}[htbp]
\centering
\includegraphics[width=7.5cm]{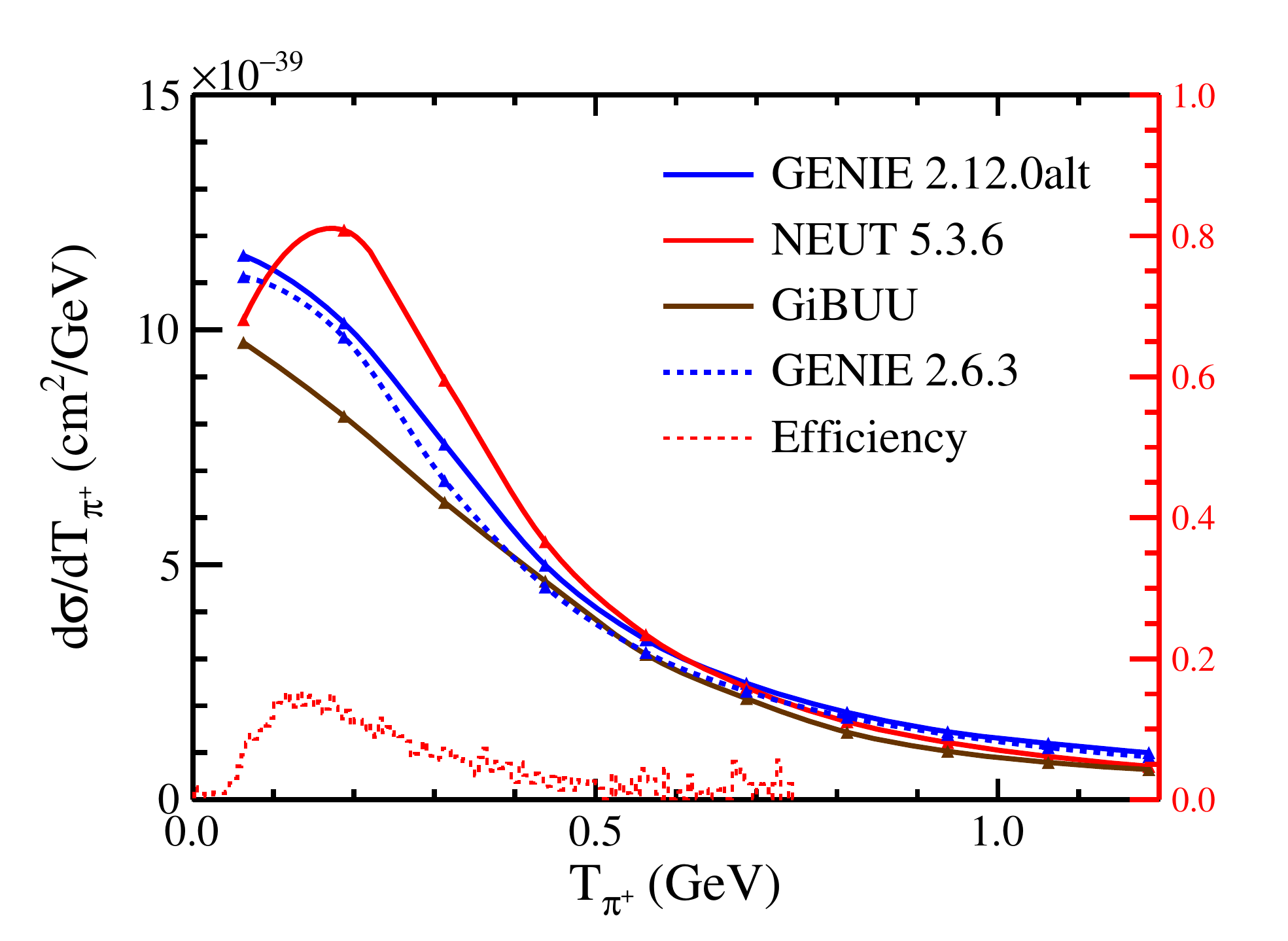} 
\includegraphics[width=7.5cm]{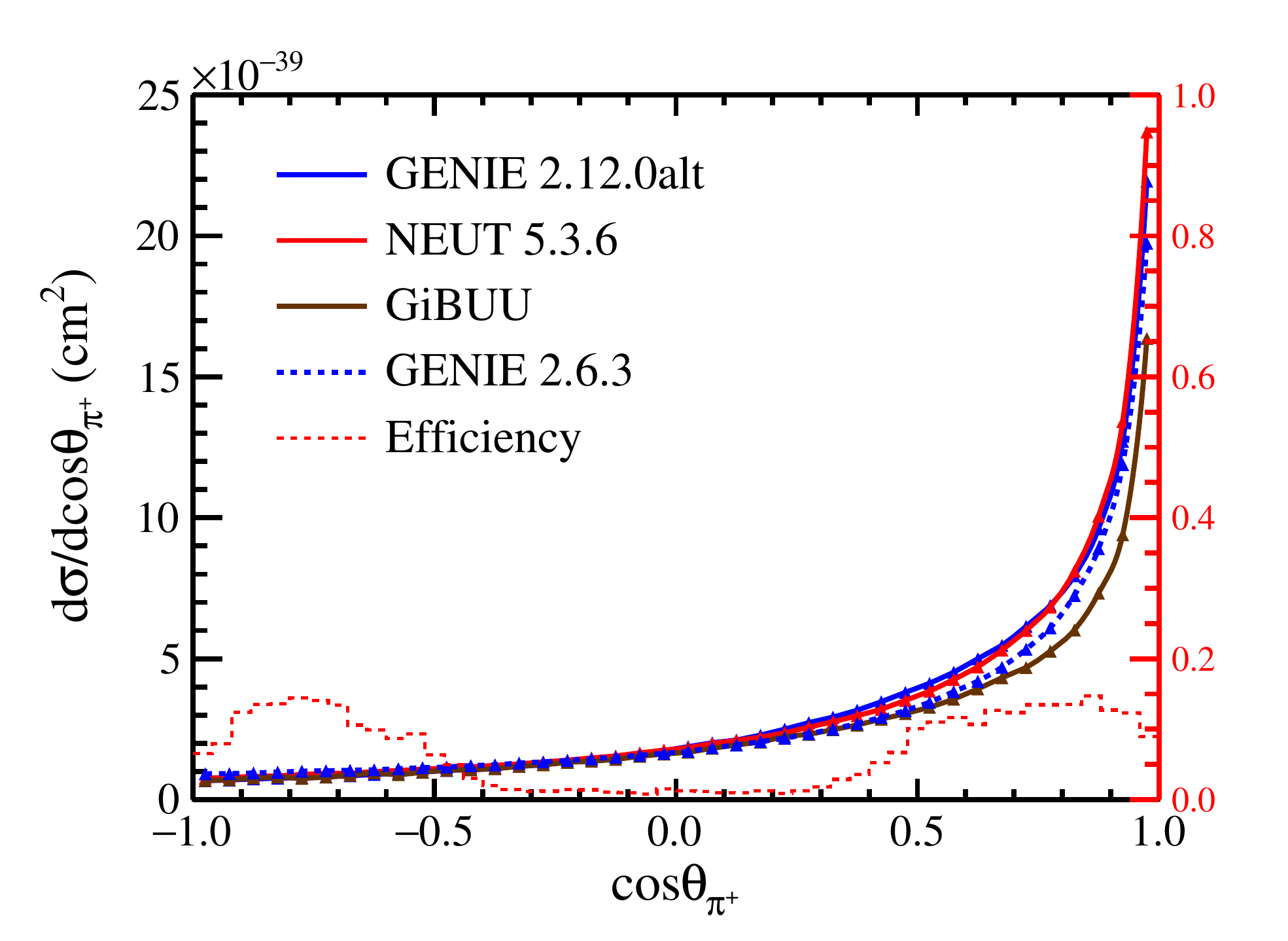}
\caption{Efficiency of the \minerva reconstruction of $T_\pi$ and $cos(\theta_\pi)$
 based on the GENIE 2.6.3. Also shown are the predictions from a selection of modern generators considered at the workshop and the generator used in the analysis (GENIE 2.6.3).}
\label{fig:pi_eff}    
\end{figure}

\begin{figure}[htbp]
\centering
\includegraphics[width=7.5cm]{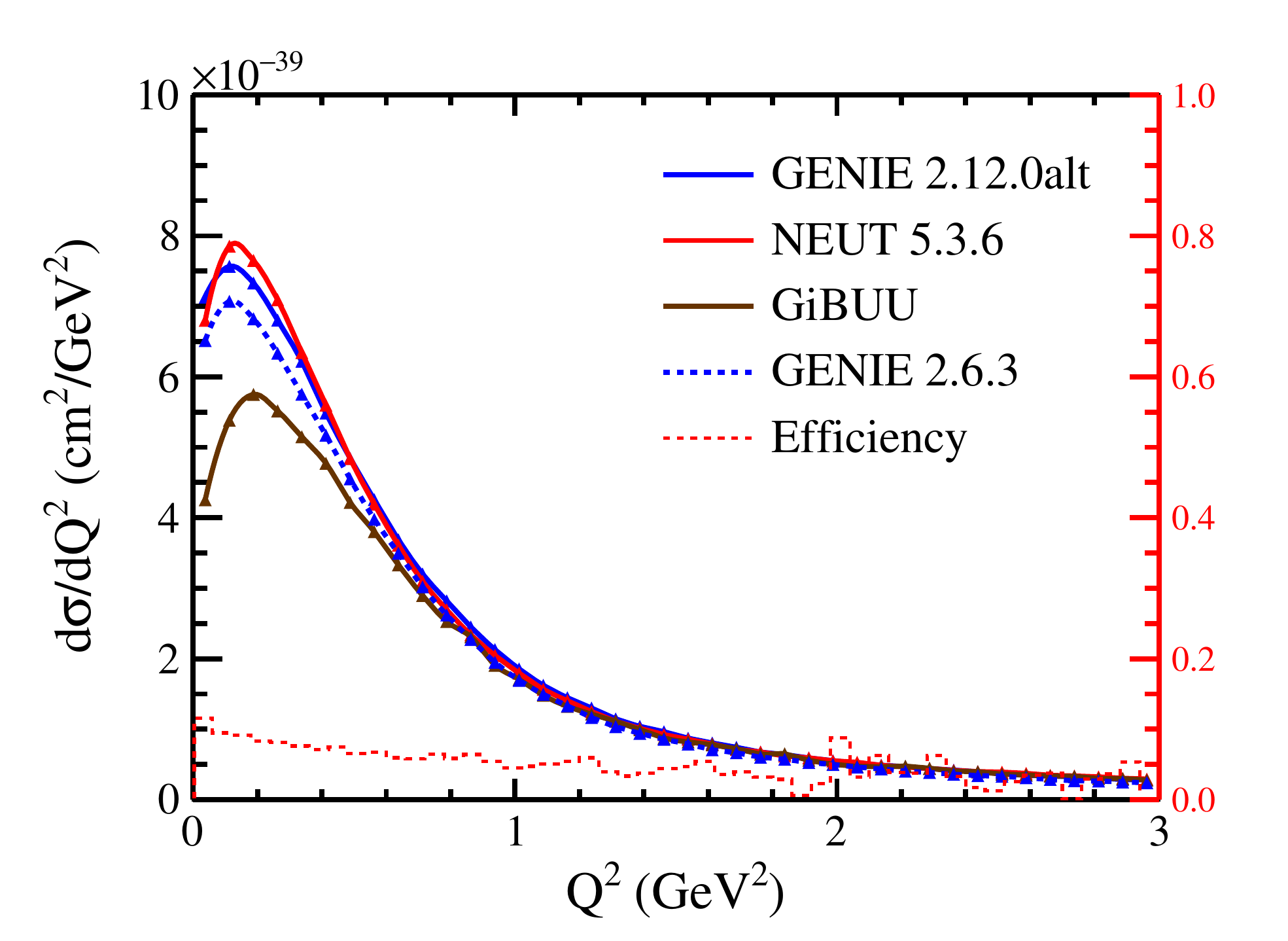} 
\includegraphics[width=7.5cm]{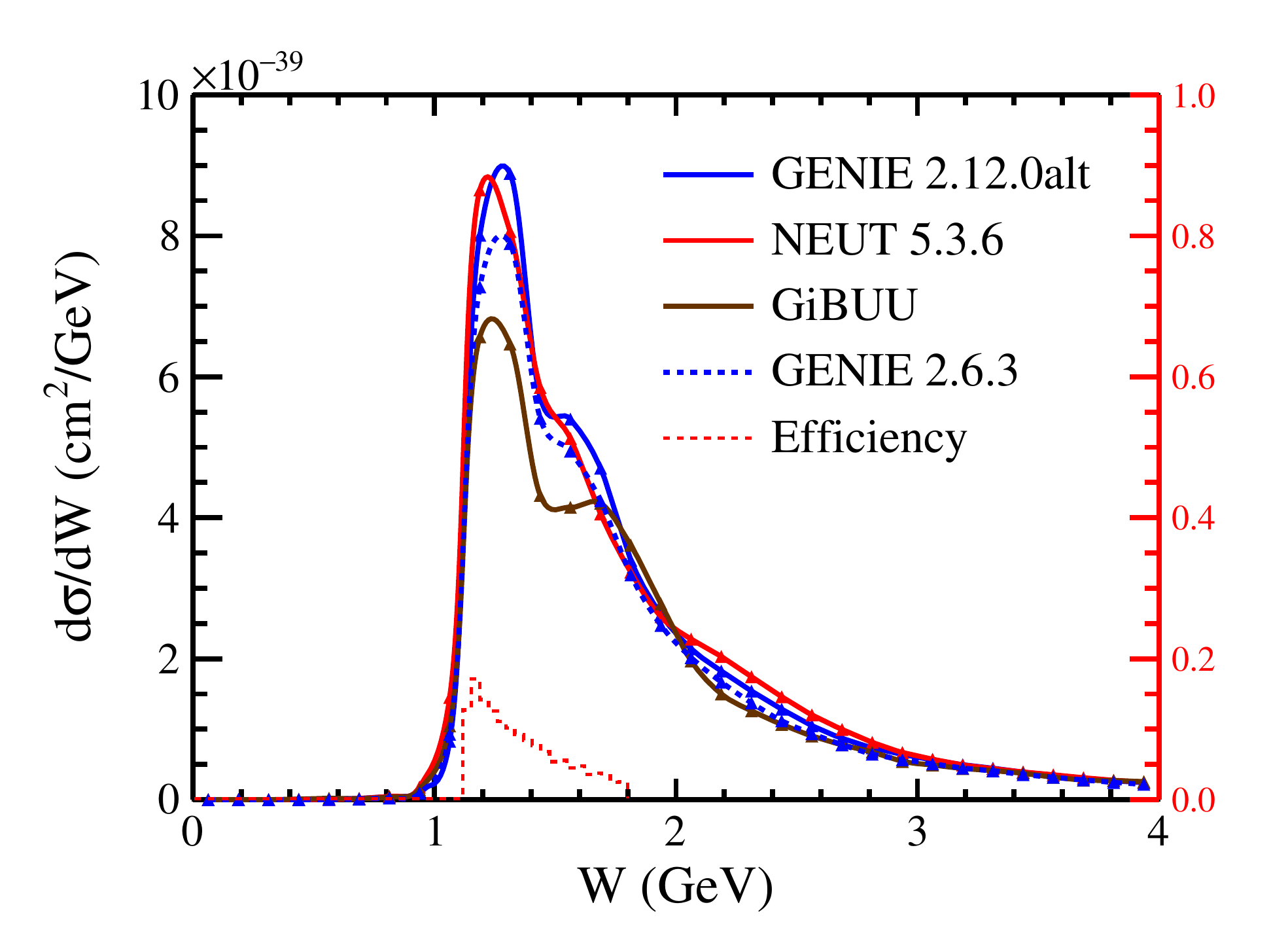}
\caption{Efficiency of the \minerva reconstruction of $Q^2$ and $W$ with NEUT and GENIE simulations. Also shown are the predictions from a selection of modern generators considered at the workshop and the generator used in the analysis (GENIE 2.6.3). }
\label{fig:q2w_pi_eff}    
\end{figure}

Fig.~\ref{fig:pi_eff} shows the \minerva $T_\pi$ and cos($\theta_\pi$) results from various Monte Carlo calculations used by the experiments overlaid with the efficiency; Fig.~\ref{fig:q2w_pi_eff} shows the same samples for physics variables $Q^2$ and $W$. Like the QE case, we see regions of low efficiency due to the response of the detector. For example, the MINERvA efficiency falls off rapidly below 50 MeV due to difficulties reconstructing short tracks as previously discussed.  The analysis cut off the $T_\pi$ distribution when the efficiency went below 1\% and used a wide bin in $\theta_\pi$ to guarantee some acceptance.  Of course, estimated errors for these data points are very large.
The model dependence of $\theta_\mu$ was tested by executing parallel analyses where the signal was all angles and only angles less than 20$\degree$.  The results differ by roughly 25\% in magnitude with a moderate change in shape; the agreement with generator predictions is very similar for the two analyses~\cite{Eberly:2014mra}.

Experiments make efforts to account for model dependence and detector uncertainties.  For pion production experiments, the set of systematic uncertainties must include all the effects discussed above in addition to a long list of other effects.





All experiments considered here, especially \minerva, must cope with 2$\pi$ background feeding down into the 1$\pi$ signal.  \minerva studied this issue~\cite{Eberly:2014mra,McGivern:2016bwh} by doing parallel analyses with signal definitions using maximum $W$ of 1.4 and 1.8 GeV.  The wider definition is much less sensitive to migration across the boundary because the cross section is much flatter at the boundary.  The $T_\pi$ and $\theta_\pi$ data show moderate change between the two signal definitions and the generator (GENIE v2.6.3) is able to account for them~\cite{Eberly:2014mra}. These data have low statistics and newer high statistics experiments must revisit this issue.

The NUANCE prediction underestimated the \mb CC$1\pi^+$ cross-section by $23\%$ on average. Notably, NUANCE also under-predicted BNL CC$1\pi^+1p$ data by a similar amount, although perhaps such experiments suffer from an error in flux normalization~\cite{Wilk-Rodr}.

\begin{figure}[htbp]
\centering
\includegraphics[width=0.495\textwidth]{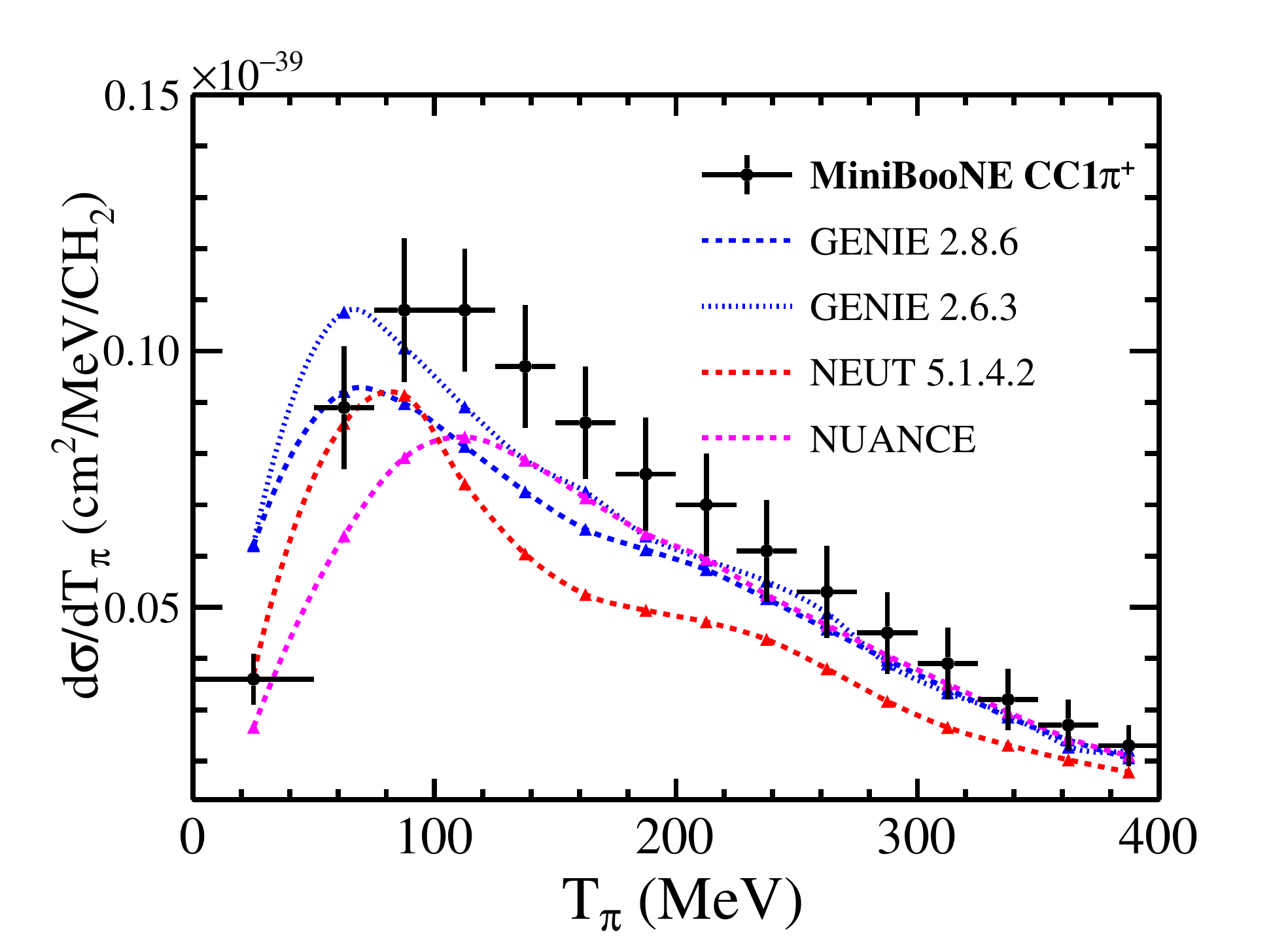} 
\includegraphics[width=0.495\textwidth]{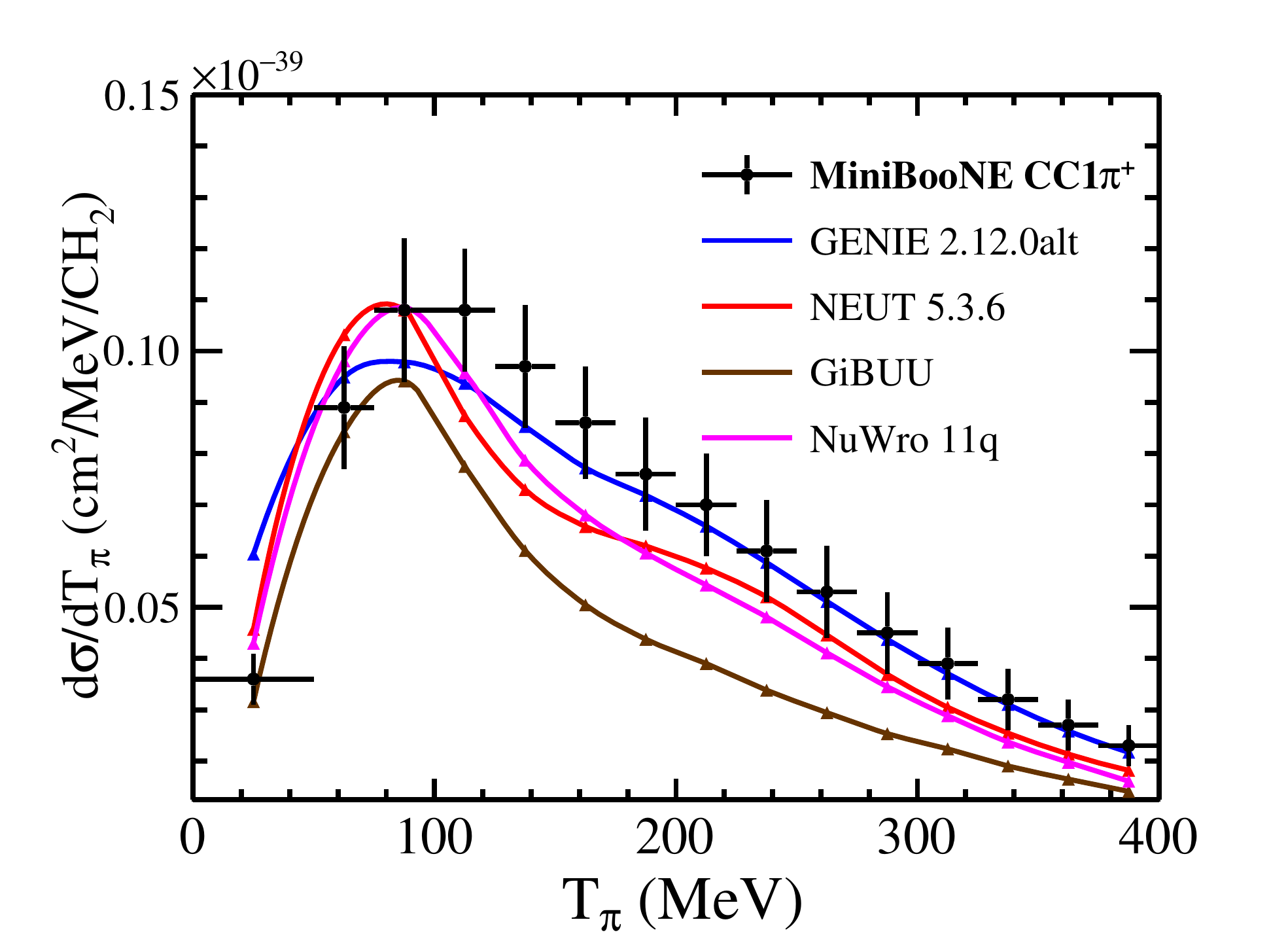}
\caption{\mb CC1$\pi$ data for $T_\pi$  with old (left) and new (right) calculations.}
\label{fig:miniboone-1pi-comp}    
\end{figure}

\begin{figure}[htbp]
\centering
\includegraphics[width=0.495\textwidth]{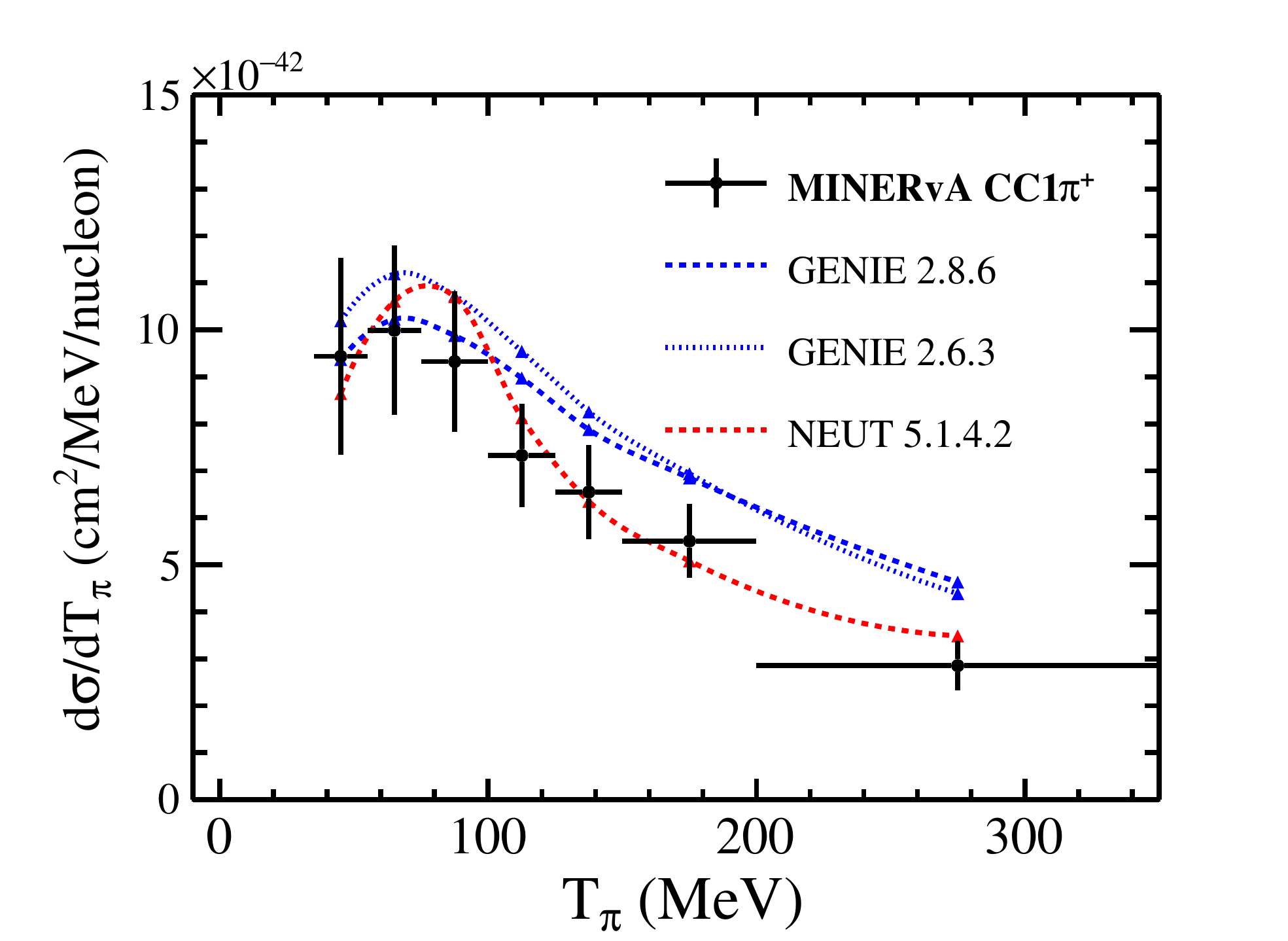} 
\includegraphics[width=0.495\textwidth]{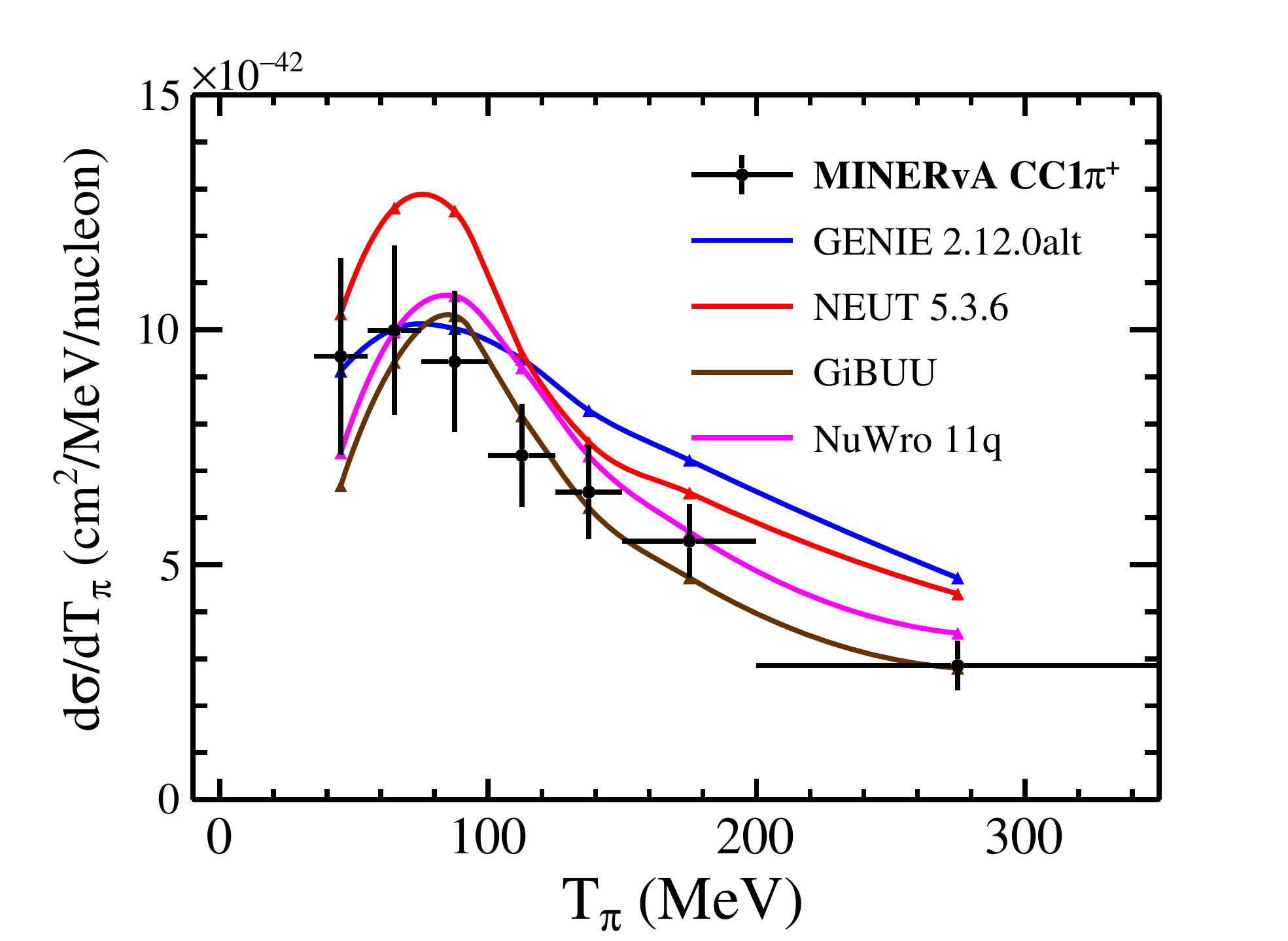}
\caption{\minerva CC1$\pi^\pm$ data for $T_\pi$ with old (left) and new (right) Monte Carlo calculations.}
\label{fig:minerva-1pi-comp}    
\end{figure}

At the time of the workshop, the \minerva 1$\pi$ measurements were being updated to use the updated flux~\cite{Aliaga:2016oaz} and improved signal definition~\cite{minerva_datarelease}.  The new signal uses a $W$ signal definition using the reconstructed value rather than the true value.  This decreases the number of events in the final sample due to smearing of signal events above the cut.  This gives approximately a 10\% reduction in the cross section.  Coupled with the 15\% increase in renormalization (also seen for the quasielastic data), there is a net 5\% increase in the cross section with little shape change.  The results for the \mb and \minerva $1\pi$ data comparison with event generator simulations are seen in Figs.~\ref{fig:miniboone-1pi-comp} and \ref{fig:minerva-1pi-comp}, respectively.  The large normalization discrepancy reported in the first \minerva publication~\cite{Eberly:2014mra} has decreased a small amount.  In addition, the new models have generally increased the cross sections, more for \minerva than for \mb.  The net result is that both the normalization and the shape discrepancy seen earlier are still present between MiniBooNE and MINERvA. 

\subsection{Model dependence}

We note the enormous amount of work by MINERvA to try to test the MiniBooNE results, including the use of the $W$ signal definition to be comparable. By using NUISANCE and a consistent set of models, we try to account for the differences between the measurements in the comparisons here. However, we still note fundamental differences in the signal model choice, and possible model dependence. 

First, the measurements have different signals. In particular, MiniBooNE and T2K veto the presence of $\pi^0$ but this is allowed in the signal definition of MINERvA. In addition, it was discovered that the signal and background categories for each measurement could be confused.  An example of this was the treatment of  CC1$\pi^-$ for the \minerva measurement.  Although the signal definition specifies both $\pi^\pm$, the \minerva selection uses a decay electron (Michel) tag, which preferentially selects $\pi^+$.  When the unfolding process is done, these events must come from Monte Carlo predictions and this adds to the model dependence.  Although the rate of CC1$\pi^-$ is low for $W<1.4$ GeV, calculations need to include this component.  We also use this as an example for analyzers for where an alternate signal definition may have been preferable. 

As in the QE-like case, care must be taken with selection cuts. MiniBooNE's reconstructed $W$ cut was used because of reconstruction considerations, not necessarily to remove high $W$, but the impact in the analysis is that this higher $W$ region is populated in the final measurement entirely by the generator used at the time (NUANCE). 
This highlights how model dependence can enter.

Both \mb and \minerva had the laudable goal of trying to provide a result that was focused on $\Delta(1232)$  production which is the dominant resonant process in all experiments with few GeV neutrinos.  This is much harder when nuclear effects must be included.  This goal must be balanced against the additional model dependence opened up.  Each experiment must assess the associated systematic errors in a reasonable way.

One largely unappreciated issue related to efficiency has to do with multi-particle final states where the efficiency is inherently highly model dependent.  This also affects 1$\pi$ measurements considered here. With two particles, the position and direction of the muon and pion both affect the efficiency; a backward pion could veto an interaction on T2K, on MINERvA a vertical muon doesn't enter the sample. In particular, the 1D projections of a given particle (e.g. pion) momentum depend on the assumed pion angular distribution and whether or not the muon was also selected. The systematic uncertainties assigned to the 1$\pi$ signal model are believed to be incomplete because we observe data disagreements with most models.  However, they are relied on to estimate the error on the efficiency propagated through the analysis. This issue poses significant challenges for future measurements of multi-particle final states.

\subsection{$1\pi$ generator considerations}
\label{sec:1picomp}
The models investigated here are not in close agreement, likely because of the added complexity in this variable as compared to the CC0$\pi$ and CCQE results in Sect.\ref{sec:qecomp}.  Models can differ due to the basic pion production mechanism (resonant or nonresonant), the nuclear model, or pion FSI.  All these aspects are difficult and are poorly characterized in previous work. This is a theory problem, not fully solved yet.  Understanding of the nucleon cross section suffers from the classic discrepancy between BNL and ANL data which is likely best fixed by a new experiment. The FSI is a complicated problem that would benefit from more theory input.  The GiBUU model has better underlying physics and excellent agreement with data.  However, the job of interfacing it to existing generators hasn't been done.  Although NEUT and NuWro have FSI with medium dependence derived from pion interaction and photoproduction data, the existing GENIE FSI model has no medium corrections.

Even though a wide variety of model choices are represented here, characterization is difficult because results come from a mix of models for the nucleon vertex, nuclear structure, and FSI.  The older calculations tend to be small compared to \mb, more accurate for \minerva.  The spread of the older calculations is about 30\%.  Newer calculations have an even wider spread, but a wider range of models is covered.  Some tuning to the \mb data is evident, but there are still problems representing the updated \minerva data.  We conclude that tensions between the data sets remain, but underlying theoretical problems still require a lot of effort. 

As noted before, NEUT and NuWro have many similarities because of overlap in authors.  Although NEUT and NuWro are very similar for the \mb kinematics, there are significant differences for the \minerva kinematics.  Despite having the best agreement with the CCQE data, the GiBUU result is well below the \mb data.  GENIE 2.8.6 is below the \mb data and a little above the \minerva data.  Addition of the Berger-Sehgal pion production model and updated form factors produces better overall agreement.

\section{Summary } 
\label{sec:summary}

The field of neutrino cross sections is in an interesting time.  A few years ago small data sets with a limited set of results were available.  In the last few years, larger datasets exploring a variety of interactions have become available.  The productive final years of \mb, a focused cross section experiment (\minerva), and a growing set of results from T2K are all impressive.  To give one example, coherent pion production measurements were historically difficult at low energies because of high backgrounds and are now commonplace.  At the same time, the challenge of making cross section measurements with broad spectrum beams has become evident, 
e.g. the difficulties in establishing a clean CCQE signal.
A novel kind of workshop was organized to bring analyzers from three experiments (\mb, \minerva, and T2K) and generator experts together for a week of exploration of the issues for experiments roughly in the neutrino energy range 0.5-5 GeV.  This encouraged candid discussion and close examination of techniques.  In lieu of data, Monte Carlo samples of various measurements were made using the cuts of the true analysis applied in software.  Thus, the distributions of important variables could be examined at various stages of the analyses. 
A parallel effort produced samples for each generator for each experiment using the published neutrino flux distributions. These could be compared in a straightforward way with a new software framework (NUISANCE). At the workshop, direct comparisons with data were prepared according to the proper experimental signal.
This gives the opportunity to confront each generator with the experimental results and to see various quantities as calculated by the different generators. 
These studies brought to light problems which should be addressed as a part of future efforts.

The first major issue discussed at the workshop was signal choice and definition in each experiment. Comparing data to any model requires a careful signal definition. It is difficult to strike a balance between the physics output desired and the related model assumptions. 

\begin{itemize}
\item We strongly support the concept of signal definition based on composition of the final state particles rather than quantities defined for the struck nucleon in the nuclear medium. \mb did the CCQE analysis both ways and the definition in terms of final state particles is now preferred, and T2K, in order to compare as much as possible to \mb, followed the same convention.  While the first \minerva $1\pi$ result~\cite{Eberly:2014mra} was defined using in-medium quantities, following \minerva analyses~\cite{McGivern:2016bwh} have adopted the convention of using final state quantities. 
\item The signal definition is linked to the response of the detector.  The convention of post-FSI signal definitions is an important advance, but does not absolve experiments of subtle model dependencies.  For the QE-like selections considered here, the T2K and \mb selections could be simple but effective as the the dominant processes are QE-like.  At the beam energies of \minerva (and DUNE), a QE selection suffers from significant additional backgrounds that require more stringent cuts; this was especially true of the QE \minerva analysis considered here which did not include a decay electron tag. 
A related consequence is that a CC$0\pi$ analysis signal definition for \minerva would include mixtures of several different processes with very different efficiencies which can bring in model dependence in the efficiency correction; the detector response to CC1$\pi^+$ with pion absorption is markedly different than CCQE for \minerva but not for \mb. The interplay between detector response and signal definition is a general concern for all analyses.
\item Experiments have put significant effort into addressing model sensitivity to the signal definition modeling. All include an extensive array of systematic errors due to parameter choices within the event generator in addition to flux and detector effects.  In T2K's case, significant work was done to produce a CC$0\pi$-like selection robust against modifications to the hadronic models available. The early \minerva CCQE analysis~\cite{Fiorentini:2013ezn} was hampered by the lack of a 2p2h model in their energy region.
Subsequent analyses on \minerva~\cite{Patrick:2018gvi} and T2K~\cite{Abe:2018pwo} have undergone extensive checks of the sensitivity of their signal analyses to mis-modeling.
\item When the hadron is detected, the issues become more complex.   The $W$ selection played an important role in the experiments reported here.  To restrict the sample to 1$\pi$, \minerva used a $W$ restriction in the signal definition.  This is a laudable goal because the $\Delta$ excitation cross section is important, but requires a tricky cut on a quantity that must be calculated (bringing in model dependence).  \mb also made a cut on $W$ to avoid confusion between $\mu$ and $\pi$ identification, but introduced model dependence by correcting for higher $W$ with a calculation.  The choice of \minerva to present results for two different $W$ cuts is important, but brings in sensitivity to new physics effects in addition to examining the cut. 
\item  In addition, a signal definition for the QE \minerva results considered here restricted values of $E_\nu$ to avoid the difficult problem on $\mu$-$\pi$ separation without a magnetic field and to avoid large flux uncertainties associated with high neutrino energies.  Again, this is a derived quantity which brings in model dependence, perhaps unnecessarily. \minerva is now abandoning the $E_\nu$ restriction~\cite{Patrick:2018gvi}. We strongly encourage experiments to avoid cuts on any inferred quantity when creating a signal definition as it only magnifies the model dependence. 
\item  In some cases the signal definition in a publication wasn't clear to members of other experiments; theorists will only find understanding more difficult to obtain.  In the \minerva 1$\pi$ measurement, the CC$\pi^-$ was in the signal but suppressed due to selection cuts favoring $\pi^+$. This results in a small correction for events that are not likely to be included in the final sample.
Signal definitions should include a clear statement of neutrino flavors considered.  The T2K and \minerva/MINOS detectors have magnetic fields so they can explicitly include $\nu_\mu$ in their signal definition.  \mb does not have a magnetic field.  They presents $\nu_\mu$  cross sections, assuming proper removal of the $\overline{\nu}_\mu$ component. While these are not enormous effects, they complicate comparisons and interpretations.
\end{itemize}


Another issue uncovered at the workshop was sensitivity to the relative kinematics of multi-particle final states. For example, in existing experiments, the efficiency of 1$\mu$+1$\pi$ topology depends on the relative angle and momenta of the two particles. The detector response has to be understood for muons which are typically forward and have a wide range of pion kinematics.
If the correlation between the pion and muon states is incorrectly modeled, then this can bias the extracted cross section.  
In general, the detector efficiency is inferred from a model specific angular and momentum distribution. Existing methods for producing systematic errors cover some of this dependence.  At the workshop, we discussed how to better visualize and mitigate this kind of problem, with the following conclusions:


\begin{itemize}
\item 
Experiments so far are comparatively simple and methods have been developed to cope with these issues.  For example, bin sizes can be adjusted to smooth out efficiency dependence. But, future experiments will have to confront increasingly complex and exclusive final states which make the estimation of efficiency a challenging problem. 
In this study, we prepared overlays of efficiency against the various models. This shows regions where model dependence may be problematic, either due to very low efficiency or rapidly changing efficiency in the detector.  

\item The impact of signal efficiency model dependence was not easily quantifiable at the workshop. The Monte Carlo samples are not always a good stand-in when one of the known problems is the inability of generators to match data well. Model dependence in the efficiency in particular, which is calculated with a single generator model, was only studied for T2K which ran full simulations of three different generators. Furthermore, some of the systematic uncertainties of the experiments' (outside the material prepared for the workshop) may cover some of these effects. Indeed, for \minerva's results, the cross section model uncertainties were relevant in the total error budget.  To give an idea of possible additional uncertainties, plots showing efficiency against various models are shown.  
\item Acceptance had a significant impact on the underlying physics reach of the individual cross section measurements. 
In the QE comparisons, \mb's flat acceptance was driven by a 4$\pi$ detector. On the other hand, both \minerva and T2K detectors have optimized acceptance in the forward direction.  \minerva accounted for this problem by using signal definitions with and without a $\theta_\mu = 20\degree$ acceptance cutoff for muons detected in MINOS. They published cross sections both with and without corrections for larger angles which aren't measured. So far, no differences in interpretation have been noted.  Where possible, applying acceptance cuts on detector-accessible variables (like $p_\mu$, or $\theta_\mu$) is recommended.  The counterbalancing effect is that the signal definition is less general.
\end{itemize}


The workshop raised important issues, but also faced logistical challenges as the first cross-collaborative workshop to focus on cross sections:
\begin{itemize}
\item We had to deal with necessarily large data sets from multiple authors.  Coordinating this work and storing the results brought significant problems, despite starting production of Monte Carlo samples three months in advance.  A central storage facility with easy access for everyone is essential for future efforts. The development of the NUISANCE framework, however, helped overcome format issues between different generator models and will be a key part of future efforts of this type.
\item Current experiments should be making plans to archive their data so that future researchers can examine it in light of future analysis improvements.  Resurrecting old analyses was a challenging task. Analyzers move on to new jobs and accessing code or Monte Carlo samples was a large problem especially for \mb analyses.    
For example, despite significant involvement from NUANCE experts, and a dedicated new person assigned to generate NUANCE event generator files, we did not succeed. It is difficult to support these kinds of efforts without the centralized structure of an active collaboration.   
\item Because of its novelty, this workshop was intentionally exploratory.  Only published results were considered to avoid commenting on unfinished work.  Even this isn't trivial, although including the PhD students in the workshop gave interesting insight into the original research.  Some methods worked and they can now be incorporated into future exercises.  Monte Carlo representations of the experiment are good for general issues but cannot reproduce details.
\item One of the major themes of the week was the uncertainties due to missing information.  Experiments all depend on Monte Carlo to fill acceptance and efficiency gaps.  By comparing various generator samples, the uncertainties can be better evaluated, but full understanding requires a very large effort (beyond the scope of this workshop).
\item Flux uncertainties are an important part of every experiment.  This was not covered in the workshop.
\end{itemize}

Despite these challenges, it was valuable to focus on the specific needs and techniques of cross section measurements in a broad format. The first real contact of this nature between the community of users who developed and interpreted these measurements turned out to be valuable.  This work can and should be expanded.

 A major purpose of this workshop was to explore the tensions between existing data sets.  Previously, this was difficult because data sets are published with a limited set of models for comparison, seldom consistent between publications.  For the first time, NUISANCE allows comparison against a series of available event generator models.  Therefore, we can provide here a consistent set of comparisons between data and generator models.  Unfortunately, NUANCE was the generator used for \mb and could not be applied in most situations.  NuWro, NEUT, and GENIE have had significant upgrades in models in the last few years.  For example, the \minerva data was published with v2.6.2 and v2.12.10 is now available.  Therefore, the effect of new nuclear, quasielastic, and pion production models are shown for \mb~\cite{MB_1pi}, \minerva~\cite{Fiorentini:2013ezn,Eberly:2014mra}, and T2K~\cite{Abe:2016tmq} data.  The \minerva data shown have also had improved flux calculations used in the data results shown here.

The older calculations cover a wide range around the data.  However, the latest calculations have much less variance, in part because they have adopted similar models, that are in better agreement with previous data.  This is especially true for the quasielastic measurements, where we see qualitative agreement with all the data sets considered. While this shows promise in understanding the underlying physics, it also highlights the critical role of signal model efficiency and associated uncertainties in the measurement. Continued validation and testing of the models, especially of the hadronic state, will better quantify and reduce this dependence.

We present comparisons of \mb and \minerva data sets for pion production through a consistent model with the correct signal definitions.
The differences first seen in the \minerva publication largely remain.  The improvement in the \minerva signal definition counteracts the flux improvement and a difference in normalization of roughly 15\% remains (depends on calculation used).  The shape of the \minerva data has changed little and the shape discrepancy remains with all generators. This remains an issue for the future.  We look forward to an independent data set from T2K at beam energy similar to \mb.  

The field benefits from multiple measurements of similar quantities, even with the differences in details reported here.  Each experiment had unique advantages in comparison with the others.  
All future measurements will hopefully benefit from the comparisons made here.  T2K and \minerva continue to produce new cross section results.  NOvA~\cite{Tsaris:2017wrk} and MicroBooNE~\cite{Devitt:2017mef} have growing data sets and are starting to produce cross sections.  Other experiments are under construction.

We recognize the rising importance of neutrino cross sections in reducing the systematic errors in neutrino oscillation experiments in the future.  The major goals of determining the neutrino mass hierarchy and lepton CP violation properties in addition to more accurate mixing angles and mass differences depend on the overall understanding of neutrino cross sections for few-GeV neutrinos.  Our goal is to highlight the importance of signal definition and careful management of generator dependence.  This will make measurements more accurate and easier to interpret.  At the same time, generator models are improving as a result of adoption of better theoretical interpretations of previous data.  Therefore, experiment and theory are building off each other to benefit all.  The interplay among experiments is an important subject and this is just the beginning. 
Effort within collaborations, or across collaborations through workshops to examine the techniques and usage of cross section measurements is necessary.  At the time this document is finished, a number of new results~\cite{Patrick:2018gvi,Abe:2018pwo} are in the process of being published which have less model dependence and better defined signals using techniques discussed here. Although significant progress is being made, another workshop addressing issues like those addressed here and other related issues is clearly needed.








\bibliography{GiBUUSection,nuwro_refs,cwret_mb1pi_refs,cwret_neut1pi_refs,sara-t2k,minervaQE,generator,nuance,missing_refs}

\end{document}